\documentclass{elsart}

\usepackage{amsfonts}
\usepackage{amsmath}
\usepackage{amssymb}
\usepackage{bm}
\usepackage{dcolumn}

\usepackage{epsfig}
\usepackage{graphicx}
\usepackage{graphics}
\usepackage[latin1]{inputenc}
\usepackage{latexsym}
\usepackage{rotating}
\usepackage{hyperref}

\usepackage{setup} 

\usepackage{fancyhdr}
 \usepackage{eucal}

\usepackage{lineno}

\journal{Physics Reports}

\begin{document}

\newcommand\be{\begin{equation}}
\newcommand\ba{\begin{eqnarray}}
\newcommand\bal{\begin{align}}
\newcommand\eal{\end{align}}
\newcommand\ee{\end{equation}}
\newcommand\ea{\end{eqnarray}}
\newcommand{\comment}{\bf}
\newcommand{\pont}{{\,^\ast\!}R\,R}

\newcommand{\bb}{\mbox{\boldmath$b$}}
\newcommand{\qq}{\mbox{\boldmath$q$}}
\newcommand{\mm}{\mbox{\boldmath$g$}}
\newcommand{\vm}{{\vec{m}}}
\newcommand{\vn}{{\vec{n}}}
\newcommand{\vl}{{\vec{l}}}

\def\hatn{{\bf \hat n}}
\def\hatnprime{{\bf \hat n'}}
\def\hatnone{{\bf \hat n}_1}
\def\hatntwo{{\bf \hat n}_2}
\def\hatni{{\bf \hat n}_i}
\def\hatnj{{\bf \hat n}_j}
\def\vecx{{\bf x}}
\def\veck{{\bf k}}
\def\hatx{{\bf \hat x}}
\def\hatk{{\bf \hat k}}
\def\hatz{{\bf \hat z}}
\def\VEV#1{{\left\langle #1 \right\rangle}}
\def\cP{{\cal P}}
\def\noise{{\rm noise}}
\def\pix{{\rm pix}}
\def\map{{\rm map}}

\def\o{\over}
\def\a{\alpha}
\def\lmat{\left( \matrix{ }
\def\rmat{ }\right) }
\def\Pl{{\rm Pl}}
\def\hp{h_+}
\def\hc{h_{\times}}
\def\half{\frac{1}{2}}
\def\a{\alpha}
\def\lmat{\left( \matrix{ }
\def\rmat{ }\right) }
\def\yboxit#1#2{\vbox{\hrule height #1 \hbox{\vrule width #1
    \vbox{#2}\vrule width #1 }\hrule height #1 }}
\def\fillbox#1{\hbox to #1{\vbox to #1{\vfil}\hfil}}
\def\ybox{\yboxit{0.4pt}{\fillbox{8pt}}\hskip-0.4pt}
\def\VEV#1{\langle{ #1} \rangle}
\newcommand{\fixme}[1]{\textbf{FIXME: }$\langle$\textit{#1}$\rangle$}
\def\sss{\scriptscriptstyle}

\def\hp{h_+}
\def\hc{h_{\times}}
\def\yboxit#1#2{\vbox{\hrule height #1 \hbox{\vrule width #1
    \vbox{#2}\vrule width #1 }\hrule height #1 }}
\def\fillbox#1{\hbox to #1{\vbox to #1{\vfil}\hfil}}
\def\ybox{\yboxit{0.4pt}{\fillbox{8pt}}\hskip-0.4pt}

\newcommand{\bomega}{\mbox{\boldmath$\omega$}}
\newcommand{\RRdual}{\mbox{\boldmath$ R \tilde R$}}
\newcommand{\Rdual}{\mbox{$\tilde R$}} 

\newcommand{\mb}[1]{\mbox{\boldmath $#1$}}
\newcommand{\ZM}{{\mbox{\tiny ZM}}}
\newcommand{\CPM}{{\mbox{\tiny CPM}}}
\newcommand{\CS}{{\mbox{\tiny CS}}}
\newcommand{\GR}{{\mbox{\tiny GR}}}
\newcommand{\AXIAL}{{\mbox{\tiny Axial}}}
\newcommand{\POLAR}{{\mbox{\tiny Polar}}}

\newcommand{\bse}{\begin{subequations}}
\newcommand{\ese}{\end{subequations}}
\newcommand{\barr}{\begin{array}}
\newcommand{\earr}{\end{array}}

\newcommand{\mP}{M_{\textrm{Pl}}}
\newcommand{\ie}{\emph{ie}}
\newcommand{\lsim}{\lesssim}
\newcommand{\gsim}{\gtrsim}

\begin{frontmatter}

\title{Chern-Simons Modified General Relativity}

\author[Haverford,PSU]{Stephon Alexander} and \ead{sha3@psu.edu}
\author[PRIN,PSU]{Nicol\'as Yunes}\ead{nyunes@princeton.edu}
\address[Haverford]{Department of Physics and Astronomy, Haverford College, Haverford, PA 19041}
\address[PRIN]{Department of Physics, Princeton University, Princeton, NJ 08544, USA}  
\address[PSU]{Institute for Gravitation and the Cosmos,
  Department of Physics, The Pennsylvania State University, University
  Park, PA 16802, USA}

\begin{abstract}



Chern-Simons modified gravity is an effective extension of general relativity that captures leading-order, gravitational parity violation. Such an effective theory is motivated by anomaly cancelation in particle physics and string theory.  In this review, we begin by providing a pedagogical derivation of the three distinct ways such an extension arises: (1) in particle physics, (2) from string theory and (3) geometrically.  We then review many exact and approximate, vacuum solutions of the modified theory, and discuss possible matter couplings. Following this, we review the myriad astrophysical, solar system, gravitational wave and cosmological probes that bound Chern-Simons modified gravity, including discussions of cosmic baryon asymmetry and inflation. The review closes with a discussion of possible future directions in which to test and study gravitational parity violation.

\end{abstract}

\begin{keyword}
Chern \sep Simons \sep string theory \sep loop quantum gravity \sep parity violation
\PACS code \sep code
\end{keyword}

\end{frontmatter}

\newpage

\tableofcontents

\newpage

\section{Introduction}

Over the last two decades we have experienced a wealth of observational data in the field of cosmology and astrophysics that has been critical in guiding physicists to our current fundamental theories. The combined CMBR~\cite{Dunkley:2008ie}, lensing~\cite{Wittman:2000tc}, large scale structure~\cite{Markevitch:1997ic} and supernovae data~\cite{Riess:2004nr} all point to an early universe scenario where at matter radiation equality, the universe was dominated by radiation and dark matter.  Soon after, the universe underwent a transition where it became dominated by a mysterious fluid, similar to a cosmologcal constant (if it is not evolving), named dark energy.   Both dark matter and dark energy might be a truly quantum gravitational effect or simply a modification of General Relativity (GR) at large distances.  

In the absence of a full quantum gravitational theory, how do we come about constructing a representative effective model? One of the most important unifying concepts in modern physics is the gauge principle, which in fact played a seminal role in the unification of the strong, weak and electromagnetic interactions  through the requirement that the action be invariant under a local $SU(3)\times SU(2)\times U(1)_{Y}$ gauge transformation.  Given the success of the gauge principle, much effort has been invested in various branches of theoretical and mathematical physics, unveiling deep connections between gauge theories and geometry. The modern route to such gauge geometrical picture is via the interpretation of Yang-Mills gauge fields as connections on a principal fibre bundle and the Riemann tensor as the curvature on the tangent bundle. Although the independent combination of the Standard Model of particle physics and GR accounts for all four observed forces, a complete unification is still lacking, in spite of their semblance as gauge theories.

Gauge principles invariably point us in the direction of a peculiar, yet generic modification to GR that consists of the addition of a Pontryagin or ``Chern-Simons'' (CS) term to the action. Due to its gauge principle roots, such an extension connects many seemingly unrelated areas of physics, including gravitational physics, particle physics, String Theory and Loop Quantum Gravity. This effective theory is in contrast with other GR modifications that are not motivated by predictive elements of a more fundamental theoretical framework.  One of the goals of this report is to explore the emergence of CS modified gravity from these theoretical lenses and to confront model-independent predictions with astrophysics, cosmology and particle physics experiments. 

Many have argued that studying an effective theory derived from String Theory or particle physics is futile because, if such a correction to the Einstein-Hilbert action were truly present in Nature, it would be quantum suppressed. In fact, String Theory does suggest that the coupling constant in front of the CS correction should be suppressed at least at the electroweak scale level or even the Planck scale level. Indeed if this were the case, the CS correction would be completely undetectable by any future experiments or observations.

Other quantities exist, however, that String Theory predicts should also be related to the Planck scale, yet several independent measurements and observations suggest this is not the case. One example of this is the cosmological constant, which according to String Theory should be induced by supersymmetry breaking. If this breakage occurs at the electroweak scale, then the value of the cosmological constant should be approximately $10^{45} \; {\textrm{eV}}^{4}$, while if it occurs at the Planck scale it should be about $10^{112} \; {\textrm{eV}}^{4}$. We know today that the value of the cosmological constant is close to $10^{-3} \; {\textrm{eV}}^{4}$, no-where near the String Theory prediction. 

A healthy, interdisciplinary rapport has since developed between the cosmology and particle communities. On the one hand, the astrophysicists continue to make more precise and independent measurements of the cosmological constant. On the other hand, the particle and String Theory communities are now searching for new and exciting ways to explain such an observable value, thus pushing their models in interesting directions.  

Similarly, a healthy attitude, perhaps, is to view the CS correction as a model-independent avenue to investigate parity violation, its signatures and potential detectability, regardless of whether {\emph{some models}} expect this correction to be Planck suppressed. In fact, there are other models that suggest the CS correction could be enhanced, due to non-perturbative instanton corrections~\cite{Svrcek:2006yi}, interactions with fermions~\cite{Alexander:2008wi}, large intrinsic curvatures~\cite{Alexander:2007vt} or small string couplings at late times~\cite{Sun:2006sd,Wesley:2005bd,Brandenberger:2001:lpi,Battefeld:2005av,Brandenberger:2006vv,Brandenberger:2007zz,Brax:2004xh,Alexander:2000:bgi,Tseytlin:1991xk,Nayeri:2005ck}.  

After mathematically defining the effective theory [Section~\ref{ABC}], we shall discuss its emergence in the Standard Model, String Theory and Loop Quantum Gravity [Section~\ref{Many-Faces_v1}]. We begin by reviewing how the CS gravitational term arises from the computation of the chiral anomaly in the Standard Model coupled to GR.  While this anomaly is cancelled in the Standard Model gauge group, we shall see that it persists in generic Yang-Mills gravitational theories.  We then present the Green-Schwarz anomaly canceling mechanism and show how CS theory arises from String Theory, leading to a Pontryagin correction to the action in four-dimensional gravity coupled to Yang-Mills theory. The emergence of CS gravity in Loop Quantum Gravity is then reviewed, as a consequence of the scalarization of the Barbero-Immirzi parameter in the presence of fermions.  

Once the effective model has been introduced and motivated, we shall concentrate on exact and approximate solutions of the theory both in vacuum and in the presence of fermions [Sections~\ref{Sec:ExactSol},~\ref{Sec:ApproxSol} and~\ref{Fermions}]. We shall see that although spherically symmetric spacetimes that are solutions in GR remain solutions in CS modified gravity, axially-symmetric solutions do not have the same fate. In the far field, we shall see that the gravitational field of a spinning source is CS modified, leading to a correction to frame-dragging. Moreover, the propagation of gravitational waves is also CS corrected, leading to an exponential enhancement/suppression of left/right-polarized waves that depends on wave-number, distance travelled and the entire integrated history of the CS coupling. 

With such exact and approximate solutions, we shall review the myriads of astrophysical and cosmological tests of CS modified gravity [Sections~\ref{Sec:AstroTests} and~\ref{SEC:CScosmology}].  We shall discuss Solar-System tests, including anomalous gyroscopic precession, which has led to the first experimental constraint of the model. We shall then continue with a discussion of cosmological tests, including anomalous circular polarization of the CMBR. We conclude with a brief summary of leptogenesis in the early universe as explained by the effective theory. 

This review paper consists of a summary of fascinating results produced by several different authors.  Overall, we follow mostly the conventions of~\cite{Carroll:1997ar,Carroll:2004st}, which are the same as those of~\cite{waldgeneral} and~\cite{Misner:1973cw}, unless otherwise specified. In particular, 
Latin letters at the beginning of the alphabet $a,b,\ldots,h$ correspond to spacetime indices, while those at the end of the alphabet $i,j,\ldots,z$ stand for spatial indices only. Sometimes $i,j,\ldots,z$ will instead stand for indices representing the angular sector in a $2+2$ decomposition of the spacetime metric, but when such is the case the notation will be clear by context. Covariant derivatives in four (three) dimensions are denoted by $\nabla_a$ ($D_i$) and partial derivatives by $\partial_a$ ($\partial_i$). The Levi-Civita tensor is denoted by $\epsilon^{abcd}$, while $\tilde{\epsilon}^{abcd}$ is the Levi-Civita symbol, with convention $\tilde\epsilon^{0123} = +1$. The notation $[A]$ stands for the units of $A$ and $L$ stands for the unit of length, while the notation ${\cal{O}}(A)$ stands for a term of order $A$. Our metric signature is $(-,+,+,+)$ and we shall mostly employ geometric with $G = c = 1$, except for a few sections where natural units shall be more convenient $h = c = 1$.

%



\section{The ABC of Chern-Simons and its Tools}
\label{ABC}

\subsection{Formulation}
\label{Formulation}
%
%
%

CS modified gravity is a $4$-dimensional deformation of GR, postulated 
by Jackiw and Pi~\cite{Jackiw:2003pm}~\footnote{Similar versions of this theory were 
previously suggested in the context of string theory~\cite{Campbell:1990fu,Campbell:1992hc}, 
and three-dimensional topological massive gravity~\cite{Deser:1981wh,Deser:1982vy}.}. The modified theory can be defined in 
terms of its action:
\be
\label{CSaction}
S := S_{\rm EH} + S_{\rm CS} +  S_{\vartheta} + S_{\rm mat},
\ee
where the Einstein-Hilbert term is given by
\be
\label{EH-action}
S_{\rm{EH}} = \kappa \int_{{\cal{V}}} d^4x  \sqrt{-g}  R, 
\ee
the CS term is given by
\be
\label{CS-action}
S_{\rm{CS}} = +\alpha \frac{1}{4} \int_{{\cal{V}}} d^4x  \sqrt{-g} \; 
\vartheta \; \pont,
\ee
the scalar field term is given by
\be 
\label{Theta-action}
S_{\vartheta} = - \beta \frac{1}{2} \int_{{\cal{V}}} d^{4}x \sqrt{-g} \left[ g^{a b}
\left(\nabla_{a} \vartheta\right) \left(\nabla_{b} \vartheta\right) + 2 V(\vartheta) \right],
\ee
and an additional, unspecified matter contribution are described by 
\be
S_{\textrm{mat}} = \int_{{\cal{V}}} d^{4}x \sqrt{-g} {\cal{L}}_{\textrm{mat}},
 \ee
 where ${\cal{L}}_{\textrm{mat}}$ is some matter Lagrangian density that does not depend on $\vartheta$. In these equations, $\kappa^{-1} = 16 \pi G$, $\alpha$ and $\beta$ are dimensional coupling constants, $g$ is the determinant of the metric, $\nabla_{a}$ is the covariant derivative associated with $g_{ab}$, $R$ is the Ricci scalar, and the integrals are volume ones carried out everywhere on the manifold ${{\cal{V}}}$. The quantity $\pont$ is the Pontryagin density, defined via
\be
\label{pontryagindef}
\pont:= R \tilde R = {\,^\ast\!}R^a{}_b{}^{cd} R^b{}_{acd}\,,
\ee
where the dual Riemann-tensor is given by
\be
\label{Rdual}
{^\ast}R^a{}_b{}^{cd}:=\frac12 \epsilon^{cdef}R^a{}_{bef}\,,
\ee
with $\epsilon^{cdef}$ the 4-dimensional Levi-Civita tensor. Formally, $\pont \propto R \wedge R$, but here the curvature tensor is assumed to be the Riemann (torsion-free) tensor. We shall discuss in Sec.~\ref{Fermions} the formulation of CS modified gravity in first-order form. 

Unfortunately, since its inception, CS modified gravity has been studied with slightly different coupling constants. We have here attempted to collect all ambiguities in the couplings $\alpha$ and $\beta$. Depending on the dimensionality of $\alpha$ and $\beta$, the scalar field will also have different dimensions. Let us for example let 
$[\alpha] = L^{A}$ , where $A$ is any real number. If the action is to be dimensionless (usually a requirement when working in natural units), it then follows that $[\vartheta] = L^{-A}$, which also forces $[\beta] = L^{2A - 2}$. Different sections of this review paper will present results with slightly different choices of these couplings, but such choices will be made clear at the beginning of each section. A common choice is $\alpha = \kappa$ and $\beta = 0$, leading to $[\alpha] = L^{-2}$ and $[\vartheta] = L^{2}$, which was used in~\cite{Jackiw:2003pm,Alexander:2004xd,Alexander:2004us,Alexander:2004wk,Alexander:2006mt,Alexander:2007qe,Alexander:2007zg,Alexander:2007vt,Alexander:2007kv,Yunes:2007ss,Grumiller:2007rv,pulsars}. On the other hand, when discussing Solar System tests of CS modified gravity, another common choice is $\alpha= - \ell/3$ and $\beta = -1$, where $\ell$ is some length scale associated with $\vartheta$~\cite{Smith:2007jm}, which then implies $[\alpha] = L$, $[\vartheta] = L^{-1}$ and $[\beta] = 1$

But is there a natural choice for these coupling constants? A minimal, practical and tempting choice is $\alpha = 1$, which then implies that $\vartheta$ is dimensionless and that $[\beta] = L^{-2}$, which suggests $\beta \propto \kappa$\footnote{When working in geometrized units, a dimensionless $\vartheta$ can still be achieved if $\beta \sim \kappa$, but here $\kappa$ is dimensionless, thus pushing all dimensions into $\alpha$, which now possess units $[\alpha] = L^{2}$}. From a theoretical standpoint, the choice of coupling constant does matter because it specifies the dimensions of $\vartheta$ and could thus affect its physical interpretation.  For example, a coupling of the form $\alpha \propto  \kappa^{-1}$ suggest $S_{CS}$ is to be thought of as a Planckian correction, since $G = \ell_{p}^{2}$, where $\ell_{p}$ is the Planck length. On the other hand, if one wishes to study the CS correction on the same footing as the Einstein-Hilbert term, then it is more convenient to let $\alpha = \kappa$ and push all units into $\vartheta$. By leaving the coupling constants unspecified with $\alpha$ and $\beta$ free, we shall be able to present generic expressions for the modified field equations, as well as particular results present in the literature by simply specifying the constants chosen in each study. 

The quantity $\vartheta$ is the so-called {\emph{CS coupling field}, which is not a constant, but a function of spacetime, thus serving as a {\emph{deformation function}}. Formally, if $\vartheta = \textrm{const.}$~CS modified gravity reduces identically to GR. This is because the Pontryagin term [Eq.~\eqref{pontryagindef}] can be expressed as the divergence
\be
\nabla_a K^a = \frac{1}{2} \pont 
\label{eq:curr1}
\ee
of the Chern-Simons topological current 
\be
K^a :=\epsilon^{abcd} \Gamma^n_{bm} \left(\partial_c\Gamma^m_{dn}+\frac{2}{3} \Gamma^m_{cl}\Gamma^l_{dn}\right)\,,
\label{eq:curr2}   
\ee
where here $\Gamma$ is the Christoffel connection. One can now integrate $S_{\textrm{CS}}$ by parts to obtain
\be
\label{CS-action-K}
S_{\textrm{CS}} =  \alpha 
\left( \vartheta \; K^{a} \right)|_{\partial {\cal{V}}}
-
 \frac{\alpha}{2} \int_{{\cal{V}}} d^4x  \sqrt{-g} \; 
\left(\nabla_{a} \vartheta \right) K^{a},
\ee
where the first term is usually discarded since it is evaluated on the boundary of the manifold\footnote{The implications of discarding this boundary term will be discussed in Sec.~\ref{Sec:Boundary}}. The second term clearly depends on the covariant derivative of $\vartheta$, which vanishes if $\vartheta = \textrm{const.}$ and, in that case, CS modified gravity reduces to GR. 
 
For any finite, yet arbitrarily small $\nabla_{a} \vartheta$, CS modified gravity becomes substantially different from GR. The quantity $\nabla_{a} \vartheta$ can be thought of as an {\emph{embedding coordinate}},  because it embeds a generalization of the standard $3$-dimensional CS theory into a $4$-dimensional spacetime. In this sense, $\nabla_{a} \vartheta$ and $\nabla_{a} \nabla_{b} \vartheta$ act as deformation parameters in the phase space of all theories. One can then picture GR as a stable fixed point in this phase space. Away from this ``saddle point,'' CS modified gravity induces corrections to the Einstein equations that are proportional to the steepness of the $\vartheta$ deformation parameter. 
 
The equations of motion of CS modified gravity can be obtained by variation of the action in Eq.~\eqref{CSaction}. Exploiting the well-known relations
\be
\delta R^b{}_{acd}=\nabla_c\delta\Gamma^b_{ad}-\nabla_d\delta\Gamma^b_{ac}
\ee
and
\be
\delta\Gamma^b_{ac} = \frac12 g^{bd}\left(\nabla_a\delta
  g_{dc}+\nabla_c\delta g_{ad}-\nabla_d\delta g_{ac}\right)\,,
\ee
one finds 
\ba
\label{variationofS}
\delta S &=& \kappa \int_{{\cal{V}}} d^4x \sqrt{-g}
  \left(R_{a b} - \frac{1}{2} g_{ab} R + \frac{\alpha}{\kappa} C_{ab} - \frac{1}{2\kappa} T_{ab} \right) \delta g^{ab} 
\nonumber \\
&+&  \int_{{\cal{V}}} d^4x \sqrt{-g}  \left[\frac{\alpha}{4}\pont + \beta \square \vartheta  - \beta \frac{dV}{d\vartheta}  \right] \delta\vartheta
\nonumber \\
&+& \Sigma_{EH} + \Sigma_{CS} + \Sigma_{\vartheta}\,
\ea
where $\square := g^{ab} \nabla_{a} \nabla_{b}$ is the D'Alembertian operator and $T_{ab}$ is the total stress-energy tensor, defined via
\be
T^{ab} = - \frac{2}{\sqrt{-g}} \left( \frac{\delta {\cal{L}}^{\textrm{mat}}}{\delta g_{ab}} + \frac{\delta {\cal{L}}^{\vartheta}}{\delta g_{ab}} \right),
\ee
where ${\cal{L}}_{\vartheta}$ is the Lagrangian density of the scalar field action, {\emph{ie.~}}the integrand Eq.~\eqref{Theta-action} divided by $\sqrt{-g}$, such that $S_{\vartheta} = \int_{{\cal{V}}} {\cal{L}}_{\vartheta} d^{4}x $. Thus, the total stress-energy tensor can be split into external matter contributions $T^{ab}_{\textrm{mat}}$ and a scalar field contribution, which is explicitly given by 
\be
\label{Tab-theta}
T_{ab}^{\vartheta} 
=   \beta  \left[  \left(\nabla_{a} \vartheta\right) \left(\nabla_{b} \vartheta\right) 
    - \frac{1}{2}  g_{a b}\left(\nabla_{a} \vartheta\right) \left(\nabla^{a} \vartheta\right) 
-  g_{ab}  V(\vartheta)  \right].
\ee

The tensor $C_{ab}$ that appears in Eq.~\eqref{variationofS} is a $4$-dimensional generalization of the $3$-dimensional Cotton-York tensor, which in order to distinguish it from the latter we shall call the {\emph{C-tensor}}\footnote{In the original work of~\cite{Jackiw:2003pm}, the C-tensor was incorrectly called ``Cotton tensor'', but the concept of a higher-dimensional Cotton-York tensor already exists~\cite{Garcia:2003bw} and differs from the definition of Eq.~\eqref{Ctensor}}. This quantity is given by
\be
\label{Ctensor}
C^{ab} := v_c
\epsilon^{cde(a}\nabla_eR^{b)}{}_d+v_{cd}{\,^\ast\!}R^{d(ab)c}\,,   
\ee
where 
\be
\label{v}
v_a:=\nabla_a\vartheta\,,\qquad
v_{ab}:=\nabla_a\nabla_b\vartheta=\nabla_{(a}\nabla_{b)}\vartheta 
\ee 
are the velocity and covariant acceleration of $\vartheta$, respectively. The last line of Eq.~\eqref{variationofS} represents surface terms that arise due to repeated integrations by parts.  Such terms play an interesting role for the thermodynamics of black hole solutions, which we shall review in Sec.~\ref{Sec:Boundary}.

The vanishing of Eq.~\eqref{variationofS} leads to the equations of motion of CS modified gravity. The equations of motion for the metric degrees of freedom (the modified field equations) are simply
\be
\label{eq:eom}
 G_{ab} + \frac{\alpha}{\kappa} C_{ab} = \frac{1}{2\kappa}  T_{ab},
\ee
where $G_{ab}=R_{ab}-\frac12 g_{ab}R$ is the Einstein tensor.  The trace-reversed form of the modified field equations 
\be
\label{eom}
R_{ab} + \frac{\alpha}{\kappa} C_{ab} = \frac{1}{2 \kappa} \left(T_{ab} - \frac{1}{2} g_{ab} T \right),
\ee
can be derived by noting that the C-tensor is in fact symmetric and traceless, where $T=g^{ab}T_{ab}$ is the trace of the total stress-energy tensor. Thus, it follows that as in GR the modified field equations must also satisfy  
\be
R = - \frac{1}{2 \kappa} T  = 0\,,
\ee
where the right-hand side holds in the absence of matter. 

The vanishing of the variation of the action also leads to an extra equation of motion for the CS coupling field, namely 
\be 
\label{eq:constraint}
\beta \; \square \vartheta = \beta \; \frac{dV}{d\vartheta} - \frac{\alpha}{4} \pont,
\ee
which we recognize as the Klein-Gordon equation in the presence of a potential and a sourcing term. 
One then sees that the evolution of the CS coupling is not only governed by its stress-energy tensor, but also by the curvature of spacetime. 
One can in fact also derive this equation from the requirement of energy-momentum conservation:
\be
\label{energy-cons}
\nabla^{a} \left(G_{ab} + C_{ab} \right) = \frac{1}{2\kappa}  \nabla^{a} T_{ab},
\ee
where the first term on the left-hand side identically vanishes by the Bianchi identities, while the second is proportional to the Pontryagin density via
\be
\label{nablaC}
\nabla_a C^{ab} = - \frac{1}{8} v^b \pont.
\ee
Equation~\eqref{eq:constraint} is then established from Eq.~\eqref{energy-cons}, provided external matter degrees of freedom satisfy $\nabla_{a} T^{ab}_{\textrm{mat}} = 0$.
Alternatively, Eq.~\eqref{energy-cons} also tells us that, provided the scalar field satisfies its evolution equation [Eq.~\eqref{eq:constraint}], then the strong equivalence principle is satisfied since matter follows geodesics determined by the conservation of its stress-energy tensor.

At this junction, one might worry if this set of equations is well-posed as an initial value problem, ie.~that given
generic initial data, there exists a unique and stable solution that is continuous on the initial data (a small change in the initial
state leads to a small change in the final state). A restricted class of modified CS theories have already been shown 
to be well-posed as a Dirichlet boundary value problem~\cite{Grumiller:2008ie}, via the construction
of a Gibbons-Hawking-York boundary term (see Sec.~\ref{Sec:Boundary}). In principle, these results can 
easily be generalized to initial or final boundaries (Cauchy hypersurfaces), by treating the case where
the normal vector to the boundary is timelike, and to generic CS field $\vartheta$, since the addition
of kinetic or potential terms should not modify the analysis of~\cite{Grumillerprivcomm,Grumiller:2008ie}.
Such conclusions thus imply that given an initial state, there exists a unique final state in CS modified gravity.

The issue well-posedness of the theory as an initial value problem, however, remains still formally open, since the above
arguments cannot necessarily be used to demonstrate that the theory is stable. Such instabilities are rooted
in the potential appearance of third time derivatives in the equations of motion. As regards to these higher-order
derivatives, notice first that for $v^{a} = {\textrm{const.}}$ such derivatives do not arise and the modified 
field equations remain second-order. Moreover, notice that even for generic $\vartheta$, third time derivatives
also vanish in a linear stability analysis, as these derivatives are multiplied by terms at least quadratic in the
metric perturbation. Non-linear stabilities, however, could arise upon the full non-linear evolution of the modified
field equations, a topic that is currently being investigated. 
 
Formally, the CS modified equations of motion presented here [Eqs.~\eqref{eq:eom} and~\eqref{eq:constraint}] represent a family of theories, parameterized by the couplings $\alpha$ and $\beta$. Of this family, two classes or formulations are particularly interesting: the {\emph{non-dynamical}} framework ($\alpha$ arbitrary, $\beta = 0$) and the {\emph{dynamical}} framework ($\alpha$ and $\beta$ arbitrary but non-zero).  These two formulations are actually two distinct theories, because in the dynamical formulation the scalar field introduces stress-energy into the modified field equations, which in turn forces vacuum spacetimes to possess a certain amount of ``scalar hair.'' On the other hand, such hairy spacetimes are absent from the non-dynamical formulation, but this one instead acquires an additional differential constraint that might overconstrain it. 

One of the benefits of introducing the coupling constants $\alpha$ and $\beta$ is that we can easily specify if we are 
considering the dynamical ($\alpha \neq 0 \neq \beta$) or the non-dynamical ($\alpha \neq 0$, $\beta = 0$) formulation. 
In this review article, we shall attempt to present as many generic expressions with $\alpha$ and $\beta$ 
unspecified as possible, but when summarizing existing results we will have
to focus on one specific formulation. On average, the non-dynamical formulation has been investigated much more
than the fully-dynamical one, which is why the presentation might seem slightly biased toward the non-dynamical
theory. One should remember that this bias is not because the non-dynamical theory is preferred, but only because
it is easier to study than the fully-dynamical scenario. It is critical then to pay close attention to the beginning of 
each section in the remainder of this review article,as we shall specify whether the results that are being presented 
correspond to the dynamical or the non-dynamical theory by specifying the choice of $\alpha$ and $\beta$ 
(an issue that is particularly relevant when discussing solutions to the modified theory in Secs.~\ref{Sec:ExactSol} 
and~\ref{Sec:ApproxSol}). 

\subsection{Non-dynamical CS gravity and the Pontryagin Constraint}

The non-dynamical framework is defined by setting $\beta = 0$ at the level of the action, such that the scalar field does not evolve dynamically, but is instead externally
prescribed. Such was the formulation introduced by Jackiw and Pi~\cite{Jackiw:2003pm}, with the particular choice $\alpha = \kappa$ and $\beta = 0$, which implies $[\vartheta] = L^{2}$. 

Within this non-dynamical model, there is a particular choice of $\vartheta$, proposed by Jackiw and Pi~\cite{Jackiw:2003pm}, that has been used extensively:
\be
\label{canon}
\vartheta = \frac{t}{\mu} \quad \rightarrow \quad v_{\mu} = \left[\frac{1}{\mu},0,0,0\right]\,.
\ee
where $\mu$ is some mass scale, such that $[\mu] = L^{-1}$. We shall refer to Eq.~\eqref{canon} as the {\emph{canonical CS coupling}}. This choice of CS scalar is popular because for certain sufficiently symmetric line elements (eg.~the Schwarzschild metric),  the $4$-dimensional C-tensor reduces exactly to the ordinary $3$-dimensional Cotton tensor.  Moreover, with this choice, spacetime-dependent reparameterization of the spatial variables and time translation remain symmetries of the CS modified action~\cite{Jackiw:2003pm}. In spite of this, there is nothing truly 
``canonical'' about this {\emph{choice}} of embedding coordinate and other interesting choices are also possible. 

Irrespective of the choice $\vartheta$, non-dynamical CS modified gravity, is a constrained theory, in the sense that all solutions must satisfy an additional differential condition, sometimes referred to as the {\emph{Pontryagin constraint}}: 
\be
\pont = 0.
\ee 
This constraint arises directly from the variation of the action in Eq.~\eqref{variationofS} with $\beta = 0$. We shall see in Secs.~\ref{Sec:ExactSol} and~\ref{Sec:ApproxSol} that this constraint imposes severe restrictions on the dynamics of solutions of the non-dynamical theory.

What does the Pontryagin constraint really mean physically? Some insight can be gained by reformulating this constraint in terms of its spinorial decomposition. Gr\"{u}miller and Yunes~\cite{Grumiller:2007rv} have realized that the trivial relation 
\be
\label{eq:ha!}
\pont=\,{^\ast\!}C\,C\,,
\ee
where $C$ is the Weyl tensor
\be
\label{Weyl}
C^{ab}{}_{cd}  :=  R^{ab}{}_{cd} - 2 \delta^{[a}_{[c} R^{b]}_{d]} + \frac13 \delta^a_{[c} \delta^b_{d]} R 
\ee
and $\,{^\ast\!}C$ its dual
\be 
{^\ast}C^a{}_b{}^{cd}:=\frac12
\epsilon^{cdef}C^a{}_{bef}\,,
\label{Cdual}
\ee
opens the door to powerful spinorial methods that allows one to map the Weyl tensor into the Weyl spinor~\cite{Penrose:1986ca}, which in turn can be characterized by the Newman-Penrose (NP) scalars
$\left(\Psi_0,\Psi_1,\Psi_2,\Psi_3,\Psi_4\right)$. Following the notation of~\cite{Stephani:2003tm}, the Pontryagin constraint translates into a reality condition on a quadratic invariant of the Weyl spinor, ${\cal{I}}$,
\be
\label{I2}
\Im \left({\cal{I}}\right) = \Im{\left(\Psi_0\Psi_4+3\Psi_2^2-3\Psi_1\Psi_3\right)}=0\,.
\ee 

The reality condition of Eq.~\eqref{I2} directly implies that any spacetime of Petrov types $III$, $N$ and $O$ automatically satisfies the Pontryagin constrain, while spacetimes of Petrov type D, II and I could violate it.  Moreover, this reality condition also directly implies that not only the Kerr solution but also gravitational perturbations thereof violate the Pontryagin constraint. This is because, although $\Psi_{1,3} = 0$ in this perturbed spacetime, $\Re (\Psi_{2}) \neq 0 \neq \Im (\Psi_{2})$ generically, which violates Eq.~\eqref{I2}~\cite{Grumiller:2007rv,Yunes:2005ve}.

Another reformulation of the Pontryagin constraint can be obtained from the electro-magnetic decomposition of the Weyl tensor (cf.~e.g.~\cite{Cherubini:2003nj}), given by
\be
\label{elecmagn}
(C_{abcd}+\frac i2 \epsilon_{abef}C^{ef}{}_{cd})u^bu^d=E_{ac}+iB_{ac}\,,
\ee
where  $u^{a}$ is a normalized time-like vector, and $E_{ac}$ and $B_{ac}$ are the electric and magnetic parts of the Weyl tensor respectively. Gr\"{u}miller and Yunes~\cite{Grumiller:2007rv}
have shown that in this decomposition~\cite{Matte:1953}, the Pontryagin constraint reduces to
\be
\label{EB}
E_{ab} B^{ab}=0\,.
\ee
Such a restriction forces certain derivatives of the Regge-Wheeler function in the Regge-Wheeler~\cite{Regge:1957rw} decomposition of the metric perturbation to vanish, which has drastic consequences for perturbations of the Schwarzschild spacetime, as we shall discuss in Sec.~\ref{Sec:ApproxSol}. 

The electromagnetic decomposition of the Pontryagin constraint leads to three possible scenarios: purely electric spacetimes $B_{ab} = 0$; purely magnetic spacetimes $E_{ab} = 0$; orthogonal spacetimes, where $E_{ab}$ is orthogonal to $B_{ab}$. In fact, Eq.~\eqref{EB} is a perfect analogue to the well-known electrodynamics condition ${^\ast\!}F\,F\propto\bm{E \cdot B}=0$. In electromagnetism, such a condition is satisfied in electrostatics ($B_{ab} = 0$), magnetostatics ($E_{ab} = 0$) and electromagnetic waves ($E_{ab} B^{ab} = 0$). The Pontryagin constraint can thus be rephrased as ``the gravitational instanton density must vanish,'' since the
quantity ${^\ast\!}F\,F$ is sometimes also referred to as the ``instanton density.''

The severe requirements imposed by the Pontryagin constraint on the space of allowed solutions, together with the arbitrariness in the choice of CS field, make the non-dynamical formulation rather contrived. First, different choices of $\vartheta$ will lead to sufficiently different solutions, each of these with different observables. Without an external prescription to decide what $\vartheta$ is, one loses the predictive power of the Einstein equations and replaces it by a family of possible solutions. Moreover, all choices of $\vartheta$ so far explored are rather unnatural or unphysical, in that they lead to a field with infinite energy, since the field's kinetic energy is constant. Such fields are completely incompatible with the dynamical framework, which implies that results arrived at in the non-dynamical framework cannot be directly extended into the dynamical scheme. Second, the Pontryagin constraint can be thought of as a selection rule, that eliminates certain metrics from the space of allowed solutions. Such a selection rule has been found to overconstrain the modified field equations, to the point that only the trivial zero-solution is allowed in certain cases~\cite{Yunes:2007ss}. 

Having said this, the non-dynamical framework has been useful to {\emph{qualitatively}} understand the effect of the CS correction on gravitational parity violation. Only recently has there been a serious, albeit limited effort to study the much more difficult dynamical formulation and preliminary results seem to indicate that solutions found in this framework do share many similarities with solutions found in the non-dynamical scheme. The non-dynamical formulation should thus be viewed as a toy-model that might help us gain some insight into the more realistic dynamical framework. 

\subsection{Dynamical CS gravity}

The dynamical framework is defined by allowing $\beta$ and $\alpha$ to be arbitrary, but non-zero constants. In fact, $\beta$ cannot be assumed to be close to zero (or much smaller than $\alpha$), because then the evolution equation for $\vartheta$ becomes singular. This formulation was initially introduced by Smith, {\emph{et.~al.~}}\cite{Smith:2007jm}, with the particular choice $\alpha = -\ell/3$ and $\beta = -1$, which implies $[\vartheta] = L^{-1}$. In this model, the CS scalar field is thus not externally prescribed, but it instead evolves driven by the spacetime curvature. The Pontryagin constraint is then superseded by Eq.~\eqref{eq:constraint}, which does not impose a direct and hard constraint on the solution space of the modified theory. Instead, it couples the evolution of the CS field to the modified field equations. 

The dynamical formulation, however is not completely devoid of arbitrariness. Most of this is captured in the potential $V(\vartheta)$ that appears in Eq.~\eqref{Theta-action}, since this is {\emph{a priori}} unknown. In the context of string theory, the CS scalar field is a moduli field, which before stabilization has zero potential (ie.~it represents a flat direction in the Calabi-Yau manifold). Stabilization of the moduli field occurs via supersymmetry breaking at some large energy scale, thus inducing an (almost incalculable) potential that is relevant only at such a scale. Therefore, in the string theory context, it is reasonable to neglect such a potential when considering classical and semi-classical scenarios. 

The arbitrariness aforementioned, however, still persists through the definition of the kinetic energy contained in the scalar field. For example, there is no reason to disallow scalar field actions of the form
\be
\label{new-S}
S_{\textrm{new} \; \vartheta} = - \frac{	1}{2} \int_{{\cal{V}}} d^{4}x \sqrt{-g} \left[\beta_{1} \left(\partial \vartheta \right)^{2}  + \beta_{2} \left(\partial \vartheta \right)^{4}  \right],
\ee
which leads to the following stress-energy tensor
\be
T_{ab}^{\textrm{new} \; \vartheta} 
=    \beta_{1} \left[1 + 2 \beta_{2} \left(\partial \vartheta \right)^{2}  \right] 
\left(\nabla_{a} \vartheta \right) \left(\nabla_{b} \vartheta \right) - \frac{1}{2} g_{ab} \left[\beta_{1} \left(\partial \vartheta \right)^{2} + \beta_{2} \left(\partial \vartheta \right)^{4} \right],
\ee
where we have used the shorthand $\left(\partial \vartheta \right)^{2}  := g^{ab} \left(\nabla_{a} \vartheta \right) \left( \nabla_{b} \vartheta \right)$. The Pontryagin constraint would then be replaced by
\be
\beta_{1} \; \square \vartheta + 2 \beta_{2} \nabla_{a} \left[ \left(\nabla^{a} \vartheta \right) \left( \partial \vartheta \right)^{2} \right] =  - \alpha \frac{\kappa}{4} \pont.
\ee
Of course, the choice of Eq.~\eqref{Theta-action} is natural in the sense that it corresponds to the Klein-Gordon action, but it should actually be the more fundamental theory, from which CS modified gravity is derived, that prescribes the scalar field Hamiltonian. In the string theory context, however, the moduli field possess a canonical kinetic Hamiltonian, suggesting that Eq.~\eqref{new-S} is the correct prescription~\cite{Alexander:2004xd}.

Another natural choice for the potential of the CS coupling is the C-tensor itself. In other words, consider the possibility of placing the C-tensor on the right-hand side of the modified field equations and treating it as simply a non-standard stress-energy contribution. Such a possibility was studied by Gr\"{u}miller and Yunes~\cite{Grumiller:2007rv} for certain background metrics, which are solutions in GR but not in CS modified gravity. The Kerr metric is an example of such a background, for which they found that the induced C-tensor stress-energy violates all energy conditions. No classical matter in the observable universe is so far known to violate all energy conditions, thus rendering this possibility rather unrealistic.

\subsection{Parity Violation in CS Modified Gravity}
\label{parity-CS}

Precisely what type of parity violation is induced by the CS correction? 
Let us first define parity violation as the {\emph{purely spatial}} reflection of the triad that defines 
the coordinate system. The operation $\hat{P} \left[ A\right] = \lambda_{p} A$ is then said to be even, 
parity-preserving or symmetric when $\lambda_{p} = +1$, while it is said to be odd, parity-violating or
antisymmetric if $\lambda_{p} = -1$. By definition, we then have that $\hat{P} \left[e^{I}_{i}\right] = - e^{I}_{i}$, 
where $e^{I}_{i}$ is a spatial triad, and thus $\hat{P} \left[e^{aijk}\right] = -e^{aijk}$. Note that parity transformations are 
slicing-dependent, discrete operations, where one must specify some spacelike hypersurface on which to operate. On the 
other hand the combined parity and time-reversal operations is a spacetime operation that is slicing independent.  

How does the CS modification transform under parity? First, applying such a transformation to the action 
one finds that $S$ is invariant (ie.~parity even) if and only if $\vartheta$ transforms as a pseudo-scalar 
$\hat{P} \left[\vartheta\right] = -\vartheta$. Applying such a transformation to the modified Einstein equations one finds 
that the C-tensor is invariant if and only if the covariant velocity of $\vartheta$ transforms as a vector $\hat{P} \left[ v_{a} \right] = + v_{a}$, or equivalently if $\vartheta$ is as a pseudo-scalar.

The transformation properties of the CS scalar are not entirely free in the dynamical formulation. Since $\vartheta$ must satisfy the evolution equation $\nabla_{a} v^{a} \propto \pont$, we see that $P\left[v^{a}\right] = + v^{a}$, and thus $\vartheta$ must be a pseudo-scalar. In the non-dynamical framework, however, one is free to choose $\vartheta$ in whichever way desired and thus the transformation properties of the action and field equations cannot be {\emph{a priori}} determined. Of course, if one is to treat the modified theory as descending from string theory or particle physics, then $\vartheta$ is required to be a pseudo-scalar as the dynamical theory also requires. 

Statements about the parity-transformation properties of a theory do not restrict the parity-properties of the solutions of the theory. A clear example can be derived from Maxwell's theory of electromagnetism. Propagating modes (electromagnetic waves) travel at the speed of light in vacuum, but in the presence of a dielectric medium, they become birefringent, leading to Faraday rotation. Even though the Maxwell action and field equations are clearly parity-preserving, solutions exist where this symmetry is not respected.  Another example can be obtained from GR, where the theory is clearly parity preserving, but solutions exist (such as the Kerr metric and certain Bianchi models) that do violate parity. 

Such symmetry considerations can be used to infer some properties of background solutions  (ie.~representations of the vacuum state) in the dynamical formulation. First, if one is searching for parity-symmetric solutions (as in the case of spherically symmetric line-elements), then $\pont = 0$, which forces $\theta$ to be constant (assuming this field has finite energy). One then sees that parity-even line-elements will not be CS corrected. On the other hand, if one is considering parity-odd spacetimes (such as the Kerr metric), then the Pontryagin density will source a non-trivial CS scalar, which will in turn modify the Kerr metric through the field equations. Such a correction will tend to introduce even more parity-violation in the solution, as we shall discuss further in Sec.~\ref{Sec:ApproxSol}.  

Clear signals of parity violation can be obtained by studying perturbations about the background solutions. As in the case of Maxwell theory, CS modified gravity has the effect of promoting the vacuum to a very special type of medium, in which left- and right- moving gravitational waves are enhanced/suppressed with propagation distance. Such an effect is sometimes referred to as ``amplitude birefringence,'' and it is analogous (but distinct) to electromagnetic birefringence (see Sec.~\ref{Sec:ApproxSol} for a more detailed discussion of amplitude birefringence). The modified theory then can be said to ``prefer a chirality,'' since it will tend to annihilate a certain polarization mode.

\subsection{Boundary Issues}
\label{Sec:Boundary}
Opposite to common belief, GR does not admit a Dirichlet boundary-value problem as formulated in the previous section. This is so because the variation of the Ricci scalar in the Einstein-Hilbert action leads to boundary terms that depend both on the variation of the metric and its first normal derivative. In order to become a well-posed, Dirichlet boundary value problem, the Einstein-Hilbert action must be supplemented by a boundary counterterm, so-called Gibbons-Hawking-York (GHY) term.  This term cancels the aforementioned boundary terms thus yielding a well-posed boundary value problem. 

The issue of non-dynamical CS modified gravity as a well-posed boundary value problem has been addressed by Grumiller, {\emph{et.~al.~}}\cite{Grumiller:2008ie} with the conventions $\alpha = \kappa$ and $\beta = 0$. As in GR, the CS action as presented in Eq.~\eqref{CS-action} does not lead to a well-posed boundary value problem and counterterms must be added. Let us then concentrate on $\Sigma_{CS}$ and, in particular, on boundary terms involving normal derivatives of the variation of the metric, neglecting irrelevant terms. In~\cite{Grumiller:2008ie} and in this section, ``irrelevant terms'' are defined as those that are bulk terms but not total derivatives, or as those that are boundary terms that vanish on the boundary.  

Let us then define the induced metric on the boundary as 
\be
h_{ab} := g_{ab} - n_{a} n_{b},
\ee
where the boundary is a hypersurface with spacelike, outward-pointing unit normal $n_{a}$. The extrinsic curvature is then 
\be
K_{ab} := h_{a}^{c} h_{b}^{d} \nabla_{c} n_{d},
\ee
which is simply the Lie derivative along $n^{a}$,  where $\nabla_{a}$ stands for the four-dimensional covariant derivative operator. Note that the variation of this quantity is given by
\be
\delta K_{ab} = \frac{1}{2} h_{a}^{c} h_{b}^{d} n^{e} \nabla_{e} \delta g_{cd},
\ee
up to irrelevant terms.

With this machinery, one then finds that the variation of the Einstein-Hilbert and CS actions lead to the following boundary terms~\cite{Grumiller:2008ie}: 
\ba
\label{var-EH}
\delta S_{EH} &=& - 2 \kappa \delta \int_{\partial {\cal{V}}} d^{3}x \; \sqrt{h} \; K,
\\
\label{var-CS}
\delta S_{CS} &=& -2 \alpha \delta \int_{\partial {\cal{V}}} d^{3}x \sqrt{h}\;  \vartheta \; CS(K),
\ea
up to irrelevant terms, where one defines~\cite{Grumiller:2008ie}
\be
\label{CSK}
CS(K) := \frac{1}{2} \epsilon^{n ijk} \; K_{i}{}^{l} \; D_{j} K_{kl}.
\ee
In Eqs.~\eqref{var-EH},~\eqref{var-CS} and~\eqref{CSK}, $K := K^{a}{}_{a}$ is the trace of the extrinsic curvature, $i,j,k$ and $n$ stand for indices tangential and normal to the hypersurface respectively, and $D_{i}$ is the covariant derivative along the boundary. Equation~\eqref{var-EH} is in fact the GHY term, while Eq.~\eqref{var-CS} is analogous to this term in CS modified gravity. Note that Eq.~\eqref{var-CS} depends only on the trace-less part of the extrinsic curvature, and thus, it can be thought of as complementary to the GHY term. 

The boundary terms introduced upon variation of the action can be cancelled by addition of the following counterterms:
\ba
\label{counter-EH}
S_{bEH} &=& 2 \kappa \int_{\partial {\cal{V}}} d^{3}x  \sqrt{h} K, 
\nonumber \\
\label{counter-CS}
S_{bCS} &=& 2 \alpha \int_{\partial {\cal{V}}} d^{3}x \sqrt{h} \; \vartheta \; CS(K).
\ea
Again, Eq.~\eqref{counter-EH} is the GHY counterterm, while Eq.~\eqref{counter-CS} is a new counterterm required in CS modified gravity in order to guarantee a well-posed boundary value problem. Interestingly, we could also have performed this analysis in terms of the CS current, using $\pont \propto v_{a} K^{a}$. Doing so~\cite{Grumiller:2008ie}
\ba
\alpha \frac{1}{4} \int_{{\cal{V}}} d^{4}x \sqrt{-g} \; \vartheta \pont 
+ 2 \int_{\partial{\cal{V}}} d^{3}x \sqrt{h} \; \vartheta CS(K) 
&=&
\nonumber \\
 - \frac{1}{2} \int_{{\cal{V}}} d^{4}x \sqrt{-g} \; v_{a} K^{a} 
+ \int_{\partial {\cal{V}}} \; \sqrt{h} \vartheta CS(\gamma), 
\ea
up to irrelevant terms,  where $CS(\gamma)$ is given by
\be
CS(\Gamma) := \frac{1}{2} \epsilon^{n ijk} \Gamma^{l}{}_{im} \left( \partial_{j} \Gamma^{m}{}_{kl} + \frac{2}{3} \Gamma^{m}{}_{jp} \Gamma^{p}{}_{kl} \right)
\ee
with $\Gamma^{i}{}_{jk}$ the Christoffel connection.

The CS counterterm presented above, however, only holds in an adaptive coordinate frame, where the lapse is set to unity and the shift vanishes. In covariant form, Grumiller, {\emph{et.~al.~}}\cite{Grumiller:2008ie} have shown that the action
\ba
S &=& \kappa \int_{{\cal{V}}} d^{4}x \sqrt{-g} \left(R + \frac{\alpha}{4 \kappa} \vartheta \pont \right)
\nonumber \\
&+& 2 \kappa \int_{\partial {\cal{V}}} d^{3}x \sqrt{h} \left(K + \frac{\alpha}{2 \kappa} \vartheta n_{a} \epsilon^{abcd} K_{b}{}^{e} \nabla_{c} K_{de} \right)
\nonumber \\
&+& \alpha \int_{\partial{\cal{V}}} d^{3}x \sqrt{h} \; {\cal{F}}\left(h_{ab},\vartheta \right), 
\ea
has a well-posed Dirichlet boundary value problem. This action is a generalization of the counterterms presented above, which holds in any frame. The last integral is an additional term that is intrinsic to the boundary and does not affect the well-posedness of the boundary value problem, yet it is essential for a well-defined variational principle when the boundary is pushed to spatial infinity~\cite{Mann:2005yr,Mann:2006bd}.

\section{The Many Faces of Chern-Simons Gravity}
\label{Many-Faces_v1}

\subsection{Particle Physics}
\label{Particle-Physics and Anomalies}
The first place we encounter the CS invariant is in the gravitational anomaly of the Standard Model.  In this chapter, we shall give a pedagogical review and derivation of anomalies that includes the gravitational one.

An anomaly describes a quantum mechanical violation of a classically conserved current.    According to Noether's theorem, invariance under a classical continuous global symmetry group G yields the conservation of a global current $j_{a}^{A}$, with $A$ labelling the generators of the group $G$:
\be
\partial_{a} j^{a A} =0.
\ee
An anomaly ${\cal{A}^{A}}$ is a quantum correction to the divergence of $j_{a}^{A}$ which renders it non-zero, $\partial_{a} j^{a A} ={\cal{A}^{A}}$.  

On the one hand, gauge theories with chiral fermions usually have global anomalies in the chiral currents, $j_{a \, 5} =\bar{\psi }\gamma_{a}\gamma_{5} \psi$.  Such anomalies do not lead to inconsistencies in the theory, but they do possess physical consequences.   Historically, precisely this type of anomaly led to the correct prediction of the decay rate of pions into photons, $\pi_{o} \rightarrow \gamma\gamma$, by including the anomalous interaction $\pi_{0} \epsilon^{a b c d}F_{ab}F_{cd}$.  

On the other hand, gauge anomalies are also a statement that the quantum theory is quantum mechanically inconsistent.  Gauge symmetries can be used to eliminate negative norm states in the quantum theory, but in order to remain unitary, the path integral must also remain gauge invariant. Quantum effects involving gauge interactions with fermions can spoil this gauge-invariance and thus lead to a loss of unitarity and render the quantum formulation inconsistent.  Therefore, if one is to construct a well-defined, unitary quantum theory and if gauge currents are anomalous, then these anomalies must be cancelled by counterterms.

A common example of a global anomaly in the Standard Model is the violation of the $U(1)$ axial current by a one-loop triangle diagram between fermion loops and the gauge field external legs.  Let us then derive the anomaly using Fujikawa's  approach in $1+1$ dimensions~\cite{Fujikawa:1979ay,Fujikawa:1980eg}, generalized to $d+1$ dimensions. Since amplitudes and currents can be generated from the path integral, Fujikawa realized that anomalies arise from the non-invariance of the fermionic measure in the path integral under an arbitrary fermionic field redefinition.  For concreteness let us consider a massless fermion coupled to electromagnetism in $3+1$ dimensions.  The action and partition function for this theory are the following:
\be 
S= \int d^{4}x ( -\frac{1}{4e^{2}} F_{ab}F^{ab} + i\bar{\psi} \gamma^{a} D_{a} \psi ) 
\ee
\be 
Z= \int DAD\psi D\bar{\psi} e^{iS[A_{a},\psi, \bar{\psi}]},
\ee
where $\psi$ is a Dirac fermion, $A^{a}$ is a gauge field, $F_{ab}$ is the electromagnetic field strength tensor, $e$ is the coupling constant or charge of the Dirac fermions and $\gamma^{a}$ are Dirac matrices, where the overhead bar stands for complex conjugation.
Such a toy theory is invariant under a chiral tranformation of the form
\be 
\psi \rightarrow e^{i \alpha \gamma_{5}\psi} = \psi + i\alpha\gamma_{5}\psi + \ldots,
\ee
where $\alpha$ is a real number and $\gamma_{5}$ is the chiral Dirac matrix. Such an invariance leads to the $U(1)$ global Noether current $j_{a}^{\AXIAL} = \bar{\psi}\gamma_{a}\gamma_{5}\psi$. In the path integration approach, current conservation is exhibited  by studying the Ward identities, which can be derived by requiring that the path integral be invariant under an arbitrary phase redefinition of the Dirac fermions.  

The non invariance of the fermionic measure is precisely the main ingredient that encodes the chiral anomaly.  
In order to study this effect, we must first define the measure precisely. For this purpose, it is helpful to expand $\psi$ in terms of orthonormal eigenstates of $ i \gamma^{a}D_{a}$ :

\be 
\gamma^{a}D_{a}\phi_{m} =\lambda_{m}\phi_{m}
\ee
and 
\be 
\psi(x) =\sum_{m} a_{m}\phi_{m}(x), 
\ee
where $a_{m}$ are Grassmann variable multiplying the c-number eigenfunctions $\phi_{m}(x)$ and $D_{a}$ is the gauge covariant derivative. The measure is then defined as
\be 
D\psi D\bar{\psi} =  \prod_{n} da_{n}d\bar{a}_{n}.
\ee

Let us now demand invariance of the partition function: 
\be 
\int DAD\psi D\bar{\psi} e^{iS[a,\psi, \bar{\psi}]} = \int DA'D\psi' D\bar{\psi}' e^{iS'[a,\psi, \bar{\psi}]}, 
\ee
where the fermions transform as follows
\be
\psi(x) \rightarrow \psi'(x) + \epsilon(x), 
\qquad
\bar{\psi}(x) \rightarrow \bar{\psi}'(x) + \bar{\epsilon}(x) 
\ee
for a chiral transformation $\epsilon(x) =  i\alpha(x)\gamma_{5}\bar{\psi(x)}$. We see then that the Lagrangian transforms as
\be 
\int d^{4}x({\bar{\psi'}}i\gamma^{a}D_{a}\psi') = \int d^{4} x [\bar{\psi}i\gamma^{a}D_{a}\psi - (\partial_{a} \alpha) {\bar{\psi}}i\gamma^{a} {\gamma_{5}}\psi]. 
\ee
Assuming that the measure is invariant, integrating by parts and varying the action with respect to $\alpha$, we recover the Ward identity
\be 
\partial_{a}<\bar{\psi}\gamma^{a}\gamma_{5}\psi> = 0,
\ee
which is nothing but the statement of axial current conservation.

Naively, we might conclude that the classical global current carries over to the quantum one, but Fujikawa~\cite{Fujikawa:1979ay,Fujikawa:1980eg} realized that such a reasoning assumes that the path integral measure is invariant. A more careful analysis then reveals that a change of variables in the measure affects the coefficient of the Dirac fermion eigenstate expansion via
\be 
a'_{m} = \sum_{n} (\delta_{mn} + B_{mn})a_{n} 
\ee
where 
\be 
B_{mn} = i\int d^{4}x \; \phi^{\dagger}_{m}\alpha \gamma_{5} \phi_{n}. 
\ee
Using the Grassmanian properties of $a_{m}$, this transformation returns the Jacobian in the measure
\be 
D\psi'D\bar{\psi}' = \left[{\textrm{det}}(1+ B)\right]^{-2} D\psi D\bar{\psi},
\ee
where ${\textrm{det}}(\cdot)$ and ${\textrm{Tr}}(\cdot)$ shall stands for the determinant and trace respectively.

The key to obtaining the anomaly resides in computing ${\textrm{det}}(1+B)$.  Expanding the determinant to first order in $\alpha$ 
\be 
{\textrm{det}}(1+B) =e^{{\textrm{Tr}}[ln(1+B)]} = e^{{\textrm{Tr}}(B)},
\ee
and hence,
\be 
\left[{\textrm{det}}(1+B)\right]^{-2} = e^{-2i\int d^{4}x \alpha(x) \sum_{n}\phi^{\dagger}\gamma_{5}\phi_{n}(x) } 
\ee
Regulating the fermion composite operator  with a cut-off $\lambda_{n}/M$
\be 
\sum_{n} \phi_{n}^{+}(x)\gamma_{5}\phi_{n}(x) \to \sum_{n} \phi_{n}^{+}(x) \gamma_{5}\phi_{n}(x)e^{\frac{\lambda_{n}^{2}}{M^{2}}}
\label{fercompop}
\ee
and using that the mode functions are eigenfunctions of $i\gamma^{a}D_{a}$, we can also write Eq.~\eqref{fercompop} as
\be 
\sum_{n}\phi_{n}^{\dagger}\gamma_{5}e^{\frac{(i\gamma^{a}D_{a})^{2}}{M^{2}}} \phi_{n} = 
< |tr[\gamma_5 e^{(i\gamma^{a}D_{a})^{2}/M^{2}}]> 
\ee
In order to simplify this expression we can use the identity $(i\gamma^{a}D_{a})^{2} = - D_{a}D^{a} +(1/2)\sigma^{ab}F_{ab}$ where $\sigma^{ab}= (i/2)[\gamma^{a},\gamma^{b}]$  which leads us to evaluate
\be <x|{\textrm{Tr}}[\gamma_{5}e^{(-D^{2} + (1/2)\sigma^{ab}F_{ab})/M^{2}}]|x>. \ee
As we take the limit $M \rightarrow \infty$ we can expand in powers of the background electromagnetic field by writing $-D^{2} = -\partial^{2} + ...$.  
We are led to the following expression:
\be <x|\e^{-\partial^{2}/M^{2}}|x> = i\int \frac{d^{4}k_{E}}{2\pi^{4}} e^{-k^{2}_{E}/M^{2}} = \frac{iM^{2}}{16\pi^{2}} .\ee
The other terms that will follow arise from bringing down powers of the background field.  

Terms with one power of the background field and with the trace of $\gamma_{5}$ vanish, since ${\textrm{Tr}} [\gamma_{5} \sigma_{ab}] =0$.  In the limit $M \rightarrow \infty$, terms that are second order in the background field also vanish, leaving:
\be {\textrm{Tr}}\big[ \gamma_{5}\frac{1}{2}(\frac{1}{2M^{2}}\sigma^{ab}F_{ab})^{2}\big] <x|e^{-\partial^{2}/M^{2}}|x> = - \frac{1}{32\pi^{2}}\epsilon^{abcd}F_{ab}F_{cd} . \ee
Putting all the non-vanishing terms above together gives us the Jacobian prefactor:
\be 
\left[{\textrm{det}}(1+ B)\right]^{-2} = e^{i\int d^{4}x \alpha(x) (\frac{1}{16\pi^{2}} \epsilon^{abcd} F_{ab} F_{cd} )} 
\ee
The partition function with this change of variables becomes
\be Z[A] = \int D\psi D\bar{\psi} e^{i\int d^{4}x (\bar{\psi}i\gamma^{a}D_{a}\psi + \alpha(x)(\partial^{a} j^{A}_{a} + (1/16\pi^{2})\epsilon^{acbd}F_{ab}F_{cd}))}\ee
When we vary the partion function with respect to $\alpha$ we get the famous ABJ anomaly~\cite{1969PhRv..177.2426A,Bell:1969ts}
\be  \label{ABJ2} \partial^{a}j_{a}^{A} = -\frac{1}{8\pi^{2}} \epsilon^{abcd}F_{ab}F_{cd}. \ee

The above derivation of the ABJ anomaly also applies for the gravitational anomaly.  Similar to Eq.~\eqref{ABJ2}, if we use the Riemann curvature tensor instead of the field strength tensor, we will obtain the gravitational ABJ anomaly:
 \be D^{a}j_{a}^{A} = -\frac{1}{384\pi^{2}} \frac{1}{2}\epsilon^{abcd}R_{abef}R_{cd}{}^{ef}. \ee
Note that the right-hand side of this equation is proportional to the Pontryagin density of Eq.~\eqref{pontryagindef}.
The gravitational ABJ anomaly can be canceled by adding the appropriate counter term in the action, which in turn amounts to including the CS modification in the Einstein-Hilbert action.

Recently, it has been shown that the CS action is also induced by other standard field theoretical means. 
In particular, \cite{Mariz:2004cv} showed that the CS action arises through Dirac fermions couplings 
to a gravitational field in radiative fermion loop corrections, while~\cite{Mariz:2007gf,Gomes:2008an} 
showed that it also arises in Yang-Mills theories and non-linearized gravity through the proper-time
method and functional integration. We refer the reader to~\cite{Mariz:2004cv,Mariz:2007gf,Gomes:2008an} for more information on these additional mechanisms that generate the CS action.

\subsection{String Theory}
\label{String-Theory and Chern Simons Gravity}
In the previous section, we derived the chiral anomaly in a $3+1$ gauge field theory coupled to fermions.  We saw that while gauge and global anomalies can exist, gauge anomalies need to be cancelled to have a consistent quantum theory.  In what follows we will show how the CS modification to general relativity arises from the Green-Schwarz anomaly  canceling mechanism in heterotic String Theory.  The key idea is that a quantum effect due to a gauge field that couples to the string induces a CS term in the effective low energy four dimensional general relativity.

Recall that the action of a free, one-dimensional particle can be described as the integral of the worldline swept out over a ``target spacetime'' $X^{a}(\tau)$, parametrized by $\tau$.  The infinitesimal path length swept out is 
\be dl= (-ds^{2})^{1/2} = (-dX^{a} dX^{b}\eta_{ab} )^{1/2} \ee
where $\eta_{ab}$ is a $9+1D$ Minkowski target space-time and the action is then
\be S= -m\int dl = -m \int d\tau (-\dot{X}^{a}\dot{X}_{a})^{1/2} \ee

We can easily extend the discussion of a point particle to a string by parametreizing the worldsheet with a target function in terms of two coordinates $(\sigma, \tau)$.   
Consider then the string, world-sheet field $X^{a}(\sigma,\tau)$ embedded in a  $D$-dimensional space-time, $G_{ab}$ that sweeps out a $1+1$ world-sheet denoted by coordinates $(\sigma, \tau)$ and world-sheet line element $ds^{2}  = h_{AB}dX^{A}dX^{B}$. The indices $A,B$ here run over world-sheet coordinates.  Analogous to the point particle, the free string action is described by 
\be S_{st}= \frac{T}{2}\int ds= \frac{T_{}}{2}\int d\sigma d\tau  \sqrt{-h} h_{AB}\partial_{a} X^{A}(\sigma, \tau) \partial_{b}X^{B}(\sigma, \tau)G^{ab}, \ee where $T$ is the string tension. 

This string is also charged under a $U(1)$ symmetry and couples to the Neveu-Schwarz two-form potential, $B_{\mu\nu}$, via
\be S_{B} = \int  d^{2}\sigma \partial_{a} X(\sigma, \tau) \partial_{b}X(\sigma, \tau)B^{ab}. \ee
It is this fundamental string field $B_{ab}$ that underlies the emergence of CS modified gravity when String Theory is compactified to $4D$.
In a seminal work, Alvarez-Gaume and Witten~\cite{AlvarezGaume:1983ig} showed that GR in even dimensions will suffer from a gravitational anomaly in a manner analogous to how anomalies are realized in the last section.   The low-energy limit of superstring theories are 10 dimensional supergravity (SUGRA) theories.  As discussed in the previous section, a triangle loop diagram between gravitons and fermions will generate a gravitational anomaly; similarly hexagon loop diagrams generate anomalies in 10 dimensions.   Remarkably, Green and Schwarz~\cite{Green:1987sp,Green:1987mn} demonstated that the gravitational anomaly is cancelled from a quantum effect of the string worldsheet $B$ field, since the string worldsheet couples to a two form, $B_{ab}$.  The stringy quantum correction shifts the gradient of the $B_{ab}$ field by a CS 3-form.  This all results in modifying the three-form gauge field strength tensor $H_{abc}$ in $10D$ supergravity.
 \be
H_{abc} = (dB)_{abc}  \rightarrow (dB)_{abc} + {1\over4}\big(\Omega_{abc}(A) - \alpha'\Omega_{abc}(\omega) \big) 
\ee
This naive shifting of $H_{abc}$ conspires to cancel the String Theory anomaly.  We refer the interested reader to Vol II of Polchinski's book~\cite{Polchinski:1998rr} for a more detailed discussion of the Green-Schwarz anomaly canceling mechanism. 

We begin our analysis from the compactification of the heterotic string to 
its 4D, ${\cal N} \,=\,1$ supergravity limit.  For concreteness we consider the 
compactification to be on six dimensional internal space (ie. a Calabi-Yau manifold).  Similar to the Kaluza-Klein idea, when we dimensionally reduce a 10 dimensional system to four dimensions, many fields (moduli) which characterize the geometry emerge.  These fields cause a moduli problem since their high energy density will overclose the universe, hence, they need to be stabilized.  The discussion of moduli stabilization is beyond the scope of this review and we point the reader interested in this field of research to the work of Gukov et. al~\cite{Gukov:2003cy}.  In what follows, we 
will assume that all moduli except the axion are stabilized and will not explicitly deal with them in our analysis.

Our starting point is the 10D Heterotic string action in Einstein frame \cite{Bergshoeff:1981um} and we ignore the coupling to fermionic fields since they are not relevant for our discussion.  In this theory the relevant bosonic field content is the 10D metric, $g_{ab}$, a dilaton $\phi$ and a set of two and three form field strength tensors $H_{3}:=H_{abc}$ and $F_{2}:=F_{ab}$ respectively.  
\be 
S=  \int d^{10}x \sqrt{g_{10}} \Big[ \, {\cal R} -{1\over 2} \partial_{a} \phi 
\partial^{a} \phi -{1\over12} e^{-\phi} \, H_{abc}H^{abc}  ~
- ~ {1\over4}e^{{-\phi\over2}} {\textrm{Tr}}(F_{ab}F^{ab})   ~ \Big],   
\ee
where 
\be 
H_{3} = dB_{2} - {1\over4}\big(\Omega_{3}(A) - \alpha'\Omega_{3}(\omega),
\big) 
\ee
$B_{2}:=B_{ab}$, and where $\Omega_{3}(A):=\Omega_{abc}(A)$ and $\Omega_{3}(\omega)=\Omega_{abc}(\omega)$ are the gauge and 
gravitational CS three-forms respectively, which in exterior calculus form are given by
\be 
\Omega_{3}(A)= Tr \big(dA\wedge A + {2\over 3}A \wedge A \wedge. 
A \big) 
\ee
We now dimensionally reduce the 10D action to 4D, ${\cal N} \,=\, 1$ supergravity 
coupled to a gauge sector by choosing a 
four-dimensional Einstein frame metric, $g^{S}_{MN}=g^{E}_{MN}e^{{\phi\over
2}}$. The 10D line element splits up into a sum of four and six dimensional spacetime line elements
\be 
ds^{2}_{10}=ds_{4}^{2} + g_{mn}dy^{m}dy^{n},
\ee
where $g_{mn}$ is a fixed metric if the internal 6-dimensions are normalized to have volume $4\alpha'^{3}$. 
The compactified effective gravitational action becomes
\be 
S_{4D}={1\over 2\kappa_{4}^{2}}\int d^{4}x\sqrt{-g}\left[ \, {\cal R} \,-\, {2\partial_{
\mu}S^{*} \partial^{\mu}S \over(S+S^{*})^{2}} \right],
\ee
where $S=e^{-\psi} + i\theta$, with $\psi$ and $\theta$ the four dimensional dilaton and model-independent axion fields respectively.
The dilaton emerges as the four dimensional Yang-Mills coupling constant $g^{2}_{YM} = e^{\psi}$, which we can assume here to be fixed, 
while the axion derives from the spacetime and internal components of $B_{ab}$.  

Let us now focus our attention on the axionic sector of the 4D heterotic string.  
The bosonic low energy effective action takes the form
\be 
S_{4d}  =  {2\over\alpha'} \int d^{4} x \sqrt{-g} \big( \, {\cal R}_{4} + A - {1\over12} e^{-\phi} \, H_{abc} \wedge *H^{abc}  
- {1\over4} e^{{-\phi\over2}} {\textrm{Tr}}(F_{ab}F^{ab}).
\ee
When we explicitly square the kinetic term of the three-form field strength tensor,
\be 
H_{abc}\wedge *H^{abc} = \big[ (dB_{2} - {1\over4}\big(\Omega_{3}(A) - \alpha'\Omega_{3}(\omega) 
\big) \big]^{2} 
\ee
we obtain the cross term $*dB_{2}\wedge \Omega_{3}$, where the dual to the three-form $dB_{2}$ is equivalent to exterior derivative of the axion  
$ d\theta = *dB_{2}$. After integrating by parts, one ends up with the sought after gravitational Pointryagin interaction
 \cite{Green:1984sg}
\be 
\label{ff} \int d^{4}x f(\theta){\cal R}\wedge {\cal R} 
\ee
where here $f(\theta) = \theta \; \cal{V}\; \rm M_{4pl} \; \alpha' $ and where $\cal{V}$ is a volume factor measured in string units and determined by the 
dimensionality of the compactification.  Additionally, integration by parts also unavoidingly introduces a kinetic term for $f(\theta)$, which we did not
write explicitly above. Alexander and Gates~\cite{Alexander:2004xd} used this construction to place a constraint on the string scale provided that the gravitational CS term was responsible for inflationary leptogenesis.

\subsection{Loop Quantum Gravity}
\label{loops}
The CS correction to the action also arises in Loop Quantum Gravity (LQG), 
which is an effort toward the quantization of GR through the postulate that 
spacetime itself is discrete~\cite{Ashtekar:2004eh,Thiemann:2001yy,Rovelli:2004tv}. In this approach, the Einstein-Hilbert 
action is first expressed in terms of certain ``connection variables'' (essentially the connection and its conjugate momenta, the triad), 
such that it resembles Yang-Mills (YM) theory~\cite{Yang:1954ek} and can thus be quantized via standard methods.  Currently, there are two 
versions of such variables: a selfdual $SL(2,\mathbb{C})$, ``Ashtekar'' connection, which must satisfy some reality 
conditions~\cite{Ashtekar:1991hf}; 
and a real $SU(2)$, Barbero connection, constructed to avoid the reality conditions of the Ashtekar one~\cite{Barbero:1994ap}.
 Both these formalisms can be computed  from the so-called Holst action, which consists of the Einstein-Hilbert term plus a new piece that 
 depends on the dual to the curvature tensor~\cite{Holst:1995pc}, but which does not affect the equations of motion in vacuum by the Bianchi 
 identities 
 
 Ashtekar and Balachandran~\cite{Ashtekar:1988sw} first analyzed parity (P) and charge-parity (CP) conservation 
 in LQG~\cite{Ashtekar:1988sw}, which led them essentially to CS theory with a constant CS parameter. 
 Since LQG resembles YM theory, its canonical variables must satisfy a Gauss-law like GR constraint 
 $D_{a} E^{a}_{I}=0$, where $D_{a}$ is a covariant derivative operator and $E^{a}_{I}$ is the triad. 
 This constraint generates internal gauge transformations in the form of triad rotations.
 
 Physical observables in any quantum theory must be invariant under both large and small, local gauge transformations. As in YM theory, the 
 latter can be associated with unitary irreducible representations of the type $e^{i n \theta}$, where $n$ is the winding number and $\theta$ is an 
 angular ambiguity parameter. Wavefunctions in the quantum theory must then be invariant under the action of these representations, but this 
 generically would lead to different wavefunctions on different $\theta$-sectors. Instead, one can rescale the wavefunctions to eliminate this 
 $\theta$ dependance, at the cost of introducing a B-field dependence on the conjugate momenta, which in turn force the Hamiltonian constraint to violate P and CP. 
 
Asthekar and Balachandran~\cite{Ashtekar:1988sw} noted that this ambiguity can be related, as in YM theory, to the 
possibility of adding to the Einstein-Hilbert action the term
 \be
 S_{\theta} = \frac{i \theta}{32 \pi^{2}} \int d^{4}x \pont,
 \ee
which is essentially the CS correction to the action when the scalar field $\vartheta = \theta$ is constant and pulled out of the integral. In this sense 
a CS-like term arises naturally in LQG due to the requirement that wavefunctions, and thus physical observables, be invariant under large gauge 
transformations. 
 
But the $\theta$-anomaly is not quite the same as CS modified gravity. After all, the above analysis is more reminiscent to the chiral anomaly in 
particle physics, discussed in Sec.~\ref{Particle-Physics and Anomalies}. 
Recently, however, the connection between LQG and CS modified gravity has been 
completed, along a bit of an unexpected path. Taveras and Yunes~\cite{Taveras:2008yf} first investigated the possibility of promoting the Barbero-
Immirzi (BI) parameter to a scalar field. This parameter is another quantization ambiguity parameter that arises in LQG and determines the 
minimum eigenvalue of the discrete area and discrete volume operators~\cite{Immirzi:1996di}. At a classical level, the BI parameter is a
multiplicative constant that controls the strength of the dual curvature correction in the Holst action~\cite{Holst:1995pc}. Taveras and Yunes 
realized that when this parameter is promoted to a field one essentially recovers GR gravity in the presence of an arbitrary scalar field at a 
classical level.  

Although the Holst action is attractive from a theoretical standpoint since it allows a construction of LQG in either Ashtekar or Barbero form, this 
action has also been shown to lead to torsion and parity violation when one couples fermions to the 
theory~\cite{Perez:2005pm,Freidel:2005sn,Randono:2005up}. This issue can be corrected, while still allowing a mapping between GR 
and the Barbero-Ashtekar formalism, by adding to the Holst action a torsion squared term, essentially transforming the Holst term to the Nieh-Yan 
invariant~\cite{Mercuri:2007ki}. When one couples fermions to the Nieh-Yan modified theory, then the resulting effective theory remains torsion 
free and parity preserving~\cite{Mercuri:2006wb}.

Inspired by the work of Taveras and Yunes~\cite{Taveras:2008yf}, 
Mercuri~\cite{Mercuri:2009zi,Mercuri:2009vk} and Mercuri and Taveras~\cite{Mercuri:2009zt} considered the 
possibility of promoting the BI parameter to a scalar field in the Nieh-Yan corrected theory. 
As in the Holst case, they found that the BI scalar naturally induces torsion, but this time
when this torsion is used to construct an effective action they found that one unavoidingly obtains CS modified gravity. 
In particular, one recovers Eq.~\eqref{CS-action} with 
$\vartheta = [3/(2 \kappa)]^{1/2} \tilde{\beta}$, with $\tilde{\beta}$ the BI scalar and $\alpha = 3/(32 \pi^{2}) \sqrt{3 \kappa}$,
while the scalar field action becomes Eq.~\eqref{Theta-action} with $\beta = 1$ and vanishing potential.

\section{Exact Vacuum Solutions}
\label{Sec:ExactSol}
One of the most difficult tasks in any alternative theory of gravity is that of finding {\emph{exact}} solutions, without the aid of any approximation scheme. In the context of string theory, Campbell, {\emph{et.~al.}}~\cite{Campbell:1990fu} showed that certain line elements, such as Schwarzschild and FRW, lead to an exact CS three-form, which thus does not affect the modified
field equations. In the context of CS modified gravity, Jackiw and Pi~\cite{Jackiw:2003pm} showed explicitly that the Schwarzschild metric remains a solution of the non-dynamical modified theory for the canonical choice of CS scalar. Shortly after, Guarrera and Hariton~\cite{Guarrera:2007tu} showed that the FRW and Reissner-Nordstrom line elements also satisfy the non-dynamical modified field equations with the same choice of scalar, verifying the results of Campbell, {\emph{et.~al.}}~\cite{Campbell:1990fu}. Recently, Grumiller and Yunes~\cite{Grumiller:2007rv} carried out an extensive study of exact solutions in the non-dynamical theory for arbitrary CS scalars, with the hope to find one that could represent a spinning black hole. All of these investigations concern {\emph{vacuum} solutions in the non-dynamical framework ($\beta = 0$), with the coupling constant choice $\alpha = \kappa$. We shall also choose these conventions here. 

\subsection{Classification of General Solutions}

Let us begin with a broad classification of general solutions in CS modified gravity.
Grumiller and Yunes~\cite{Grumiller:2007rv} classified the space of solutions, a $2$-dimensional representation of which is shown in Fig.~\ref{sol-space}, into an {\emph{Einstein space}}, ${\cal{E}}$, and a {\emph{CS space}}, ${\cal{CS}}$. Elements of the former satisfy the Einstein equations, while the elements of the later satisfy the CS modified field equations. The intersection of ${\cal{E}}$ with ${\cal{CS}}$, ${\cal{P}} := {\cal{E}} \cap {\cal{CS}}$, defines the {\emph{Pontryagin space}}, whose elements satisfy both the Einstein and the CS modified field equations independently.
\begin{figure}
\begin{center}
\includegraphics[scale=0.4,clip=true,angle=-90]{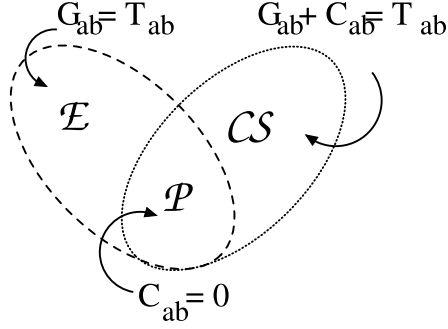}
\caption{\label{sol-space} 
  Space of solutions of Einstein gravity ${\cal{E}}$ and CS modified gravity ${\cal{CS}}$. In this figure, we have set $2 \kappa = 1$ for simplicity of presentation.}
\end{center}
\end{figure}
From the above definitions we can now classify solutions in CS modified gravity. Elements in ${\cal{P}}$ are {\emph{GR solutions}}, because they satisfy the Einstein equations and possess a vanishing C-tensor and Pontryagin density.  Elements in ${\cal{CS}} \; \backslash \; {\cal{P}}$ are {\emph{non-GR solutions}}, because they are not Ricci-flat but they do satisfy the Pontryagin constraint and the CS modified field equations. 

A full analytic study of exact solutions has been possible only regarding spacetimes with sufficient symmetries that allow for the modify field equations to simplify dramatically. For such scenarios, however, the search for CS GR solutions have lead mostly to either Minkowski space or the Schwarzschild metric. 
This can be perhaps understood by considering the vacuum sector of ${\cal{P}}$, where the C-tensor becomes
\be
\label{Cvacuum}
C^{ab}|_{\rm R_{ab}=0} = v_{cd}{\,^\ast\!}R^{d(ab)c} =
v_{cd}{\,^\ast\!}C^{d(ab)c}=0\,, 
\ee
where $C_{abcd}$ and ${\,^\ast\!}C$ are the Weyl tensor and its dual respectively [Eqs.~\eqref{Weyl} and~\eqref{Cdual}]. Such a condition implies the Weyl tensor must be divergenceless via the contracted Bianchi identities, which leads to three distinct possibilities:
\begin{enumerate}
\item The (dual) Weyl tensor vanishes. In vacuum, elements of ${\cal{P}}$ are also Ricci flat, so this possibility leads uniquely to Minkowski space.
\item The covariant acceleration of $\vartheta$ vanishes. Such a restriction imposes strong constraints on the geometry  (cf.~e.g.~\cite{Stephani:2003tm}), leading either to flat space or to the existence of a null Killing vector.
\item The contraction of the covariant acceleration and the dual Weyl tensor vanishes.
\end{enumerate}
Elements of ${\cal{P}}$ are very special, possessing a large number of symmetries and Killing vectors. On the other hand, elements of ${\cal{CS}} \; \backslash \; {\cal{P}}$ cannot possess too many symmetries, which explains why it has been so difficult to find them.

\subsection{Spherically Symmetric Spacetimes}

Consider first the most general, spherically symmetric spacetime, whose metric can be decomposed as the warped product of two $2$-dimensional metrics~\cite{Gerlach:1979rw,Gerlach:1980tx}: a Lorentzian one $g^{}_{\alpha \beta}(x^\gamma)$ ($\alpha, \beta, \ldots = t,r$) with some coordinates $x^\gamma$;  and the metric on the $2$-sphere $\Omega^{}_{ij}(x^{i})$ ($i,j,\ldots = \theta,\phi$) with some coordinates $x^i$.  Such a line element can be written in the following $2+2$ form:
\be
\label{spherical}
ds^2=g_{\alpha\beta}(x^\gamma)\,dx^\alpha dx^\beta + \Phi^2(x^\gamma)\, d\Omega^2_{{\cal S}^2}\,,
\ee
where $d\Omega^2_{{\cal S}^2}$ is a line element of the round 2-sphere  and the warped factor is the square of the scalar field $\Phi(x^\gamma)$, often called  ``areal'' radius. 

A spherically symmetric line element [eg.~Eq.~\eqref{spherical}] always leads to a vanishing Pontryagin density, $\pont = 0$, and to a decoupling of the modified field equations~\cite{Grumiller:2007rv}:
\be
\label{decoupled}
R_{ab}=0\,,\qquad C_{ab}=0\,.
\ee
For the metric in Eq.~\eqref{spherical}, the only non-vanishing components of these tensors are $R_{\alpha\beta}$, $R_{ij}$ and $C_{\alpha i}$, provided $\vartheta$ belongs to the generic family~\cite{Grumiller:2007rv,Yunes:2007ss} 
\be
\label{possibletheta2}
\vartheta = F(x^\gamma) + \Phi(x^\gamma)\, G(x^i)\,.
\ee
In the non-dynamical framework, these results imply that spherically symmetric line elements are always pushed to ${\cal{P}}$. Similar conclusions also hold for spherically symmetric line-elements in non-vacuum spacetimes.

In the dynamical framework, one must also solve the evolution equation for the CS scalar, which here becomes a wave-like equation without a source. In the absence of a potential, this wave-like equation need not necessarily have well-defined decaying solutions that will lead to finite energy contained in the scalar field. If this is the case, the scalar field is forced to be a constant, which reduces the modified theory to GR.  

The study of spherically symmetric line elements naturally leads to the study of Birkhoff's theorem in CS modified gravity. This theorem states that the most general, spherically symmetric solution to the vacuum Einstein equations is the Schwarzschild line element. For spherically symmetric line elements and the CS scalar of Eq.~\eqref{possibletheta2}, the non-dynamical CS modified field equations decouple and the C-tensor identically vanishes, which suffices to guarantee that Birkhoff's theorem still holds in the non-dynamical formulation~\cite{Yunes:2007ss}.

In spite of the clear persistence of Birkhoff's theorem in the non-dynamical formulation, this theorem does not in fact hold in the dynamical framework. In this scheme, the dynamical field equations contain a scalar-field stress-energy contribution that will unavoidingly lead to non-vacuum (ie.~hairy) solutions. Due to the presence of such a dynamical scalar field, spherical symmetry need not lead to staticity, and in fact, spherically symmetric spacetimes will in general be dynamical. Static solutions are, however, still present in dynamical CS modified gravity provided the CS scalar is a constant~\cite{slow-rot}.

The study of the spherically symmetric spacetimes in CS modified gravity leaves us with two important lessons: 
\begin{itemize}
\item The existence of specific solutions depends sensitively on the
  choice of the scalar field.
\item The satisfaction of the Pontryagin constraint is a necessary but
  not a sufficient condition for the C-tensor to vanish.
\end{itemize}
In fact, it is simple to construct a CS scalar, such as a trigonometric function of spacetime, and show that for such a scalar the C-tensor does not vanish, thus rendering the Schwarzschild metric not a solution of CS modified gravity, in spite of the vanishing of the Pontryagin density.

\subsection{Why is the Kerr Metric Not a Solution in CS Modified Gravity?}
\label{Pont-Kerr}
Consider the Kerr metric in Boyer-Lindquist coordinates $(t,r,\theta,\phi)$: 
\ba
\label{kerr}
ds^2 &=& -\frac{\Delta-a^2\sin^2\Theta}{\Sigma}dt^2-\frac{4aMr\sin^2\Theta}{\Sigma}dtd\phi 
\nonumber \\
&+& 
\frac{(r^2+a^2)^2-a^2\Delta\sin^2\Theta}{\Sigma}\sin^2\Theta d\phi^2+\frac{\Sigma}{\Delta}dr^2+\Sigma d\Theta^2
\ea
where $\Sigma=r^2+a^2\cos^2\Theta$ and $\Delta=r^2+a^2-2Mr$. When CS modified gravity was proposed, Jackiw and Pi~\cite{Jackiw:2003pm} realized that the Kerr metric would not be a solution of the modified theory because the Pontryagin density is not vanishing:
\ba
\label{Pkerr}
\pont &=& 96 \frac{a M^2r}{\Sigma^6}\cos{\Theta} \left(r^2 - 3 a^2
  \cos^2{\Theta} \right) \left(3 r^2 - a^2 \cos^2{\Theta} \right)\,.
\ea
This statement is also true in the dynamical frameworks, because the Pontryagin density will induce a non-constant CS scalar that will lead to a non-vanishing C-tensor. One can show that the parity-odd quantity in Eq.~\eqref{Pkerr}, is also non-vanishing for the Kerr-Newman and Kerr-NUT spacetimes~\cite{Guarrera:2007tu,Grumiller:2007rv}, but it is is satisfied in certain interesting physical limits, namely $a \to 0$ (the Schwarzschild limit) and $M \to 0$ (the pp-wave limit). 

Just because the Kerr metric is not a solution in CS modified gravity does {\emph{not}} imply that a rotating BH
solution is absent in the modified theory.  
Modifications of the Kerr metric that do satisfy the CS modified field equations can be obtained be either considering approximate solutions or by studying the dynamical framework. For example, in the far field limit, $M/r \ll 1$, the Pontryagin constraint is satisfied to ${\cal{O}}(M/r)^{3}$ and approximate modified solutions can be derived. On the other hand, in the dynamical formalism Eq.~\eqref{Pkerr} serves as a source term that drives the evolution of the CS scalar, which in turn sources corrections in the metric~\cite{slow-rot}. 

An example of the latter can be found in the quantum-inspired studies of Campbell~\cite{Campbell:1990ai,Campbell:1991rz,Campbell:1992hc}, Reuter~\cite{Reuter:1991cb} and Kaloper~\cite{Kaloper:1991rw}. For example, in~\cite{Campbell:1990ai,Reuter:1991cb}, $\phi \;F_{ab} \; {}^{\ast}F^{ab}$ is added to the Einstein-Maxwell-Klein-Gordon action, with $\phi$ a scalar (axion) field and $F_{ab}$ the photon strength field tensor. The equations of motion for the scalar field acquire a source (the expectation value of a chiral current), when one treats the photon quantum mechanically. Upon accounting for the one-loop fluctuations of the four-vector potential~\cite{Dolgov:1987yp,Dolgov:1988qx}, one essentially finds Eq.~\eqref{eq:constraint}, which on a Kerr background can be solved to find~\cite{Reuter:1991cb} 
\be
\label{KR-axion}
\phi = \frac{5}{8} \frac{\alpha}{\beta} \frac{a}{M} \frac{\cos{\theta}}{r^{2}} + {\cal{O}}\left(\frac{M^{3}}{r^{3}}\right).
\ee
The scalar field presents a $r^{-2}$ fall-off and a dipolar structure, identical to the Kalb-Ramond axion~\cite{Campbell:1990ai}. This field leads to a non-trivial C-tensor that corrects the Kerr line element, thus disallowing this metric as a solution of the modified theory.  Nonetheless, the axion in Eq.~\eqref{KR-axion} could (and has) been used to construct correction to the Kerr metric~\cite{slow-rot}, as we shall discuss in Sec.~\ref{Sec:ApproxSol}.

\subsection{Static and Axisymmetric Spacetimes}
Let us now consider static and axisymmetric line elements in vacuum and in the non-dynamical framework. Both stationary and static, axisymmetric spacetimes possess a timelike $\left(\partial_{t}\right)^{a}$ and an azimuthal $\left(\partial_{\phi}\right)^{a}$ Killing vector, but the difference is that static metrics contain no cross-terms of type $dt d\phi$. The most general such metric is diffeomorphic to~\cite{waldgeneral}
\be
\label{ds2}
ds^2 = -V dt^2 + V^{-1} \rho^2 d\phi^2 + \Omega^2 \left(d\rho^2 + \Lambda dz^2\right),
\ee
where $V(\rho,z)$, $\Omega(\rho,z)$ and $\Lambda(\rho,z)$ are undetermined functions of two coordinates, $\rho$ and $z$. 

Consider first the canonical choice of CS scalar. The modified field equations once more decouple, as in Eq.~\eqref{decoupled} and the spherically symmetric case, which implies all static and axisymmetric solutions are elements of ${\cal{P}}$, identically satisfying the Pontryagin constraint. Due to this, we can make choose $\Lambda=1$ and put the metric into Weyl class~\cite{waldgeneral,Stephani:2003tm}: 
\be
\label{dsWeyl}
ds^2=-e^{2U}dt^2+e^{-2U}\left[e^{2k}(d\rho^2+dz^2)+\rho^2d\phi^2\right]\,,
\ee
where $U(\rho,z)$ and $k(\rho,z)$ replace the functions $V$ and $\Omega$. 

The vanishing of the Ricci tensor reduces to
\be
\label{EE-cyl}
\Delta U=0\,,\quad k_{,\rho} = \rho (U_{,\rho}^2-U_{,z}^2)\,,\,\,
k_{,z}=2\rho U_{,\rho} U_{,z}\,, 
\ee
where $\Delta = \partial^2/\partial\rho^2+1/\rho \partial/\partial_\rho+\partial^2/\partial z^2$ is the flat space Laplacian in cylindrical coordinates. The function $k$ can be solved for through a line integral once $U$ is determined~\cite{Stephani:2003tm}. 

The vanishing of the C-tensor reduces to the vanishing of the contraction of the dual Weyl tensor and the covariant acceleration of the CS coupling field. For a canonical $\vartheta$, one finds~\cite{Grumiller:2007rv} 
\be
\Gamma^t{}_{\rho t}{\,^\ast\!}C^{t(ab)\rho} +
\Gamma^t{}_{zt}{\,^\ast\!}C^{t(ab)z} = 0\,,
\ee 
which leads to a set of nonlinear PDEs for $U$, in addition to the Laplace equation of Eq.~\eqref{EE-cyl}. 

The set of equations imposed by Eq.~\eqref{decoupled} can be solved exactly, yielding either flat space, the Schwarzschild solution or the following two solutions:
\be
\label{eq:Per1}
ds^2 = -\frac 1z dt^2 + z dz^2 + z^2(d\rho^2+\rho^2d\phi^2)
\ee
and
\ba
ds^2 &=& -\left(\frac{2m}{z}-1\right)dt^2 +
\left(\frac{2m}{z}-1\right)^{-1} dz^2 
+  z^2(d\rho^2+\sinh^2{\!\rho}\, d\phi^2)
\label{eq:Per2}
\ea
These solutions, however, contain undesirable, non-physical features, such as the existence of a naked singularity at $z=0$ [Eq.~\eqref{eq:Per1}], or the existence of a spacelike Killing vector  $k^a=(\partial_t)^a$ in the ``outside'' region $z>2m$ [Eq.~\eqref{eq:Per2}].

The above results can also be obtained by noticing that the spatial sector of $C_{ab}$ reduces identically to the $3$-dimensional Cotton-York tensor for static and axisymmetric line elements, which by the decoupling of the field equations must vanish exactly~\cite{Grumiller:2007rv}. This implies the metric must be spatially conformally flat. Luk{\'a}cs and Perj{\'e}s~\cite{Lukacs:1982} have shown that such solutions to the Einstein equations reduce to Minkowski, Schwarzschild or the line elements of Eqs.~\eqref{eq:Per1} and~\eqref{eq:Per2}, as described above.

The above results hold for any CS scalar in the family
\be
\label{v-gen-static}
\vartheta = \vartheta_1(t,\rho,z) + \vartheta_2(\rho,z,\phi).
\ee
For even more general CS scalars, as would arise for instance in the dynamical formulation, the modified field equations do not necessarily decouple, and thus, new and interesting solutions could arise. Unfortunately, when this is the case, the system of partial differential equations becomes too difficult to study analytically and has not really been analyzed in detail.

\subsection{Stationary and Axisymmetric Spacetimes}
Consider now stationary and axisymmetric spacetimes~\cite{Stephani:2003tm}:
\be
ds^2 = -V \left(dt - w d\phi\right)^2 + V^{-1} \rho^2 d\phi^2 +
\Omega^2 \left(d\rho^2 + \Lambda dz^2 \right)\,,
\ee
where $V$, $\Omega$, $\Lambda$ and $w$ depend on $\rho$ and $z$, with the latter identified with angular velocity of rotation about the Killing axis. 

Let us first consider the canonical choice of CS scalar, for which the modified field equations decouple once more. The Einstein equations can be used to set $\Lambda$ to unity and put the line element in Lewis-Papapetrou-Weyl form~\cite{Stephani:2003tm}
\be
\label{LPW}
ds^2=-e^{2U}(dt-wd\phi)^2+e^{-2U}\left[e^{2k}(d\rho^2+dz^2)+\rho^2d\phi^2\right] \,,
\ee 
where $U$ and $k$ replace $V$ and $\Omega$.

The vanishing of the Ricci tensor reduces to a set of partial differential equations similar to Eq.~(\ref{EE-cyl}), with a non-trivial source that depends on $w$ and an additional equation for this function. The vanishing of the C-tensor, however, does not lead to a vanishing Cotton-York tensor, and thus, it does not necessarily imply spatial conformal flatness. Grumiller and Yunes~\cite{Grumiller:2007rv} have argued that it seems unlikely that other non-trivial and physically interesting solutions besides the static ones could arise, because $R_{ab} = 0$ and $C_{ab} = 0$ is a strongly over-constrained differential system. This is an example of how the non-dynamical theory can lead to an overconstrained system of equations when searching for physically relevant solutions, but a strict proof remains elusive.

The argument presented above holds for any $\vartheta$ in the family
\be
\label{stationary-axi-theta}
\vartheta = \frac{t}{\mu} + \frac{\phi}{\nu}\,,
\ee
with constants $\mu$ and $\nu$. When the CS coupling is not of the form of Eq.~\eqref{stationary-axi-theta}, then the field equations do not decouple, the arguments presented above do not hold and the system become more difficult to analyze analytically. 

In the non-dynamical framework, another route to exact solutions for generic CS scalars is through the Pontryagin constraint. For axisymmetric spacetimes, $\pont \neq 0$, and this leads to a complicated set of partial differential equations for $w$, $U$ and $k$. Solutions of $\pont = 0$ have been found to be of Petrov type II~\cite{Grumiller:2007rv}, which correspond to the Van Stockum class~\cite{Stephani:2003tm}
\be
ds^2 = \rho \Omega dt^2 - 2 \rho dt d\phi + \frac{1}{\sqrt{\rho}}
\left(d\rho^2 + dz^2\right)\,,
\label{eq:vSle}
\ee
where $\Omega = \Omega(\rho,z)$ is arbitrary and there is no $d\phi^{2}$ component. 

The reduced Van-Stockum class of metrics leads to a complete decoupling of the modified field equations, except for the $dt^{2}$ component. The vanishing of the C-tensor can be achieved if $\vartheta = \vartheta(\rho,z)$, while the $dt^{2}$ component of the field equations can be solved to find
\ba
\label{not-flat}
\Omega &=& c, \qquad \vartheta = \vartheta(\rho,z),
\\
\label{not-GR}
\Omega &=& c + \frac{d}{\sqrt{\rho}}, \qquad \vartheta = \frac{2}{3} \sqrt{\rho}\, z + \tilde{\vartheta}(\rho),
\ea
where $c$ and $d$ are constants. Equation~\eqref{not-flat} is non-flat, possessing a non-vanishing Riemann tensor and a third Killing vector $t\partial_t-\phi\partial_\phi+ct\partial_\phi$, but it satisfies $R_{ab} = 0$ and $C_{ab} = 0$ independently (ie.~it belongs to ${\cal{P}}$). Equation~\eqref{not-GR} is Riemann and Ricci non-flat ($R_{abcd} \neq 0$, $R_{ab} = -C_{ab} \neq 0$), belonging to ${\cal{CS}} \; \backslash \; {\cal{P}}$. This last solution can be interpreted as a  BH solution in the {\emph{mathematical}} sense only, provided $\Omega$ vanishes for some $\rho$ and a Killing horizon appears, because it allows for closed timelike curves outside the Killing horizon~\cite{Grumiller:2007rv}.

We see then that physically relevant, stationary and axisymmetry, exact solutions have not been found in non-dynamical CS gravity, even when considering generic CS scalars fields. We say relevant solutions here, because non-physical ones have been found, but they either contain naked singularities or closed time-like curves. 
In fact, as an example of the latter, recently~\cite{Furtado:2009ji} has shown that
the Godel line element $ds^{2} = a^{2} (dt^{2} - dx^{2} + e^{2x} dy^{2}/2 - dz^{2} + 2 e^{x} dt dy)$, 
which is a subclass of the metrics considered in this section, satisfies the non-dynamical modified field equations with the CS scalar $\vartheta = F(x,y)$, for some arbitrary function $F$ . 
Such a solution is of class {\cal{P}} because the C-tensor automatically vanishes with such choices. 
All evidence currently points at the modified field equations being over-constrained by the Pontryagin condition and the field equations. Although this evidence is strong, no proof currently exists to guarantee that no solution can be found. Moreover, in the dynamical framework, the relaxation of the Pontryagin constrain suffices to
allow the existence of Kerr-like solutions~\cite{slow-rot}.  

\subsection{PP-Waves and Boosted Black Holes}
Consider line elements that represent exact gravitational wave solutions (pp-waves~\cite{Jordan:1960}):
\be
\label{pp}
ds^2=-2dvdu-H(u,x,y)du^2+dx^2+dy^2,
\ee
where $H$ is the only free function of $u$, $x$ and $y$. The Aichelburg-Sexl limit~\cite{Aichelburg:1971dh} of various BHs is in fact an example of such a line element. This limit consists of ultrarelativistically boosting the BH, while keeping its energy finite, by simultaneously taking the mass to zero as the boost speed approaches that of light~\cite{Lousto:1992th,Balasin:1995tb}.

For such metrics, the Pontryagin constraint vanishes identically, which at first seems in contradiction with the results 
of Jackiw and Pi~\cite{Jackiw:2003pm}, who showed that generic linear GW perturbations lead to  
$\pont = {\cal{O}}(h^{2})$. In fact, these two results are actually consistent, because the line elements considered here
and in~\cite{Jackiw:2003pm} are intrinsically different and cannot be related via diffeomorphism. The fact that the
Pontryagin density vanishes for the metric of Eq.~\eqref{pp} is related to the fact that the CS velocity field
$v_{a} = \partial_{a} \vartheta$ is a null Killing vector $v^{a}v_{a} = 0$ for such spacetimes~\cite{Grumiller:2007rv}. 

The modified field equations decouple, except for the $du^{2}$ component. Consider first solutions that live in ${\cal{P}}$ [Eq.~\eqref{decoupled}], which require the vanishing of the C-tensor, thus forcing   
\be
\label{red_theta}
\vartheta=\lambda(u) v + \tilde{\vartheta}(u,x,y)\,.
\ee
The vanishing of the Ricci tensor forces
\ba
\label{lift}
\Delta H &=& 0\,, 
\\
\label{generation}
2 H_{,yy} \tilde\vartheta_{,xy} &=& 
H_{,xy} (\tilde\vartheta_{,yy}-\tilde\vartheta_{,xx})\,.
\ea
Therefore, for any $H$ that solves the Laplace equation in Eq.~\eqref{lift}, we can find a $\tilde{\vartheta}$ such that Eq.~\eqref{generation} is also satisfied.  Such a scheme allows one to {\emph{lift}} any pp-wave solution of the vacuum Einstein equations to a pp-wave solution of non-dynamical CS modified gravity of class ${\cal{P}}$ through a choice of $\vartheta$ that satisfies Eq.~\eqref{red_theta} and~\eqref{generation}~\cite{Grumiller:2007rv}.

Consider now solutions that live in ${\cal{CS}} \; \backslash \; {\cal{P}}$. All non-$uu$ components of the modified field equations decouple, and thus the vanishing of the C-tensor is satisfied by scalar field of the form of Eq.~\eqref{red_theta}. Choosing $\lambda(u) = 0$ for simplicity, we find that the $uu$ component of the modified field equations reduces to a third order PDE
\ba
\label{ole}
(1+\tilde\vartheta_{,y}\partial_{,x}-\tilde\vartheta_{,x}\partial_{,y})\Delta  
H
&=& 
(\tilde\vartheta_{,xx}-\tilde\vartheta_{,yy})H_{,xy}-(H_{,xx}-H_{,yy})\tilde\vartheta_{,xy}\,.   
\ea
Simplifying this scenario further by choosing $\tilde\vartheta=a(u)x+b(u)y+c(u)$, Eq.~\eqref{ole} reduces to the Poisson equation $\Delta H=f$, whose source satisfies a linear first order PDE $bf_{,x}-af_{,y}-f=0$, with general solution [assuming $b(u)\neq 0$]
\be
\label{f}
f(u,x,y)=e^{x/b(u)} \phi\left[a(u)x+b(u)y\right]
\ee
where $\phi$ is an arbitrary function. With such a source and  two supplementary boundary conditions, we can now solve the Poisson equation and specify the full modified pp-wave solution~\cite{Grumiller:2007rv}. Thus, generic ${\cal{CS}} \backslash {\cal{P}}$ solutions do exist in non-dynamical CS modified gravity and can be found via the algorithm described above.

\subsection{Non-Axisymmetric Solutions and Matter}

In the non-dynamical framework, axisymmetry seems to limit the existence of solutions for a certain class of coupling functions. However, if either axisymmetry, the non-dynamical behavior of $\vartheta$ or the vacuum content assumption is relaxed, it is possible that solution in fact do exist. 

Consider losing axisymmetry first. The general idea here is to add new degrees of freedom in the metric that could compensate for the overconstraining of the decoupling of the modified field equations. Such idea is in fact inspired from approximate far-field solutions found in non-dynamical modified CS gravity, which indeed require the presence of additional, non-vanishing, gravitomagnetic metric components. Up to the writing of this review, the only attempts to find such solutions have failed~\cite{Grumiller:2007rv}, due to the incredible complexity of the differential system.

Consider next spacetimes with matter content. The Kerr BH is a ``vacuum''  solution in GR, but it does possess a distributional energy momentum tensor~\cite{Balasin:1994kf}. Moreover, in string theory and cosmology~\cite{Alexander:2004us}, CS modified gravity arises from matter currents, so the inclusion of such degrees of freedom might in principle be important. In the dynamical scheme, for example, one could lift any GR solution to a solution of CS modified gravity by requiring that 
\ba
R_{ab}-\frac12 g_{ab}R &=&8\pi T_{ab}^{\rm
  mat}\,,
  \label{eq:bg}
  \\
  C_{ab} &=& 8\pi T_{ab}^{\vartheta}\,,
\label{eq:ind}
\ea
where $T_{ab}^{\rm mat}$ stands for the stress-energy of matter degrees of freedom (such as the distributional one associated with the Kerr solution), while $T_{ab}^{\vartheta}$ is the energy-momentum of the CS coupling. One would now have solve the system of PDEs associated with Eq.~\eqref{eq:ind} for the background which satisfies Eq.~\eqref{eq:bg}. Such a task, however, would imply also solving the equation of motion for the CS scalar, thus reducing this analysis to the study of exact solutions in dynamical CS modified gravity, which has not yet been performed.

\section{Approximate Vacuum Solutions}
\label{Sec:ApproxSol}
Approximate schemes have been employed to solve the CS modified field equations in different limits. The first attempt along this lines was that of Alexander and Yunes~\cite{Alexander:2007zg,Alexander:2007vt}, who performed a far-field, PPN analysis of non-dynamical CS modified gravity ($\alpha = \kappa$, $\beta = 0$) with canonical $\vartheta$. This study was closely followed by that of Smith, {\emph{et.~al.~}}\cite{Smith:2007jm} who carried out a far-field investigation of non-dynamical solutions representing the gravitational field outside a homogenous, rotating sphere, taking careful account of the matching between interior and exterior solutions. Konno,  {\emph{et.~al.~}}\cite{Konno:2007ze} investigated the slow-rotation limit of stationary and axisymmetric line elements in non-dynamical CS modified gravity with non-canonical CS scalar fields. Finally, gravitational wave solutions of CS modified gravity have been studied by a number of authors, both in Minkowski spacetime and in an FRW background~\cite{Jackiw:2003pm,Alexander:2004us,Alexander:2007qe,Alexander:2004wk,Alexander:2007kv,pulsars}. The last two set of studies were carried out in the non-dynamical formalism with the choices $\alpha = \kappa$ and $\beta = 0$. Little is known about GW propagation or generation in dynamical CS modified gravity, although
early efforts are being directed on that front~\cite{slow-rot,Sopuerta:2009iy}.


\subsection{Formal Post-Newtonian Solution}
\label{PN-sol}
The PN approximation has seen tremendous success to model full general relativity in the slow-motion, weakly gravitating regime (for a recent review see~\cite{Blanchet:2002av}). This approximation is used heavily to study Solar System tests of alternative theories of gravity in the PPN framework, as well as to describe gravitational waves from inspiraling compact binaries, which could be observed in gravitational wave detectors in the near future. For these reasons, it is instructive to study the PN expansion of CS modified gravity, before submerging ourselves in other approximate solutions. 

The PN approximation is essentially a slow-motion and weak-gravity scheme in which the field equations of some theory are expanded and solved perturbatively and iteratively. As such, this scheme makes use of multiple-scale perturbation theory~\cite{Bender,Kevorkian,Yunes:2005nn,Yunes:2006iw}, where the perturbation parameters are the self-gravity of the objects (an expansion in powers of Newton's constant $G$)  and their typical velocities $v$ (an expansion in inverse powers of the speed of light). For example, matter densities $\rho$ are dominant over pressures $p$ and specific energy densities $\Pi$, while spatial derivatives are dominant over temporal ones. 

The PN framework also requires the presence of external matter degrees of freedom, {\emph{ie.~}}bodies that are self-gravitating and slowly-moving. Such objects can be described in a point-particle approximation~\cite{Blanchet:2002av}, or alternatively with a perfect fluid stress-energy tensor~\cite{Will:1999dq}:
\be
T^{ab} = \left( \rho + \rho \Pi + p \right) u^{a} u^{b} + p g^{ab},
\ee
where $u^{a}$ is the object's four-velocity. In GR, the internal structure of the gravitating objects can be neglected to rather large PN order~\cite{damour-effecing}, and thus one can effectively take the radius of the fluid balls to zero, which reproduces the point-particle result. This statement is that of the effacing principle~\cite{damour-effecing}, which is the view we shall take in the next section when we study CS modified gravity in the PPN framework. However, care must be taken, since the effacing principle need not hold in alternative theories of gravity. In fact, as we shall see later on, the effacing principle must be corrected in CS modified gravity due to modifications to the junction conditions~\cite{Lanczos:1922,Lanczos:1924,Darmois:1927,Lichnerowicz,Misner:1964,Barrabes:1991ng,Israel:1966nc,Misner:1973cw,poisson}. 

Perturbation theory, and thus the PN approximation, requires the use of a specific background and coordinate system. In the traditional PN scheme, one linearizes the field equation with the metric $g_{ab} = \eta_{ab} + h_{ab}$, where $\eta_{ab}$ is the Minkowski background, since cosmological effects are usually subdominant. Moreover, a Lorentz gauge is usually chosen $h_{b a,}{}^{a} = h_{,b}/2$, which allows one to cast the field equations as a wave equation with non-trivial, non-linear source terms. One can show that to first order in the metric perturbation, the linearized CS modified field equations in the non-dynamical formalism, with canonical CS scalar and in the Lorentz gauge, can be written as~\cite{Alexander:2007zg,Alexander:2007vt}  
\be
\label{formal-DE}
\square_{\eta}{\cal{H}}_{ab} = -16 \pi \left(T_{ab} -  \frac{1}{2} g_{ab} T \right) + {\cal{O}}(h)^2,
\ee
where the superpotential ${\cal{H}}_{ab}$ captures the CS modification to the PN expansion to the modified field equations  and it is given by
\be
\label{effective-metric}
{\cal{H}}_{ab} := h_{ab} + \dot{\vartheta} \; \tilde\epsilon^{0 c d  }{}_{(a} h_{b) d,c},
\ee
where $\tilde{\epsilon}^{abcd}$ stands for the Levi-Civita tensor density.
The $00$ component, the symmetric spatial part and the trace of the superpotential are equal to that of the metric perturbation, because the Levi-Civita symbol forces the CS correction to vanish.  The formal solution to the modified field equations then reduces to
\be
\label{formal-sol}
{\cal{H}}_{ab} = -16 \pi \; \square_{\eta}^{-1} \left(T_{ab} -
  \frac{1}{2} g_{ab} T \right) + {\cal{O}}(h)^2,
\ee
where the inverse D'Alembertian operator stands for a Green function integral. This formal solution is in fact identical to that of the PN expansion of GR in the limit $\dot{\vartheta} \to 0$. 

This formal solution can also be cast into a more practical form by perturbatively solving for the metric perturbation~\cite{Alexander:2007zg,Alexander:2007vt}. Let us then make the ansatz 
\be
\label{ansatz}
h_{ab}  = h_{ab}^{(GR)} + \dot{\vartheta} \; \zeta_{ab} +
{\cal{O}}(h)^2,
\ee
where $h_{ab}^{(GR)}$ is the GR solution, which is $\vartheta$-independent and satisfies
\be
h_{ab}^{(GR)} := -16 \pi \; \square_{\eta}^{-1} \left(T_{ab} -
  \frac{1}{2} g_{ab} T \right),
\ee
and where $\zeta_{ab}$ is an unknown function that is first order in $\dot{\vartheta}$. When we combine Eqs.~(\ref{formal-sol}),~(\ref{effective-metric}) and~\eqref{ansatz} we find
\be
\zeta_{ab} + \dot{\vartheta} \tilde\epsilon^{0 c d}{}_{(a} \zeta_{b) d,c} = 16 \pi \tilde\epsilon^{0}{}_{c}{}^{d}{}_{(a} \partial^{c} \square_{\eta}^{-1} \left(T_{b) d} - \frac{1}{2} g_{b) d} T \right),
\ee
which can be solved for to find
\be
\zeta_{ab} = 16 \pi \tilde\epsilon^{0}{}_{c}{}^{d}{}_{(a} \partial^{c} \square_{\eta}^{-1}
\left(T_{b) d} - \frac{1}{2} g_{b) d} T \right),
\ee
where we have neglected terms second order in $\vartheta$. The metric perturbation then reduces to
\be
\label{formal-sol2}
h_{ab}  = -16 \pi \; \square_{\eta}^{-1} \left(T_{ab} -
  \frac{1}{2} \eta_{ab} T \right) 
+
16 \pi \dot{\vartheta} \tilde\epsilon^{k \ell i}
 \square_{\eta}^{-1} \left(\delta_{i (a} T_{b) \ell,k} -
   \frac{1}{2} \delta_{i (a} \eta_{b) \ell} T_{,k} \right),
\ee
where $i,j,k$ stand for spatial indices only. 

\subsection{Parameterized Post-Newtonian Expansion}
\label{PPN}

Solar system tests of alternative theories of gravity are best performed within the PPN framework. This framework was first proposed by Eddington, Robertson and Schiff~\cite{Will:1993ns,Schiff:1960gi}, but it matured with the work of Nordtvedt and Will~\cite{Nordtvedt:1968qs,1972ApJ...177..775N,1971ApJ...163..611W,1973ApJ...185...31W} (for a review see e.~g.~ \cite{Will:1993ns}). PPN theory proposes the construction of a model-independent {\emph{super-metric}} that represents the PN approximate solution to a family of gravity theories, parameterized by PPN parameters. Solar system experiments can then measure these parameters, thus selecting a particular member of this family. Currently, many of these parameters have been experimentally determined with tight error bars, all of which are consistent with GR~\cite{Will:2005va}. 

The PPN framework allows for tests of alternative theories of gravity through such PPN parameters. Given an alternative theory, one must first construct its PN solution and then compare it to the PPN super-metric. Through this comparison, one can read off how the PPN parameters depend on fundamental parameters of the alternative theory. But since PPN parameters have been experimental constrained, one can propagate these constraints to the fundamental parameters of the alternative theory under consideration, thus obtaining an automatic Solar System test.

Solar System tests in the PPN framework require the PN solution to the modified field equations. In this framework, however, it is not sufficient to leave the solution expressed in terms of the inverse D'Alembertian operator, but instead it must be parameterized in terms of PPN potentials, which are simply Green function integrals over the stress energy tensor. Moreover, the PN expansion must be carried out to slightly different orders in $v$ for different components of the metric, so as to obtain a consistent Lagrangian formulation of the theory\footnote{For example, such order counting is necessary in order to calculate the gravitational deflection of light consistently to first order.} . In the PPN framework, it suffices to compute $g_{00}$ to ${\cal{O}}(v^4)$, $g_{0i}$ to ${\cal{O}}(v^{3})$ and $g_{ij}$ to ${\cal{O}}(v^{2})$. Finally, the Lorentz gauge differs slightly from the PPN gauge, related via an infinitesimal gauge transformation, where the latter is perturbatively defined via~\cite{Will:1993ns}
\ba
h_{jk,}{}^{k} - \frac{1}{2} h_{,j} &=& {\cal{O}}(4),
\qquad 
h_{0k,}{}^{k} - \frac{1}{2} h^{k}{}_{k,0} = {\cal{O}}(5), 
\ea
where $i,j,k$ stand for spatial indices only in the remaining of this section, $h^{k}{}_{k}$ is the spatial trace of the metric perturbation and the symbol ${\cal{O}}(A)$ stands for terms of ${\cal{O}}(\epsilon^A)$, with $\epsilon$ the standard PN expansion parameter of ${\cal{O}}(1/c)$~\cite{Alexander:2007zg,Alexander:2007vt}.

With this machinery at hand, we can now perturbatively expand the trace-reversed CS modified field equations. For the remaining of this chapter, we shall concentrate on the non-dynamical formalism of CS modified gravity, with the canonical choice of CS scalar. Expressions for the linearized Ricci and C-tensors to ${\cal{O}}(4)$ in an arbitrary gauge and for generic $\vartheta$ are long and unilluminating, so we shall not present them here, but they can be found in~\cite{Smith:2007jm}. On the other hand, in the PPN gauge and with the canonical choice for $\vartheta$, the linearized Ricci tensor becomes
\ba
R_{00} &=& - \frac{1}{2} \nabla^2 h_{00} - \frac{1}{2} h_{00,i}
h_{00,}{}^{i} + \frac{1}{2} h^{ij} h_{00,ij} + {\cal{O}}(6),
\nonumber \\
R_{0i} &=& - \frac{1}{2} \nabla^2 h_{0i} - \frac{1}{4} h_{00,0i} + {\cal{O}}(5),
\nonumber \\
R_{ij} &=& - \frac{1}{2} \nabla^2 h_{ij} + {\cal{O}}(4),
\ea
while the linearized C-tensor is given by
\ba
\label{expanded-Cotton}
C_{00} &=& {\cal{O}}(6),
\nonumber \\
C_{0i} &=& - \frac{1}{4} \dot{\vartheta} \tilde{\epsilon}^{0kl}{}_{i} \nabla^2
h_{0l,k} + {\cal{O}}(5), 
\nonumber \\
C_{ij} &=& - \frac{1}{2} \dot{\vartheta} \tilde{\epsilon}^{0kl}{}_{(i}
\nabla^2 h_{j) l,k} + {\cal{O}}(4),
\ea
where $\nabla = \eta^{ij} \partial_i \partial_j$ is the flat-space Laplacian. As in Sec.~\ref{PN-sol}, we find two distinct corrections due to the CS modification: one to the transverse-traceless part of the spatial metric and the
other to the vector metric perturbation.

Let us now solve the linearized CS modified field equations iteratively and perturbatively. To ${\cal{O}}(2)$, the $00$ component of the metric is not modified by the C-tensor and the field equation becomes
\be
\label{h00}
\nabla^2 h_{00} = -8 \pi \rho,  
\ee
because $T=-\rho$. Equation~(\ref{h00}) is the Poisson equation, whose solution in terms of PPN potentials is
\be
\label{h00-sol}
h_{00} = 2 U + {\cal{O}}(4),
\ee
where $U$ is the Newtonian potential~\cite{Alexander:2007zg,Alexander:2007vt}. We shall not present these potentials here, but they can be found in~\cite{Will:1993ns}.

To this same order, the $0i$ component does not provide any information, while the $ij$ sector leads to the following field equations:
\be
\label{hij}
\nabla^2 h_{ij} +  \dot{\vartheta} \tilde{\epsilon}^{0kl}{}_{(i}
\nabla^2 h_{j) l,k} = -8 \pi \rho \delta_{ij},
\ee
where $\delta_{ij}$ is the Kronecker delta. This equation can be rewritten in terms of the superpotential of Sec.~\ref{PN-sol} as
\be
\label{effective-H}
\nabla^2 {\cal{H}}_{ij} = -8 \pi \rho \delta_{ij},
\ee
whose solution is~\cite{Alexander:2007zg,Alexander:2007vt}
\be
\label{effective-H-solved}
{\cal{H}}_{ij} = 2 U \delta_{ij} + {\cal{O}}(4).
\ee
Following the same procedure as in Sec.~\ref{PN-sol}, we can use the decomposition of Eq.~\eqref{ansatz} to find that 
\be
\label{ij-O2}
\zeta_{ij} +  \dot{\vartheta} \tilde{\epsilon}^{0kl}{}_{(i}
\zeta_{j) l,k} =  0.
\ee
The second term on the left-hand side of Eq.~\eqref{ij-O2} is a second order correction in $\dot\vartheta$ and can thus be neglected, which then renders $h_{ij} = {\cal{H}}_{ij}$ to ${\cal{O}}(2)$, where the latter is given in Eq.~\eqref{effective-H-solved}. The physical reason why the CS modification does not correct the spatial sector of the metric is related to the source studied here, together with the PPN gauge. In fact, we shall see in Sec.~\ref{GWsols} that when one studies gravitational wave propagation in vacuum, the spatial sector of the metric is indeed modified. Moreover, if we were to study the ${\cal{O}}(4)$ corrections to the spatial sector of the metric, we would probably find non-vanishing CS corrections, but such a study has not yet been carried out.

To next order, ${\cal{O}}(3)$, the only relevant field equations are related to the gravitomagnetic sector of the metric. The field equations become~\cite{Alexander:2007zg,Alexander:2007vt}
\be
\label{h0i}
\nabla^2 h_{0i} + \frac{1}{2} h_{00,0i} + \frac{1}{2} \dot{\vartheta}
\tilde\epsilon^{0kl}{}_{i} \nabla^2h_{0l,k} 
= 16 \pi \rho v_i,
\ee
which with the lower-order solution of Eq.~\eqref{h00-sol} and the superpotential of Eq.~\eqref{effective-metric} becomes
\be
\label{DE-h0i}
\nabla^2 {\cal{H}}_{0i} + U_{,0i} = 16 \pi \rho v_i, 
\ee
whose solution is
\be
\label{H0i-sol}
{\cal{H}}_{0i} = -\frac{7}{2} V_i - \frac{1}{2} W_i,
\ee
where $V_{i}$ and $W_{i}$ are PPN vector potentials. Combining Eq.~\eqref{effective-metric} with Eq.~\eqref{ansatz}, we can solve Eq.~\eqref{H0i-sol} to find~\cite{Alexander:2007zg,Alexander:2007vt}
\be
h_{0i} =  -\frac{7}{2} V_i - \frac{1}{2} W_i + 2 \dot{\vartheta} \left(\nabla
  \times V\right)_i + {\cal{O}}(5),
\ee
where $\left(\nabla \times A\right)^i = \tilde\epsilon^{ijk} \partial_j A_k$ is the standard curl operator of flat space

Finally, to next order, ${\cal{O}}(4)$, we need only analyze the $00$ component of the modified field equations. Since the $00$ component of the $C$-tensor does not contribute to the field equations to this order, the modified field equations reduce exactly to those of GR, as also does their solution. We therefore find that the PPN solution to the CS modified field equations is given by
\ba
\label{CS-full-metric}
g_{00} &=& g_{00}^{(GR)} + {\cal{O}}(6), 
\nonumber \\
g_{0i} &=& g_{0i}^{(GR)} + 2 \dot{\vartheta} \left(\nabla \times V\right)_i + {\cal{O}}(5),
\nonumber \\
g_{ij} &=& g_{ij}^{(GR)}+ {\cal{O}}(4),
\ea
where $g_{ab}^{(GR)}$ is the PPN solution of GR. This solution in fact satisfies the Pontryagin constraint $^{\star}R R = 0$ to leading order because the contraction of the Levi-Civita symbol with two partial derivatives vanishes. Thus, the equations of motion for the fluid can be obtained directly from the covariant derivative of the stress-energy tensor (the strong equivalence principle holds).

By comparing this PPN solution to the super-metric of the PPN framework, we can now read off the PPN parameters of CS gravity~\cite{Alexander:2007zg,Alexander:2007vt}. Doing so, one finds that all PPN parameters of CS gravity are identical to those of GR ($\gamma = \beta = 1$, $\zeta = 0$ and $\alpha_1 = \alpha_2 = \alpha_3 = \xi_1 = \xi_2 = \xi_3 = \xi_4 = 0$), except for the new term in $g_{0i}$. This CS correction cannot be mimicked by any standard PPN parameter, and thus, in order to model parity-violating theories an additional PPN parameter $\chi$ must be introduced, namely
\be
g_{0i} = g_{0i}^{(PPN)} +  \chi \left(M \nabla \times V\right)_{i},
\ee
where we have multiplied the curl operator by the total mass $M$, in order to make $\chi$ a proper dimensionless parameter. In CS modified gravity, this parameter is simply $\chi = 2 \dot{\vartheta}/M$. 

Canonical CS modified gravity can then be tested by experimentally constraining $\chi$, which thus directly  places a constraint on the canonical CS scalar $\dot{\vartheta} = 1/\mu$. The only requirement for such a contribution to be non-vanishing is that the PPN vector potential $V^{i}$ be non-vanishing.  This is satisfied by non-static sources, {\emph{ie.~}}objects that are either moving or spinning relative to the PPN rest frame. Just because the correction to the gravitomagnetic sector of the metric is non-vanishing, however, does not imply that physical observables, such as the Lense-Thirring effect, will also be corrected. In fact, as we shall see in Sec.~\ref{PPN-tests}, for such corrections to emerge, the curl of the gravitomagnetic correction to the metric must be non-vanishing, which holds only for moving point particles. 

A caveat should be discussed at this point. As we already alluded to, the point-particle approximation holds in GR for many sources, such as black holes, neutron stars or other regular stars, because the effacing principle holds and Birkhoff's theorem also holds. In CS gravity, however, the former does not necessarily hold and the junction conditions are modified. One should then also be aware that a homogenous solution to the linearized field equations might be lacking from this analysis. Such a solution is most likely oscillatory in nature and would not affect the average behavior of the correction, probably acquiring importance only when the latter vanishes. In the next section we shall see how such a boundary solution arises and contributes significantly for spinning bodies, for which the homogeneous correction to the Lense-Thirring effect vanishes.

\subsection{Rotating Extended Bodies and the Gravitomagnetic Analogy}
\label{far-field}

The analysis presented so far has dealt with the non-dynamical formalism, concentrating primarily on point particle sources. Approximate solutions, however, can be developed for extended objects as well. As already described, in GR the effacing principle guarantees that for certain sources both the extended and point particle approach coincide up to rather high PN order. This result somewhat relies on the junction conditions of GR, which here are modified due to the presence of the C-tensor in the field equations. Moreover, the PN scheme is not the only formalism in which far field solution can be investigated. In particular, the gravitoelectromagnetic analogy~\cite{1986bhmp.book.....T,Mashhoon:2003ax}  is also useful to obtain far field solutions, since it allows one to use machinery from the theory of electromagnetism. 

Smith, {\emph{et.~al.~}}\cite{Smith:2007jm} performed the first detailed analysis of rotating, extended objects in CS modified gravity in the gravitoelectromagnetic scheme, modeled via the stress-energy tensor 
\be
     T_{ab} = 2 t_{(a} J_{b)} - \rho t_{a} t_{b},
\ee
where the current $J_a := (-\rho, \vec{J})$ and $t^{a} = [1,0,0,0]$ is a time-like unit vector\footnote{Recall that indices are raised and lowered with the Minkowski metric in the far-field approximation.}. Such a choice of stress-energy is equivalent ot that of a pressureless perfect fluid in the limit as the fluid balls tend to particles. Moreover, Smith, {\emph{et.~al.~}}\cite{Smith:2007jm} focused on the dynamical formulation of CS modified gravity,  with the conventions $\alpha = -\ell/3$ and $\beta = -1$. Such a study is forbidden in the non-dynamical formalism, because the Pontryagin constraint does not allow for rotating solutions. In the dynamical scheme, however, such solutions are allowed, provided the Pontryagin density is balanced by the dynamics of the scalar field. 

In spite of working in the dynamical theory, Smith,  {\emph{et.~al.~}}\cite{Smith:2007jm} chose a canonical $\vartheta$, thus implicitly treating this field as non-dynamical. Such a choice of scalar field is formally inconsistent with dynamical CS modified gravity, since the non-vanishing Pontryagin constraint will force spatial variations on $\vartheta$.  Nonetheless, this choice could become a good approximation, if one is only concerned with far-field solutions. This is because, as we have shown in Eq.~\eqref{Pkerr} of Sec.~\ref{Pont-Kerr}, the exterior Pontryagin density for a rotating source scales as $\pont \approx a M^{2}/ r^{7}$, which then forces 
\be
\beta \square \vartheta = -24 \alpha \frac{a M^{2}}{r^{7}} = 8 \ell \frac{a M^{2}}{r^{7}},
\ee
with uncontrolled remainders of ${\cal{O}}(M/r)^{4}$. Thus, spatial variations will induce modifications to the canonical $\vartheta$ that are at most of ${\cal{O}}(M/r)^{3}$, which is beyond the order considered in this section. A full dynamical study of this problem has yet to be carried out. 

Let us now return to the gravitoelectromagnetic formalism, in which certain components of the metric perturbation are identified with a four-vector potential, namely 
\be
\label{vec_pot_def}
A_{a} := - \frac{1}{4} \bar{h}_{a b} t^{b},
\ee
where $\bar{h}_{ab}$ is the trace-reversed metric perturbation $\bar{h}_{ab} := h_{ab} - \frac{1}{2} \eta_{ab} h$ and $h = \eta^{ab} h_{ab}$ is the trace of the latter. In analogy with electromagnetism, one can now define electric and magnetic fields via 
\be
     E^i = \partial_i A_0 - \partial_0 A_i, \qquad
     B^i = \epsilon^{0ijk}\partial_j A_k,
\ee
where $i,j$ stand for spatial indices only. The geodesic equations can then be written in terms of these fields as
\be
F^{i} =  - m E^{i} - 4 m \left(v \times B\right)^{i},
\ee
where the cross product is that of flat Euclidean space, from which one can read off the Lorentz force acceleration $a^{i} = F^{i}/m$.

We can now apply this formalism to the linearized CS modified field equations. In the Lorentz gauge, $A^{a}_{,a} = 0$, the modified field equations become~\cite{Smith:2007jm}
\ba
    \vec{\nabla} \cdot \vec{B} &=& 0,
    \\
     \vec{\nabla} \times \vec{E} &=& - \frac{\partial \vec{B}}{\partial t}, 
     \\
      \vec{\nabla} \cdot \vec{E} &=& 4 \pi G (\rho + T_{00}^{\vartheta})
      \label{gauss}
      \\    
      \vec{\nabla} \times \vec{B} - \frac{\partial \vec{E}}{\partial t} - \frac{1}{m_{\rm cs}}\Box \vec{B} &=& 4 \pi G \vec{J},
       \label{amp}
\ea
where $m_{\rm cs}$ is a characteristic mass scale defined via
\be
\label{CS-mass}
m_{\rm cs} :=  \frac{2 \kappa}{\alpha \dot{\vartheta}} = - \frac{6 \kappa}{\ell \dot{\vartheta}}.
\ee
In the above equations, the first two arise from the definitions of the four-vector potential, while the last two (Gauss's law in Eq.~\eqref{gauss} and Amp\`ere's law in Eq.~\eqref{amp}) are modified in CS modified gravity, the former arising from the time-time component of the modified field equation and the latter from the gravitomagnetic sector.  Smith,  {\emph{et.~al.~}}\cite{Smith:2007jm} neglect $T_{00}^{\vartheta}$ because they associate it with the energy density of the scalar field, which they argue must be uniform throughout the Solar System, not larger than the mean cosmological energy density and negligible relative to $\rho$. Moreover, from Eq.~\eqref{Tab-theta}, we see that $T_{00}^{\vartheta}$ is at least quadratic in $\dot{\vartheta}$ and can thus be neglected, provided $3 \beta/(2 \rho \mu^{2}) \ll 1$. 

Let us now specialize the above treatment to that of homogenous rotating source, with mass current
\be
     \vec{J} = \rho \left[\vec{\omega} \times \vec{r}\right] \Theta(R-r),
\ee
where $R$ is the radius of the rotating body, $\rho$ is its density, $\vec{\omega}$ is its angular velocity,  $r$ is the distance from the origin, and $\Theta$ is the Heaviside step function. One can now use this mass current in Amp\`ere's law to solve this equation for $A^{i}$, imposing continuity and finiteness at the origin. The actual expressions for the four-vector potential can be found in~\cite{Smith:2007jm}, but its associated gravitomagnetic field is given by $\vec{B} =  \vec{B}_{\rm GR}+ \vec{B}_{\rm CS}$, where the GR piece is simply
\be
     \vec B_{\rm GR}  = \frac{4 \pi G \rho R^2}{15} 
\begin{cases}
     \left(5-3\frac{r^2}{R^2}\right) \vec\omega + 3\frac{r^2}{R^2} \hat r \times (\hat r \times \vec
     \omega), \ &r
        \leq R, \\
     \frac{R^3}{r^3}
     \left[2 \vec\omega + 3 \hat r \times (\hat r \times \vec
     \omega)\right],\ &r \geq R, 
\end{cases}
\label{BfullGR}
\ee
while the CS correction is given by
\be
     \vec B_{\rm CS} = 4 \pi G \rho R^2\left\{ D_1(r)\, \vec\omega
     +  D_2(r)\, \hat r \times \vec\omega 
     + D_3(r) \, \hat r \times (\hat r \times
     \vec\omega) \right\}.
\label{BfullCS}
\ee
The function $D_{1,2,3}(r)$ are actually functionals of spherical Bessel functions of the first $j_{\ell}(x)$ and second $y_{\ell}(x)$ kind, given explicitly in~\cite{Smith:2007jm}. Remarkably, the gravitoelectromagnetic analogy allows for a solution to the linearized modified field equations for an extended rotating source, which possesses both a poloidal and toroidal gravitomagnetic field. In other words, the metric does not only contain gravitomagnetic components co-aligned with the spin axis of the rotating extended body, but also along other axis, perpendicular to the plane defined by $\vec{\omega}$ and $\vec{r}$. Such terms cannot indeed be removed via a coordinate transformation since $B_{\phi}^{i}$ is generically non-vanishing. 

The solution for the metric perturbation found in the gravitoelectromagnetic analogy for extended sources differs slightly from that found in the PN scheme for point particles. Some of these differences arise because in the latter the point particles were allowed to possess non-vanishing angular and linear momentum. The main difference, however, is not due to the stress-energy tensor studied, but to the fact that extended and point particle treatments are not equivalent in CS modified gravity. In the former, an additional oscillatory behavior is needed, encoded in the appearance of Bessel functions, in order to guarantee continuity across the surface of the sphere. Such terms are necessary because the C-tensor modifies the junction conditions of GR~\cite{Lanczos:1922,Lanczos:1924,Darmois:1927,Lichnerowicz,Misner:1964,Barrabes:1991ng,Israel:1966nc,Misner:1973cw,poisson}.  

The gravitoelectromagnetic analogy also sheds some light on the study of exact solutions to CS modified gravity, reviewed in Sec.~\ref{Sec:ExactSol}. The CS correction to Amp\`ere's law, Eq.~(\ref{amp}) changes the character of the differential system, from a first-order one to a second-order one. Such a result is in fact expected from the structure of the modified field equations [Eq.~\eqref{eq:eom}], since the C-tensor depends on third-order derivatives of the metric. One then could also expect that approximate solution to CS modified theory need not be easily obtained as deformations of GR solutions. This is because, in principle, the CS correction could produce dramatic changes to the dynamical behavior of the solution (ie.~the solution, for example, could be an element of ${\cal{CS}} \; \backslash \; {\cal{P}}$ in the nomenclature of Sec.~\ref{Sec:ExactSol}, which would hold little resemblance to GR solutions). CS modified theory and GR, however, have both now been sufficiently tested to allow for a perturbative treatment of the modified theory, at least in the Solar System. Whether such treatment is allowed in more non-linear, strong field scenarios remains to be studied further.

\subsection{Perturbations of the Schwarzschild Spacetime}

Black hole perturbation theory has been incredibly important in GR, leading for example to a better understanding of subtle issues related to BH physics and the extraction of gravitational waves in dynamic spacetimes. Moreover, through the discovery of quasinormal ringing, new tests of GR have been proposed (see eg.~\cite{lrr-2006-3} for a review). These accomplishments alone suggest a study of BH perturbation theory in CS modified gravity could lead to interesting results. Such a study was carried out by Yunes and Sopuerta~\cite{Yunes:2007ss} in the non-dynamical formalism with $\alpha = \kappa$ and $\beta = 0$. 

Let us begin by introducing the basics of BH perturbation theory, as developed in GR. Consider then the following perturbed metric 
\be
\label{BHPT-split}
g_{ab} = \bar{g}_{ab} + h_{ab}\,,
\ee
where $\bar{g}_{ab}$ is the background Schwarzschild metric and $h_{ab}$ is a generic metric perturbation. Henceforth, objects associated with the background will be denoted with an overhead bar. The spherical symmetry of the background allows one to expand the metric perturbation in tensor spherical harmonics, thus separating the angular $i=\{\theta,\phi\}$ and temporal-radial $\mu=\{t,r\}$ dependence in the perturbed field equations and yielding a $1+1$ PDE system. This system can be further simplified by distinguishing between polar and axial parity harmonics: polar (axial) modes acquire a $(-1)^{l}$ [$(-1)^{l+1}$] prefactor upon parity transformations. The aforementioned simplification comes about because in GR the perturbative field equations decouple into two subsystem that can be classified by their parity.  


With these considerations in hand, we split the metric perturbation into $h_{ab} = h^{\mbox{\small a}}_{ab} +
h^{\mbox{\small p}}_{ab}$, where each of these pieces is decomposed in tensor spherical harmonics: the temporal-radial sector becomes
\be
h^{\mbox{\small a}}_{\mu \nu} = 0, 
\qquad
h^{\mbox{\small p}}_{\mu \nu} = 
\sum_{\ell,m} h^{\mbox{\small p},\ell m}_{\mu \nu} = 
\sum_{\ell,m} h_{\mu \nu}^{\ell m}\, Y^{\ell m},
\ee
the temporal-angular sector is given by
\be
h^{\mbox{\small a}}_{\mu i} = 
\sum_{\ell,m} h^{\mbox{\small a},\ell m}_{\mu i} =
\sum_{\ell,m} h_{\mu}^{\ell m} S_{i}^{\ell m}, 
\qquad
h^{\mbox{\small p}}_{\mu i} = 
\sum_{\ell,m} h^{\mbox{\small p},\ell m}_{\mu i} = 
\sum_{\ell,m} p_{\mu}^{\ell m}\, Y^{\ell m}_{i},
\ee
and the angular-angular sector becomes
\be
h^{\mbox{\small a}}_{i j} = 
\sum_{\ell,m} h^{\mbox{\small a},\ell m}_{i j} =
\sum_{\ell,m} H^{\ell m} S_{i j}^{\ell m}, 
\qquad
h^{\mbox{\small p}}_{i j} = 
\sum_{\ell,m} h^{\mbox{\small p},\ell m}_{i j} = 
\sum_{\ell,m} r^{2} \left(K^{\ell m} Y_{ij}^{\ell m} + G^{\ell m} Z_{ij}^{\ell m} \right),
\ee
asterisks denote components given by symmetry and where $Y^{\ell m}$ are standard scalar spherical harmonics [see~\cite{Sopuerta:2006wj} for conventions], $Y_{i}^{\ell m}$ and $S_{i}^{\ell m}$ are polar and axial vector spherical harmonics respectively, and $Y^{\ell m}_{ij}$, $Z_{ij}^{\ell m}$ and $S_{ij}^{\ell m}$ are polar, polar and axial, tensor spherical harmonics. Vector spherical harmonics are defined for $\ell \geq 1$ via
\be
Y^{\ell m}_i\equiv Y^{\ell m}_{:i}\,, \qquad
S^{\ell m}_j\equiv \epsilon^{}_i{}^j\,Y^{\ell m}_j\,.
\label{vec-sph-harm}
\ee
while tensor spherical harmonics are defined for $\ell \geq 2$ via
\be
Y_{ij}^{\ell m} := Y^{\ell m}\Omega^{}_{ij}\,,
\qquad
Z^{\ell m}_{ij} := Y^{\ell m}_{:ij}+\frac{\ell(\ell+1)}{2}Y^{\ell m}
\Omega_{ij}\,, 
\qquad
S^{\ell m}_{ij} := S^{\ell m}_{(i:j)}\,,
\ee
where $\Omega_{ij}$ and $\epsilon^{ij}$ are the metric and Levi-Civita tensor on the $2$-sphere respectively, while colon stands for covariant differentiation on the $2$-sphere. All metric perturbations are functions of $t$ and $r$ only.

The perturbative field equations can be decoupled in GR in terms of complex master functions, known as the Cunningham-Price-Moncrief (CPM) master function~\cite{Cunningham:1978cp} and the Zerilli-Moncrief (ZM) master function~\cite{Zerilli:1970fj,Moncrief:1974vm}. The former is given by
\be
\Psi^{\ell m}_{\CPM} = -\frac{r}{\lambda^{}_\ell}\left( 
h^{\ell m}_{r,t} -  h^{\ell m}_{t,r} + \frac{2}{r}h^{\ell m}_t \right)\,, 
\label{PsiCPM_sch},
\ee
where $\lambda_{\ell} =  (\ell+2)(\ell-1)/2$, while the latter is given by
\be
\Psi^{\ell m}_{\ZM} = \frac{r}{1+\lambda^{}_\ell}\left\{ K^{\ell m}
  + (1+\lambda^{}_\ell)G^{\ell m}
+ \frac{f}{\Lambda^{}_\ell}\left[ f h^{\ell
  m}_{rr}-r  K^{\ell m}_{,r} -
  \frac{2}{r}(1+\lambda^{}_\ell)p^{\ell m}_r \right] \right\} \,, 
\label{PsiZM_sch}
\ee
where $f = 1 - 2 M/r$ is the Schwarzschild factor and $\Lambda_{l} = \lambda^{}_\ell + 3M/r$. 

The perturbative field equations can be written in terms of these master function, obtaining the so-called master equations
\be
\left[-\partial^{2}_{t^2} + \partial^{2}_{r^{2}_{\star}}
 - V^{\POLAR/\AXIAL}_\ell(r)\right]\Psi_{\CPM/\ZM}^{\ell m} = 0 \,,
\label{masterequations} 
\ee
where $r^{}_{\star}$ is the {\em tortoise} coordinate $r^{}_{\star} = r + 2M\ln\left[r/(2M)-1\right]$ and $V_{\ell}^{\POLAR/\AXIAL}$ are potentials, which depend on $r$, parity and harmonic number~\cite{Sopuerta:2006wj}. In principle, the right-hand side of the master equations is not zero, but depends on some matter sources, which we neglect here since we are searching for vacuum metric perturbations. Moreover, although the master functions are gauge invariant, the analysis is simplified if one picks a gauge. Henceforth we choose the Regge-Wheeler gauge in which $H^{\ell m} = 0 = G^{\ell m} = p_{\mu}^{\ell m}$. 

With all this machinery, we can now study BH perturbation theory in CS modified gravity. Yunes and Sopuerta~\cite{Sopuerta:2006wj} studied the non-dynamical formalism, in which the Pontryagin plays a critical role. In terms of perturbation functions this constraint reduces to
\be
\pont = \frac{96 M}{r^6} \left[h_t^{\ell m} + \frac{r}{2} \left(
 h_{r,t}^{\ell m} -  h_{t,r}^{\ell m} \right) \right]
\ell \left( \ell +1 \right) Y^{\ell m}\,.  
\label{pconsch}
\ee
Although this constraint is automatically satisfied by the background, it is not satisfied for generic metric perturbations. Remarkably, this precise combination of metric perturbations can be written in terms of the CPM function exactly:
\be
\pont = - \frac{24 M}{r^6} \frac{\left(\ell + 2 \right)!}{\left(\ell -
    2 \right)!} \Psi_{\CPM}^{\ell m} Y^{\ell m}\,. 
\label{deltapont}
\ee
The Pontryagin constraint then forces the CPM function to vanish for all harmonics $\ell \geq 2$, which is not equivalent to requiring that all axial perturbations also vanish. One then arrives at the constraint
\be
\label{pont2}
h_{r,t}^{\ell m} =  h_{t,r}^{\ell m} - \frac{2}{r} h_{t}^{\ell
  m}\,. 
\ee
The set of allowed solutions is thus reduced by this constraint, which in fact tends to lead to an overconstrained system of perturbative field equations, as we shall see next. It is also in this sense that the non-dynamical framework leads to an overconstrained system of partial differential equations. 

Let us now concentrate on the perturbative field equations in CS modified gravity. In order to simplify calculations, Yunes and Sopuerta~\cite{Sopuerta:2006wj} chose a scalar field of the form $\vartheta = \bar{\vartheta}(t,r)$, which possesses the same symmetries as the background. After harmonically decomposing the perturbative field equations, one finds~\cite{Sopuerta:2006wj} 
\ba
\label{decoup1}
{\cal{G}}^{\ell m}_{\mu \nu}[\mb{U}^{\ell
  m}_{\POLAR}] &=& - {\cal{C}}^{\ell m}_{\mu \nu}[\mb{U}^{\ell m}_{\AXIAL}]\,, 
\qquad
{\cal{G}}^{\ell m}_{\mu}[\mb{U}^{\ell
  m}_{\POLAR}] = -  {\cal{C}}^{\ell m}_{\mu}[\mb{U}^{\ell m}_{\AXIAL}]\,, 
\\
{\cal{G}}^{\ell m}[\mb{U}^{\ell
  m}_{\POLAR}] &=& -  {\cal{C}}^{\ell m}[\mb{U}^{\ell m}_{\AXIAL}]\,, 
\qquad
{\cal{H}}^{\ell m}[\mb{U}^{\ell
  m}_{\POLAR}] = -  {\cal{D}}^{\ell m}[\mb{U}^{\ell m}_{\AXIAL}]\,,
\label{decoup1bis}
\\
{\cal{H}}^{\ell m}_{\mu}[\mb{U}^{\ell m}_{\AXIAL}] &=& - {\cal{D}}^{\ell m}_{\mu}[\mb{U}^{\ell
  m}_{\POLAR}]\,,
\qquad
{\cal{I}}^{\ell m}[\mb{U}^{\ell m}_{\AXIAL}] = -  {\cal{E}}^{\ell m}[\mb{U}^{\ell
  m}_{\POLAR}]\,.
\label{decoup2}
\ea
where $\mb{U}^{\ell m}_{\POLAR}$ and $\mb{U}^{\ell m}_{\AXIAL}$ denote polar and axial metric perturbations respectively 
\be
\mb{U}^{\ell m}_{\POLAR} = (h^{\ell m}_{\mu \nu}\,, p^{\ell m}_{\mu}\,,
K^{\ell m}, G^{\ell m}) \,,
\qquad
\mb{U}^{\ell m}_{\AXIAL} = (h^{\ell m}_{\mu}\,, H^{\ell m} )\,,
\ee
and where the right hand sides of Eq.~\eqref{decoup1} and~\eqref{decoup2} also depend on derivatives of $\vartheta$. The functionals in Eqs.~\eqref{decoup1} and~\eqref{decoup2} are long and unilluminating, so we shall omit them here, but they are presented in full detail in~\cite{Sopuerta:2006wj}. Perhaps not too surprisingly, the parity-violation induced by CS modified gravity breaks the axial-polar decoupling expected in GR. Instead, we now find that modes with opposite parity are coupled and cannot in general be treated separately. 

Due to the non-decoupling of the perturbative field equations, a generic study of their solution is a quixotic task. Yunes and Sopuerta~\cite{Sopuerta:2006wj} investigated several specific cases, from which one can then extrapolate generic conclusions. Let us then first consider the canonical choice of $\vartheta$ and single-handed metric perturbations, {\emph{ie.~}}purely polar or purely axial perturbations. In either of these cases, one can show that the Pontryagin constraint leads to an overconstrained PDE system, and thus, if one set of perturbations vanishes, then all metric perturbations must vanish. Moreover, these conclusions do not only hold for the canonical choice of $\vartheta$ but also for many other members of the family $\bar{\vartheta}(t,r)$~\cite{Sopuerta:2006wj}. However, a generic result for arbitrary $\vartheta$ or arbitrary axial and polar metric perturbations has not yet been produced, due to the incredible complexity of the perturbative equations.  

The overconstraints of the non-dynamical formalism are mainly due to the Pontryagin condition, which is relaxed in the dynamical formalism. Yunes and Sopuerta~\cite{Sopuerta:2006wj} have studied BH perturbation theory in the dynamical formalism, where $\vartheta$ and $h_{ab}$ are treated as independent perturbation parameters. Consider then the split
\ba
\label{metricpert}
g^{}_{ab} & = & \bar{g}^{}_{ab} + \epsilon\,h^{}_{ab}\,, \\
\label{theta-chosen}
\vartheta & = & \tau(\bar{\vartheta} + \epsilon\,\delta\vartheta)
      =   \tau\bar\vartheta + \tau\epsilon
\sum^{}_{\ell\geq 1, m} \tilde{\vartheta}^{\ell m} \, Y^{\ell m}\,,
\ea
where here $\epsilon$ and $\tau$ are bookkeeping, independent perturbation parameters. The expansion for $\vartheta$ is a bivariate expansion, where $\bar{\vartheta}$ is some background value for the CS coupling and is spherically symmetric, while $\tilde{\vartheta}^{\ell m}$ are harmonic coefficients of the $\vartheta$ perturbation.

In the dynamical formalism, the Pontryagin constraint is replaced by an evolution equation for the scalar field, which to leading order is simply $\bar{\square}\bar{\vartheta} = 0$, where we have set the potential to zero. To first order in $\epsilon$, the evolution equations become
\be
- \epsilon \frac{\alpha}{4}\delta(\pont)
= \epsilon \tau \beta \left\{\bar{\square}\delta\vartheta - \left[\bar\vartheta^{}_{,ab}
+ \left(\ln\sqrt{-\bar{g}}\right)^{}_{,a}\bar\vartheta^{}_{,b} \right]h^{ab}
- 
\bar\vartheta^{}_{,a}h^{ab}{}^{}_{,b} + \frac{1}{2} h_{,b}
\bar{g}^{ab} \bar{\vartheta}_{,b} \right\} \,. \label{eqdeltatheta}
\ee
where $\delta(\pont)$ is the functional coefficient of $\pont$ to ${\cal{O}}(\epsilon)$ given explicitly in~\cite{Sopuerta:2006wj}. Similarly, to leading order in $\epsilon$ the perturbative field equations also become modified, with corrections that arise from the expansion of the stress-energy tensor of $\vartheta$. By relaxing the Pontryagin constraint, the PDE system ceases to be overconstrained and might allow generic metric oscillations. However, since $\tau \ll 1$, the magnitude of the CPM master function will be constrained to be small, which could lead to interesting observational consequences, for example in the emission of energy by gravitational wave radiation. A full dynamical study of the solution to these equations has not yet been carried out. 

\subsection{Slowly Rotating Kerr-like Black Holes}

Instead of studying arbitrary perturbations, let us now concentrate on perturbations that represent slow-rotation deformations of the Schwarzschild spacetime. By searching for such solutions, we might gain some insight on how to extend the Kerr metric in CS modified gravity. In the previous section, we saw that a CS scalar of the form $\vartheta = \vartheta(t,r)$ will lead to an overconstrained system of equations, so in this section we shall explore more general choices. 

The slow-rotation limit of stationary and axisymmetric line elements in non-dynamical CS modified gravity ($\alpha = \kappa$ and $\beta = 0$) was first studied by Konno, {\emph{et.~al.~}}\cite{Konno:2007ze}. Consider then the following general, stationary and axisymmetric line element
\ba
\label{ds2-konno}
ds^{2} &=& -f \left[1 + h(r,\theta) \right] + f^{-1} \left[ 1 + m(r,\theta) \right] 
\nonumber \\
&+& r^{2} \left[1 + k(r,\theta) \right] \left[ d\theta^{2} + \sin^{2}\theta \left(d\phi - \omega(r,\theta) dt \right)^{2} \right] 
\ea
where $f = 1 - 2 M/r$ is the Schwarzschild factor, $h,m,k,\omega$ are unknown functions of $r$ and $\theta$. The metric perturbations are assumed to be linear in the perturbation parameter: $J/M^{2}$, the ratio of the angular momentum of the rotating compact object to the squared of its mass. With this line element, the Pontryagin constraint leads to a condition on the function $\omega$, namely $\omega_{,r\theta} + 2 \cot \theta \omega_{,r} = 0$, which leads to the solution $\omega = \bar{\omega}(r)/\sin^{2} \theta$. Such a result is so far independent of the choice of CS scalar. 

The modified field equations can be now linearized in the unknown functions and solved, given some choice for $\vartheta$. When a canonical scalar field is chosen, Konno, {\emph{et.~al.~}}\cite{Konno:2007ze} showed that the linearized equations force these functions to vanish and thus a rotating solution cannot be found to first order. Note that this result is not in disagreement with the discussion in Sec.~\ref{PPN} and~\ref{far-field}, since there the far-field solutions found cannot be put in the form of Eq.~\eqref{ds2-konno}.

A solution to the linearized CS modified field equation can in fact be obtained in the slow-rotation limit, provided we explore other choices for $\vartheta$.  Konno, {\emph{et.~al.~}}\cite{Konno:2007ze} showed that with the scalar 
\be
\label{Konno-scalar}
\vartheta = r \cos{\theta}/\lambda_{0},
\ee
a solution can indeed be found and it is given by 
\ba
h(r,\theta) &=& m(r,\theta) = k(r,\theta) = 0,
\nonumber \\
\bar{\omega}(r) &=& \frac{D_{1}}{r^{2}} f + \frac{D_{2}}{r^{3}} \left[ r^{2} - 2 M r - 4M^{2} + 4 M r f \ln(r - 2 M)\right]
\label{Konnoomega}
\ea
which leads to the Schwarzschild metric, plus a new term in the $t\phi$ sector of the metric, namely, 
\be
\label{KMTsol}
g_{t\phi} = D_{3} f + \frac{D_{4}}{r} \left[ r^{2} - 2 M r - 4 M^{2} + 4 M r f \ln(r - 2M) \right],
\ee
where $D_{i}$ are constants of integration that are assumed linear in $J$. Note, however, that if $D_{4} \neq 0$, the gravitomagnetic sector of the metric naively looks as if it could diverge in the limit as $r \to \infty$ increases. On closer inspection, however, one finds that invariants and physically relevant observables do not diverge. For example, the scalar invariant $R^{abccd}R_{abcd} \propto 48 M^{2}/r^{6} - 4 D_{4}^{2}/(r^{4} \sin^{2} \theta)$, which indeed vanishes at spatial infinity, where the divergence at $\theta = 0$ or $\pi$ presumably arises due to the first-order linear perturbation scheme~\cite{Konno:2008np}. The quantity $\lambda_{0}$ in $\vartheta$ is a constant with units of inverse length, which curiously does not appear in the solution for $\bar{\omega}$. This is because the embedding coordinate can be factored out and does not enter into the linearized modified field equations to leading order in the angular momentum. 

Interestingly, the above solution cannot be interpreted as a small deformation of the Kerr line element. That is, there is no choice of $D_{i}$ for which $g_{t \phi}$ can be considered a small deformable correction to Kerr. Such an observation implies that the frame-dragging induced by such a metric will be drastically different from that predicted by the Kerr line element, in fact sufficiently so to allow for an explanation of the anomalous velocity rotation curves of galaxies which we shall discuss further in Sec.~\ref{Sec:AstroTests}. At the same time, however, Solar System experiments have already measured certain precessional effects in agreement with the GR prediction~\cite{Ciufolini:2007wx,Ciufolini:2004rq,Ciufolini:2004gp}, and thus a drastically different frame-dragging prediction might be in contradiction with these Solar System tests.

Recently, Yunes and Pretorius~\cite{slow-rot} have extended and generalized this result. They showed that in fact the solution in Eq.~\eqref{Konnoomega} is preserved for any $\vartheta$ in the family
\be
\vartheta^{\textrm{gen}} = A_{0} + A_{x} r \cos{\phi} \sin{\theta} + A_{y} r \sin{\phi}  \sin{\theta}  + A_{z} r  \cos{\theta},
\ee
where $A_{i}$ are constants. In fact, we can rewrite this CS coupling field as $\vartheta = \delta_{ab} A^{a} x^{b}$, where $x^{a} = [1,x,y,z]$ and
$\delta_{ab}$ is the Euclidean metric. Note, however, that the stress-energy tensor associated with any member of this family [including Eq.~\eqref{Konno-scalar}] is constant, and thus the energy associated with such a field is infinite. Because of this, the solution found here cannot be extended to the dynamical framework. 

Moreover, Yunes and Pretorius~\cite{slow-rot} also found another solution to the slow-rotation limit of the modified field equations, if one considers a generic CS scalar field in the non-dynamical framework: 
\ba
\vartheta &=& \bar{f}(r,\phi) + r \bar{g}(\phi) + r \bar{h}\left(C_{1} \phi - t\right) +  r \bar{k}(\theta,\phi)
\nonumber \\
&+& r \int \frac{dr}{r} \left[ -\partial_{r} \bar{f}(r,\phi)+ \frac{1}{r} \bar{f}(r,\phi) + \frac{1}{r} \bar{j}(r)\right] , 
\nonumber \\
\label{w-good}
\bar{\omega} &=& -\frac{C_{1}}{r^{2}} f,
\ea
where $\bar{f}$, $\bar{g}$, $\bar{h}$, $\bar{j}$ and $\bar{k}$ are arbitrary functions and $C_{1}$ is another integration constant. 
This arbitrary function can be chosen such that the new CS scalar possesses a sufficiently fast decaying stress-energy tensor with non-infinite energy.
For example, if $\bar{f} = \bar{g} = \bar{h} = \bar{k} = 0$ and $\bar{j}=-3 j_{0}/r^{2}$, then $\vartheta = j_{0}/r^{2}$, for constant $j_{0}$, and thus 
Eq.~\eqref{w-good} is compatible with the dynamical framework. 

The existence of two independent solutions to the modified field equations in the non-dynamical framework and in the slow-rotation limit suggests that
there is a certain non-uniqueness in the framework encoded in the arbitrariness of the choice of $\vartheta$. This scenario is to be contrasted with the 
dynamical framework, where the CS scalar is uniquely determined by its evolution equation and there is no additional freedom (except for that encoded
in initial conditions).

In view of this problems with the non-dynamical framework, 
Yunes and Pretorius~\cite{slow-rot} studied the same scenario but in the full dynamical framework. 
A new approximation scheme is employed on top of the slow-rotation requirement, which essentially demands that the CS correction be a small deformation of the Kerr line element, ie.~the CS coupling is assumed small relative to the GR one.  In this way, one finds the solution\footnote{Notice that the solution for $\vartheta$ is identical to that found by Campbell~\cite{Campbell:1990ai} and Reuter~\cite{Reuter:1991cb} and discussed in Eq.~\eqref{KR-axion}, except that there the backreaction of this field on the metric was ignored.}~\cite{slow-rot}:
\ba
\label{hairy-sol}
ds^{2} &=& ds^{2}_{\textrm{Kerr}}  + 
\frac{5}{4} \frac{\alpha^{2}}{\beta \kappa} \frac{a}{r^{4}}\left( 1 + \frac{12}{7} \frac{M}{r} + \frac{27}{10} \frac{M^{2}}{r^{2}} \right) \sin^{2}{\theta}\ d\phi dt,
\nonumber \\
\vartheta &=&  \frac{5}{8} \frac{\alpha}{\beta} \frac{a}{M} \frac{\cos(\theta)}{r^2} \left(1 + \frac{2 M}{r} + \frac{18 M^2}{5 r^2} \right),
\ea
where $ds^{2}_{\textrm{Kerr}}$ is the slow-rotation limit of the Kerr metric,
$M$ is the BH mass and $J = a \cdot M$ is the BH angular momentum to leading order. This solution is valid to second order in the slowly-rotation expansion parameter $a/M$, as well as in the strength of the coupling $\alpha/(\beta \kappa M^{4})$.  
Notice that this solution, derived under the dynamical formulation, is perfectly well-behaved at spatial infinity, remaining
asymptotically flat. 

Equation~\eqref{hairy-sol} is the first\footnote{Shortly after publication of this result, Konno, {\emph{et.~al.~}} employed slightly different methods to verify that the solution found by Yunes and Pretorius indeed satisfies the modified field equations~\cite{Konno:2009kg}.} rotating BH solution in dynamical CS modified gravity, and can be thought of as a small deformation of a Kerr black hole with additional CS scalar ``hair'' of finite energy. Although this is a ``hairy'' solution, Sopuerta
and Yunes~\cite{Sopuerta:2009iy} have shown that the solution is still entirely described by the mass and the angular momentum of the source. The no-hair theorem is, however, violated in that the relation between higher-multipoles and the mass quadrupole and current dipole is CS modified at $\ell = 4$ multipole
due to the CS correction in Eq.~\eqref{hairy-sol} of the gravitomagnetic sector. 

The solution found in the dynamical theory presents interesting parity properties. Since the dynamics of the CS scalar field are determined by the Pontryagin
density, this field is parity-violating (ie.~it is a pseudo-scalar). Both the Kerr metric and the CS correction to it are also parity-violating, but the later is induced 
by a curvature-scalar field interaction, instead of due to the Kerr distributional stress-energy. In fact, the parity violation introduced by the CS correction 
becomes dominant in regions of high curvature.   

Moreover, the dynamical solution presented above shows remarkable similarities with some of the far-field results found in the non-dynamical framework. 
In particular, it is only the gravitomagnetic sector of the metric that is CS modified, thus implying that the frame-dragging effect will be primarily corrected. 
Notice, however, that the CS correction is highly suppressed by large inverse powers of radius, which suggest that weak-field tests will not be able to
constrain the dynamical framework. We shall investigate this possibilities further in Sec.~\ref{Sec:AstroTests}.

\subsection{Gravitational Wave Propagation}
\label{GWsols}

GW solutions have been studied by a large number of authors \cite{Jackiw:2003pm,Alexander:2004us,Alexander:2004wk,Alexander:2007kv,pulsars,Guarrera:2007tu,Sopuerta:2009iy}, but mostly in the non-dynamical theory, 
which we shall concentrate on here. The first GW investigation in non-dynamical 
CS modified gravity was carried out by Lue, Wang and Kamionkowski~\cite{Lue:1998mq}, who studied the effect
of GWs in the cosmic microwave background (see Sec.~\ref{SEC:CScosmology}). 
Jackiw and Pi~\cite{Jackiw:2003pm} also studied GWs 
in CS modified gravity, concentrating on the generation of such waves and the power carried by them in the modified theory. Shortly after, such waves were used to explain baryogenesis during inflation~\cite{Alexander:2004us} and to calculate the super-Hubble power spectrum~\cite{Alexander:2004wk}. The generation of GWs was also studied in the dynamical formalism~\cite{Guarrera:2007tu} through the construction of an effective stress-energy tensor and the Isaacson scheme~\cite{Sopuerta:2009iy}. Recently, GW tests have been proposed to constrain CS gravity with space-borne~\cite{Alexander:2007kv} gravitational wave interferometers. 

We shall here discuss GW solutions in non-dynamical CS modified gravity and postpone any discussion of GW generation to the next section. Moreover, we shall not discuss in this section cosmological power spectra, since these will be summarized in Sec.~\ref{SEC:CScosmology}. GW propagation in CS gravity has only been studied in the non-dynamical formalism with $\beta = 0$ and $\alpha = \kappa$. Let us begin with a discussion of GW propagation in a FRW background. Consider then the background
\be
ds^{2} = a^{2}(\eta) \left[ - d\eta^{2} + \left( \delta_{ij} + h_{ij} \right) d\chi^{i} d\chi^{j} \right],
\label{FRW-metric}
\ee
where $a(\eta)$ is the conformal factor, $\eta$ is conformal time and $\chi^{i}$ are comoving coordinates. The quantity $h_{ij}$ stands for the gravitational wave perturbation, which we take to be transverse and traceless (TT), $h := h^{i}{}_{i} = \delta^{ij} h_{ij} = 0$ and $\partial_{i} h^{ij} = 0$. One can show that a coordinate system exists, such that the gravitational wave perturbation can be put in such a TT form. For the remainder of this section, $i,j,k$ stand for spatial indices only.

With such a metric decomposition, one can linearize the action to find the perturbed field equations. In doing so, one must choose a functional form for the CS scalar and we shall here follow Alexander and Martin~\cite{Alexander:2004wk}, who chose  $\vartheta = \vartheta(\eta)$. One can show that the linearized action (the Einstein-Hilbert piece plus the CS piece) to second order in the metric perturbation yields
\ba
\label{action}
S_{EH} + S_{CS} = \frac{\kappa}{4} \int_{{\cal{V}}} &d^4x& 
\left[    a^2(\eta)
\left(   h^i{}_{j,\eta} \; h^j{}_{i,\eta} - h^i{}_{j,k} \; h^j{}_{i,}{}^{k} \right)
\right.
\nonumber \\
&-& \left.
\vartheta_{,\eta} \; \tilde{\epsilon}^{ijk} \; \left( h^q{}_{i,\eta} h_{kq,j\eta} - h^q{}_{i,}{}^{r}  h_{kq,rj} \right) 
\right] + {\cal{O}}(h)^{3}
\ea
where $\alpha = \kappa$, commas in index lists stand for partial differentiation, $\eta$ is conformal time and $\tilde{\epsilon}^{ijk} = \tilde{\epsilon}^{\eta ijk}$. Variation of the linearized action with respect to the metric perturbation yields the linearized field equations, namely~\cite{Alexander:2004wk}
\be
\label{GW-FE}
\bar{\square} h^{j}{}_{i} := \frac{1}{a^{2}} \tilde{\epsilon}^{p k (j} \left[ \left(\vartheta_{,\eta \eta} - 2 {\cal{H}} \vartheta_{,\eta} \right) h_{i)k,p \eta} + \vartheta_{,\eta} \bar{\square} h_{i)k,p} \right],
\ee
where $\bar{\square}$ is the D'Alembertian operator associated with the background, namely
\be
\bar{\square} f = f_{,\eta \eta} + 2 {\cal{H}} f_{,\eta} - \delta^{ij} f_{,ij},
\ee
with $f$ some function of all coordinates and the conformal Hubble parameter ${\cal{H}} := a_{,\eta}/a$. One could have, of course, obtained the same linearized field equations by perturbatively expanding the C-tensor. 

One can see from Eq.~\eqref{GW-FE} that the evolution of GW perturbations is governed by second and third derivatives of the GW tensor.  Jackiw and Pi~\cite{Jackiw:2003pm} were the first to point out that for the canonical choice of $\vartheta$ the GW evolution is governed by the D'Alembertian of flat space only, if we neglect corrections due to the expansion history of the Universe (ie.~if this vanishes, then the linearized modified field equations for the GW perturbation are satisfied to linear order). Such a result implies there are two linearly independent polarizations that propagate at the speed of light. 

Let us now concentrate on gravitational wave perturbations, for which one can make the ansatz
\be
h_{ij} = \frac{{\cal{A}}_{ij}}{a(\eta)} \exp\left[-i \left(\phi(\eta) - \kappa n_{k} \chi^{k} \right) \right],
\ee
where the amplitude $\mathcal{A}_{ij}$, the unit vector in the direction of wave propagation $n_k$ and the conformal wavenumber $\kappa>0$ are all constant. It is convenient to decompose the amplitude into definite parity states, such as
\be
\mathcal{A}_{ij} = \mathcal{A}_{R}e^{R}_{ij} + \mathcal{A}_{L}e^{L}_{ij}
\ee
where the circular polarization tensors $e^{R,L}_{ij}$ are given in terms of the linear ones $e^{+,\times}_{ij}$ by~\cite{Misner:1973cw} 
\ba
e_{kl}^{R} &=& \frac{1}{\sqrt{2}}\left(e^{+}_{kl} + ie^{\times}_{kl}\right) \\
e_{kl}^{L} &=& \frac{1}{\sqrt{2}}\left(e^{+}_{kl} - ie^{\times}_{kl}\right).
\ea
These polarization tensors satisfy the condition
\be
\label{eigen}
n_i\epsilon^{ijk} e_{kl}^{\text{R,L}} = i
\lambda_{\text{R,L}} \left(e^j{}_l\right)^{\text{R,L}}, 
\ee
where $\lambda_{R} = +1$ and $\lambda_{L} =-1$

With this decomposition, the linearized modified field equations [Eq.~\eqref{GW-FE}] reduce to
\ba
\label{eq:disp}
\left[ i\phi^{\text{R,L}}_{,\eta\eta}+\left(\phi^{\text{R,L}}_{,\eta}\right)^2 +\mathcal{H}_{,\eta}+\mathcal{H}^2 - \kappa^2 \right] 
&& \left(1- \frac{\lambda_{\text{R,L}}\kappa \vartheta_{,\eta}}{a^2}\right) = 
\\ \nonumber 
&& \frac{i \lambda_{\text{R,L}} \kappa}{a^{2}}
\left(\vartheta_{,\eta\eta} - 2\mathcal{H} \vartheta_{,\eta}\right) \left(\phi_{,\eta}^{\text{R,L}}-i\mathcal{H}\right)
\ea
Before attempting to solve this equation for arbitrary $\vartheta(\eta)$ and $a(\eta)$, it is instructive to take the flat-space limit, that is $a \to 1$, and thus, $\dot{a} \to 0$. Assuming further that time derivatives of the CS scalar do not scale with ${\cal{H}}$, one finds that the above equation reduces to~\cite{Alexander:2004wk,Alexander:2007kv}
\be
\label{flatGW-FE}
\left( i \ddot{\phi}_{\text{R,L}} + \dot{\phi}_{\text{R,L}}^2 - k^2 \right) 
\left(1- \lambda_{\text{R,L}} k \dot{\vartheta} \right) =
i \lambda_{\text{R,L}} k \; \ddot{\vartheta} \; \dot{\phi}_{\text{R,L}},
\ee
where $k$ is the physical wavenumber $3$-vector, $t$ stands for cosmic time and overhead dots stand for partial differentiation with respect to time. Let us further assume that the GW phase satisfies $\phi_{,tt}/\phi_{,t}^{2} \ll 1$, which is the standard short-wavelength approximation, as well as  $\ddot{\vartheta}= \vartheta_{0} = \textrm{const.}$ Then the above equation can be solved to first order in $\vartheta$ to find
\be
\phi(t) = \phi_{0} + k t + \frac{i \lambda_{\textrm{R,L}} \vartheta_{0}}{2} kt + {\cal{O}}(\vartheta)^{2}, 
\ee
where $\phi_{0}$ is a constant phase offset and the uncontrolled remainder ${\cal{O}}(\vartheta)^{2}$ stands for terms of the form $k^{2 } \dot{\vartheta}^{2}$ or $k \dot{\vartheta} \ddot{\vartheta}$. The imaginary correction to the phase then implies an exponential enhancement/suppression effect of the GW amplitude, as this propagates in CS modified gravity. Recall that here we are interested in the propagation of GWs, which is why the right-hand side of Eq.~\eqref{flatGW-FE} implicitly omits the stress-energy tensor. We shall see in the next section that if there is a stress-energy tensor, then the CS correction depends both on the first and second derivatives of the CS scalar. Lastly, if we had not assumed that $\ddot{\vartheta} = \textrm{const.}$ then the solution would have become
\be
\phi(t) = \phi_{0} + k t + \frac{i \lambda_{\textrm{R,L}}}{2}  k \dot{\vartheta}(t) + {\cal{O}}(\vartheta)^{2},
\ee
which still exhibits the exponential suppression/enhancement effect. 

The exponentially growing modes could be associated with instabilities in the solution to gravitational wave propagation. One must be careful, however, to realize that the results above have been obtained within the approximation $k^{2 } \dot{\vartheta}^{2} \ll 1 \gg  k \dot{\vartheta} \ddot{\vartheta}$. Thus, provided $\dot\vartheta$ is smaller than the age of the universe, then the instability time scale will not have enough time to set in. For larger values of $\dot{\vartheta}$, the approximate solutions we presented above break down and one must account for higher order corrections. A final caveat to keep in mind is that these results are derived within the non-dynamical formulation of the theory; GW solutions in the fully dynamical theory are only now being actively investigated. 

Let us now return to the field equation for the phase in an FRW background [Eq.~\eqref{eq:disp}]. The solution to this equation is now complicated by the fact that the scale factor also depends on conformal time, and thus, one cannot find a closed form solution prior to specifying the evolution of $a(\eta)$. Let us then choose a matter-dominated cosmological model, in which $a(\eta) = a_{0} \eta^{2} = a_{0}/(1 + z)$, where $a_{0}$ is the value of the scale factor today and $z$ is the redshift. It then follows that the conformal Hubble parameter is simply given by ${\cal{H}} = 2/\eta = 2 (1 + z)^{1/2}$. With this choice, one can now compute the CS correction to the accumulated phase as the plane-wave propagates from some initial conformal time to $\eta$, namely~\cite{Alexander:2007kv}
\be
\label{eq:dPhi1}
\Delta\phi_{(\text{R,L})} = i\lambda_{\text{R,L}} k  H_{0} \int_{\eta}^1 
\left[ \frac{1}{4}  \vartheta_{,\eta\eta}(\eta) -
\frac{1}{\eta} \vartheta_{,\eta}(\eta)
\right]\, \frac{d\eta}{\eta^{4}} + \mathcal{O}(\vartheta)^{2},
\ee
which one can check reduces to the flat space result of above in the right limit. Note that the exponential enhancement/suppression effect now depends on an integrated measure of the evolution of the CS scalar and the scale factor. 

The CS correction to the GW amplitude derives from a modification to the evolution equations of the gravitational perturbation, but it also leads to important observational consequences. One of these can be understood by considering a GW generated by a binary black hole system in the early inspiral phase. The GW produced by such a system can be described as follows
\be
h_{\textrm{R,L}} = \sqrt{2} \frac{\cal{M}}{d_{L}} \left(\frac{{\cal{M}} k_{0}(t)}{2} \right)^{2/3} \left(1 + \lambda_{\textrm{R,L}} \cos{\iota} \right)^{2} \exp\left[ -i \left( \Psi(t) + \Delta\phi_{\textrm{R,L}} \right) \right],
\ee
where $d_{L} = a_{0} \eta \left(1 + z\right)$ is the luminosity distance to the binary's center of mass, $\Psi(t)$ is the GW phase described by GR, $k_{0}(t)$ is the instantaneous wave number of the gravitational wavefront passing the detector and ${\cal{M}}$ is the comoving chirp mass, which is a certain combination of the binary mass components. The inclination angle $\iota$, the angle subtended by the orbital angular momentum and the observer's line of sight, can be isolated as 
\be
\frac{h_{\textrm{R}}}{h_{\textrm{L}}}  = \frac{1 + \cos{\iota}}{1 - \cos{\iota}} \exp\left[\frac{2 k(t)}{H_{0}} \zeta \right]
 = \frac{1 + \cos{\bar{\iota}}}{1 - \cos{\bar{\iota}}},
\ee
where we have defined 
\be
\label{zeta}
\zeta := H_{0}^{2}  \int_{\eta}^1  \left[ \frac{1}{4}  \vartheta_{,\eta\eta}(\eta) - \frac{1}{\eta} \vartheta_{,\eta}(\eta) \right]\, \frac{d\eta}{\eta^{4}}.
\ee
We see then that from an GW observational standpoint, the CS correction leads to an {\emph{apparent inclination angle}} 
$\bar{\iota}$, which effectively modifies the actual inclination angle by a factor that depends on the integrated history of the CS correction:
\begin{equation}
\cos \bar{\iota} = \frac{\sinh\left(\frac{k_0(t) \xi(z)}{H_{0}} \right) + \cosh\left(\frac{k_0(t) \xi(z)}{H_{0}}\right) \cos{\iota}}{\cosh\left(\frac{k_0(t) \xi(z)}{H_{0}}\right) + \sinh\left(\frac{k_0(t) \xi(z)}{H_{0}}\right)  \cos{\iota}} 
\sim \cos{\iota} + \frac{k_0(t) \xi(z)}{H_{0}}  \; \sin^{2}{\iota} + {\cal{O}}\left(\xi^{2}\right).
\end{equation}
We see then that the CS correction effectively introduces an {\emph{apparent}} evolution of the inclination angle, which tracks the gravitational wave frequency.

The interpretation of the CS correction as inducing an effective inclination angle should be interpreted with care. In GR, if a gravitational wave propagates along the line of sight, such that the actual inclination angle is ({\emph{eg.}}~$0$ or $\pi$), then the amplitude is a maximum. In CS gravity, however, the amplitude can be either enhanced or suppressed, depending on whether the wave is right- or left-circularly polarized. When the CS effect suppresses the GW amplitude, one can think of this as an effective modification of the inclination angle away from the maximum. However, when the CS effect enhances the amplitude, there is no {\emph{real}} inclination angle that can mimic this effect ({\emph{ie.}}~the effective angle would have to be imaginary). 

The evolution equation for the gravitational wave perturbation depends sensitively on the scale factor evolution [see, eg.~Eq.~\eqref{GW-FE}]. Alexander and Martin~\cite{Alexander:2004wk} have investigated gravitational wave solutions when the scale factor presents an inflationary behavior. Suffice it to say in this section that Eq.~\eqref{GW-FE} can be recast in the form of a parametric oscillator equation, with a non-trivial effective potential. In certain limits appropriate to inflation, one can solve this differential equation in terms of Whittaker functions, which can be decomposed into products of trigonometric functions and exponentials. In essence, the solutions present the same structure as that of a matter-dominated cosmology. Once the gravitational wave modes have been computed, one can proceed to calculate the power spectrum, but these results will be discussed further in Sec.~\ref{Sec:Inflation}.

\subsection{Gravitational Wave Generation} 
\label{GWgen}

The issue of GW generation by dynamical matter sources in CS modified gravity was first studied by Jackiw and Pi~\cite{Jackiw:2003pm}. Once more, this problem has been treated only in the non-dynamical formalism ($\beta = 0$ and $\alpha = \kappa$) and with the canonical choice of $\vartheta$, although very recently the much more difficult problem of GW generation
in dynamical CS modified gravity has begun to be investigated~\cite{Sopuerta:2009iy}. 

In the presence of a stress-energy tensor, the modified field equations linearized about a Minkowski background ($g_{ab} = \eta_{ab} + h_{ab}$) become
\be
\square_{\eta} h^{j}{}_{i} + \dot{\vartheta} \tilde{\epsilon}^{p k (j}  \square_{\eta} h_{i)k,p}  - \ddot{\vartheta} \tilde{\epsilon}^{p k (j} \dot{h}_{i)k,p}  = - \frac{1}{\kappa} T^{i}{}_{j},
\ee
where $t$ is the standard time coordinate of Minkowski spacetime, $\tilde{\epsilon}^{p j k} = \tilde{\epsilon}^{0 p j k}$, the D'Alembertian operator is $\square := - \partial_{t}^{2} + \delta^{ij} \partial_{i} \partial_{j}$ and overhead dots stand for partial differentiation with respect to time. One can derive this equation simply from Eq.~\eqref{GW-FE} by taking the limit $a \to 1$ and ${\cal{H}} \to 0$ and reinserting the stress-energy tensor $T_{ij}$, which must now be TT, since so is $h_{ij}$ implicitly. 

Although the explicit solution to the above equation has not yet been computed for a general CS scalar field, this problem has been studied for the canonical choice of $\vartheta$. With the assumption that $\ddot{\vartheta} = 0$, we then obtain, to first order, the solution found in Eq.~\eqref{formal-sol2}, which for a TT stress energy can be recast as
\be
\square_{\eta} h_{ij} = - \frac{1}{\kappa} \bar{\bar{T}}_{ij},
\ee
where the effective stress-energy has been defined as
\be
\bar{\bar{T}}_{ij} := T_{ij} - \dot{\vartheta} \tilde\epsilon^{kl}{}_{(i} T_{j)l,k}.
\ee
We see then that GWs generation in non-dynamical CS gravity with a canonical $\vartheta$ is nothing but GR GWs in the presence of such an effective stress-energy tensor~\cite{Jackiw:2003pm}.

For concreteness, let us assume the stress-energy tensor represents a monochromatic source, with definite frequency $\omega$, which is radiating GWs in the $\hat{z}$-axis with wave-vector $k$. The only non-vanishing components of the stress-energy tensor are then $T_{xx} = - T_{yy}$ and $T_{xy} = T_{yx}$ and it's Fourier transform shall be denoted $T_{+}$ and $T_{\times}$ respectively. We then find that the effective stress-energy becomes
\be
\label{effec-T}
\tilde{T}_{ij} = 
\left( 
\begin{array}{c c}
T_{+} - i k \dot{\vartheta} T_{\times} & T_{\times} + i k \dot{\vartheta} T_{+} \\
T_{\times} + i k \dot{\vartheta} T_{+} &  -T_{+} + i k \dot{\vartheta} T_{\times}\\
\end{array}
\right),
\ee
which exhibits the natural mixture of polarizations of CS GWs.

A natural next step is to compute the power carried by such CS modified GWs per unit angle $d\Omega$. Since CS GW theory for a canonical $\vartheta$ is identically equivalent to GR with an effective stress-energy tensor, it follows that the GW power is given by 
\be
\label{power}
\frac{d P_{\textrm{R,L}}}{d\Omega} = 16 \kappa G^{2} \; \omega^{2} \; \tilde{T}^{\ast}{}_{ij} \tilde{T}^{ij}.
\ee
One is tempted to insert the solution for the effective stress-energy found in Eq.~\eqref{effec-T} into Eq.~\eqref{power}. This effective stress-energy, however, is insufficient, since it is only a solution to first-order in $\dot{\vartheta}$ and the expression for the power emitted requires a second-order solution. Jackiw and Pi~\cite{Jackiw:2003pm} have found this solution, which reduces to Eq.~\eqref{effec-T} multiplied by $(1 - k^{2} \dot{\vartheta}^{2})^{-1}$. One then finds that the power emitted is given by~\cite{Jackiw:2003pm}
\ba
\frac{d P_{\textrm{R,L}}}{d\Omega} &=& 32 \kappa G^{2} \omega^{2} \left(1 - k^{2} \dot{\vartheta}^{2} \right)^{-2} \left[ \left(1 + k^{2} \dot{\vartheta}^{2} \right) \left(T_{+}^{2} + T_{\times}^{2} \right) 
\right.
\nonumber \\
&+& \left.  2 i k \dot{\vartheta} \left(T_{+} T_{\times}^{\ast} - T_{\times} T_{+}^{\ast} \right) \right],
\ea
which when linearized to second order in $\dot{\vartheta}$ becomes
\be
\frac{d P_{\textrm{R,L}}}{d\Omega} \sim  32 \kappa G^{2} \omega^{2}  \left[ \left(1 - k^{2} \dot{\vartheta}^{2} \right) \left(T_{+}^{2} + T_{\times}^{2} \right) + 2 i k \dot{\vartheta} \left(T_{+} T_{\times}^{\ast} - T_{\times} T_{+}^{\ast} \right) \right].
\ee
The power carried by circularly polarized GWs ($T_{+} = i T_{\times}$) is corrected by CS gravity only to second order in $\dot{\vartheta}$~\cite{Jackiw:2003pm}. 

The results presented above hold only for the canonical choice of $\vartheta$. Had we allowed the second derivative of the CS scalar to
be non-vanishing, we would have found a linear correction to the power carried by CS GWs. In the dynamical formalism, this effect is 
more clear as the CS scalar also carries energy-momentun away from the system~\cite{Sopuerta:2009iy}. Such modifications to
radiation-reaction would affect the inspiral and merger of binaries, as suggested in~\cite{Sopuerta:2009iy}, which leads to a powerful
test of the dynamical formulation that we shall discuss in Sec.~\ref{Sec:AstroTests}.

\section{Non-Vaccum Solutions and Fermionic Interactions}
\label{Fermions}
The first-order formulation of CS modified gravity was first studied by Cantcheff~\cite{Cantcheff:2008qn}, who realized that the modified theory would lead to non-vanishing torsion for a canonical CS scalar. Such an idea was later generalized to arbitrary $\theta$ in the non-dynamical formalism and the torsion tensor was specialized to Earth's gravitational field~\cite{Alexander:2008wi}. Such a formulation naturally allows for the coupling of fermions to the modified theory, thus permitting the study of non-vacuum spacetimes~\cite{Alexander:2008wi}. We shall summarize these results here, beginning with a description of the first-order formalism and Einstein-Cartan theory, following mainly~\cite{Carroll:2004st,Romano:1991up,Randono:2007mr}. We then continue with a discussion of the first-order formulation of CS modified gravity. 


\subsection{First-Order Formalism}

Consider a $4$-dimensional manifold ${\cal{M}}$ with an associated $4$-dimensional metric $g_{ab}$. At each point on this manifold, let there be a tetrad $e^{I}_{a}$, so that the metric can be recast as $g_{ab} = e_{a}^{I} e_{b}^{J} \eta_{IJ}$, where $\eta_{IJ}$ is the Minkowski metric. Internal indices range $I,J = (0,1,2,3)$, just as spacetime indices do, and often, we will suppress spacetime ones and the tetrad shall be written $e^{I}$. We can raise or lower spacetime and internal indices with the metrics $g_{ab}$ and $\eta_{IJ}$ respectively. 

The introduction and differentiation of internal versus spacetime indices is crucial to the first order formalism. Riemannian fields, like the metric tensor, exist on the {\emph{base manifold}} ${\cal{M}}$ and have a finite dimension, but gauge fields can be infinite dimensional and so they must exist on a different vector space. For example, the tetrad $e^{I}$ and the spin connection $\omega^{KL}$ are $1$-forms on the base manifold, while the curvature tensor associated with it, $F^{KL}$, is a $2$-form on the base manifold. On the other hand, the fermion field $\psi$ is a $0$-form on the internal vector space. A {\emph{fiber bundle}} is defined as the union of the base manifold and the internal vector space, with each fiber a different copy of the internal vector space. One can think of the Lie group associated with the fiber bundle as glueing all fibers and the base manifold together~\cite{Carroll:2004st}. 

The recovery of spacetime indices is sometimes achieved via the wedge product operator, defined via%
\be
\left(A \wedge B \right)_{ab} := \frac{\left(p + q\right)!}{p! q!} A_{ [a_{1} \ldots a_{p}} B_{b_{1} \ldots b_{q}]}
\ee
with $A$ and $B$ $p$- and $q$-forms respectively. Note that we shall here follow the convention that spacetime indices always appear after internal ones. The wedge product operator satisfies the following chain rule
\be
D_{(\omega)} \left( A\wedge B \right) = \left(D_{(\omega)} A\right) \wedge B + (-1)^{p} A \wedge \left(D_{(\omega)} B \right),
\ee
and the following commutativity relation
\be
A \wedge B = (-1)^{pq} B \wedge A.
\ee
Also note that since the wedge product acts on spacetime indices only, it exists on the base manifold and not
on the internal space. 

Now that the metric and tetrad have been defined, let us introduce the generalized covariant derivative operator $D$ and the spin connection $\omega^{IJ}$. Given a tensor $A^{KL}{}_{a}$ we can define the covariant derivative as
\ba
D_{(\omega)} A^{KL} &:=& dA^{KL} + \omega^{KM} \wedge A_{M}{}^{L} + \omega^{LM} \wedge A^{K}{}_{M},
\nonumber \\
\\
D_{(\omega)} A_{KL} &:=& dA_{KL} - \omega_{K}{}^{M} \wedge A_{ML} - \omega_{L}{}^{M} \wedge A_{KM},
\nonumber \\
\ea
where the exterior derivative operator $d$ acts on spacetime indices only:
\be
dA^{KL} := 2 \partial_{[a} A^{KL}{}_{b]}.
\ee
The commutator of covariant derivatives allow us to define the curvature tensor associate with $\omega^{IJ}$, namely
\be
F^{KL} = d\omega^{KL} + \omega^{K}{}_{M} \wedge \omega^{ML},
\ee
which reduces to the Riemann tensor if the spin connection is metric compatible and torsion-free, {\emph{ie.~}}if the connection is the Christoffel one.
One can then show after some calculation that 
\be
\delta_{\omega} F^{IJ} = D_{(\omega)} \delta \omega^{IJ}.
\ee

The spin connection has a certain degree of freedom that can be fixed by demanding that it be internally metric compatible $D_{(\omega)} \eta^{IJ} = 0$. This condition forces the connection to be completely antisymmetric on its internal indices $\omega^{(IJ)} = 0$. We can then define the torsion tensor as
\be
\label{def-T}
T^{I} := D_{(\omega)} e^{I} = de^{I} + \omega^{I}{}_{M} \wedge e^{M},
\ee
which is equivalent to $T^{I}{}_{ab} = 2 D_{[a} e_{b]}^{I}$. If we reinstate spacetime indices, we find  
\be
\label{def-T2}
T^{a}{}_{bc} = 2 C^{a}{}_{[bc]},
\ee 
where $C^{a}{}_{bc}$ is the antisymmetric part of the connection, or contorsion tensor.  Note that internal metric compatibility is not equivalent to a torsion-free condition.

The contorsion tensor can be defined purely in terms of wedge products. Let us then decompose the spin connection into a symmetric and tetrad compatible piece $\Gamma^{I}{}_{J}$ and an antisymmetric piece $C^{I}{}_{J}$, the contorsion tensor. 
The torsion tensor is then given in terms of the contorsion via
\be
T^{I} = C^{I}{}_{J} \wedge e^{J},
\ee
which can be inverted to find
\be
\label{cont-T}
C_{IJK} = - \frac{1}{2} \left(T_{IJK} + T_{JKI} + T_{KJI} \right).
\ee
The contorsion tensor is fully antisymmetric on its first two indices, while the torsion tensor is fully antisymmetric on its last two indices.  Equation~\eqref{def-T2} can also be obtained from Eq.~\eqref{def-T} in spacetime indices, if we use the transformation law from spin to spacetime connection:
\be
\label{chg-basis}
\omega_{ab}{}^{e} = e_{I}^{e} \; \omega_{K}{}^{I}{}_{a} \; e^{K}_{b} - e_{I}^{e} \; \partial_{a} e_{b}^{I},
\ee
which can be established from $D_{(\Gamma)} e^{I} = 0$. Eq.~\eqref{chg-basis} is sometimes referred to as ``the tetrad postulate''. 

The curvature tensor can be expressed purely in terms of the Riemann tensor (that depends only on $\Gamma^{I}{}_{J}$) and terms that depend on the contorsion tensor:
\be
F^{IJ} = R^{IJ} + D_{(\Gamma)} C^{IJ} + C^{I}{}_{M} \wedge C^{MJ},
\ee
where $D_{(\Gamma)}$ is the connection compatible with the symmetric connection. The Bianchi identities in first-order form become
\be
D_{(\omega)} T^{I} = R^{I}{}_{K} \wedge e^{K},
\qquad
D_{(\omega)} R^{IJ} = 0. 
\ee
%

\subsection{First-Order Formulation of CS Modified Gravity}

Let us now apply the formalism of the previous section to GR and to the modified theory. The Einstein-Hilbert action can be recast in terms of forms as
\be
S_{EH} = \frac{\kappa}{2} \int_{{\cal{V}}} \epsilon_{IJKL} e^{I} \wedge e^{J} \wedge F^{KL}.
\ee
One can convert this into Eq.~\eqref{EH-action} by rewriting the curvature tensor as $F^{IJ} = (1/2) F^{IJ}{}_{KL} e^{K} \wedge e^{L}$ and using the identity
\be
e^{I} \wedge e^{J} \wedge e^{K} \wedge e^{L} = - \tilde{\sigma} \; \epsilon^{IJKL},
\ee
where $\tilde{\sigma} = \sqrt{-g} \; d^{4}x $, as well as the Kronecker-Delta relations
\ba
\epsilon^{a b c d} \epsilon_{a b e f} &=& - 4 \; \delta^{[c}_e
\delta^{d]}_f, 
\\
\tilde{\eta}^{a b c d} \epsilon_{a b e f} &=& + 4 \; \sqrt{-g} \; 
\delta^{[c}_e \delta^{d]}_f,  
\\
\tilde{\eta}^{a b c d} \tilde{\eta}_{a b e f} &=& + 4 \; \delta^{[c}_e
\delta^{d]}_f.   
\ea
Note that full internal index contractions are identically equal to spacetime index contractions, if the quantities contracted are tensors. 

Similarly, we can attempt to recast the CS action in terms of forms via
\be
\label{CS-action-1st}
S_{CS} = \frac{\alpha}{2} \int_{{\cal{V}}} \vartheta \; F \wedge F,
\ee
where the integrand reduces to $R \wedge R = R_{IJ} \wedge R^{IJ}$ for a symmetric connection. Again, Eq.~\eqref{CS-action-1st} can be converted into Eq.~\eqref{CS-action} in the same way as above, which then establishes that $\tilde{\sigma} \pont = 2 R \wedge R$ for a symmetric connection.  We can now integrate by parts to obtain the CS action in terms of the Pontryagin current, but to do so we first must realize that $F \wedge F = d \Omega_{3}$, where the Chern-Simons $3$-form is defined as~\cite{Alexander:2004us,Alexander:2004xd,Alexander:2007qe}
\be
\Omega_{3} = \omega^{IJ} \wedge d \omega_{IJ} + \frac{2}{3} \omega^{I}{}_{J} \wedge \omega^{J}{}_{K} \wedge \omega^{K}{}_{I}.
\ee
We then find that Eq.~\eqref{CS-action-1st} can be rewritten as
\be
S_{CS} = - \frac{\alpha}{2} \int_{{\cal{V}}} d\vartheta \wedge \Omega_{3},
\ee
where we have neglected the boundary contribution. Converting this expression to spacetime indices, one can check that one recovers Eq.~\eqref{CS-action-K} if the connection is symmetric. In fact, the dual to this $3$-form is actually the Pontryagin current in Eq.~\eqref{eq:curr2} in disguise, while $^{\ast}(d\Omega_{3}) = \;^{\ast}FF/2$.

We can now vary the action with respect to all degrees of freedom to obtain the modified field equations. The variation of the action with respect to the tetrad yields
\be
\label{1st-FE1}
F_{ab} - \frac{1}{2} g_{ab} F = \frac{2}{\kappa} T_{ab},
\ee
since the CS current does not depend on the tetrad, but only the spin connection. Equation~\eqref{1st-FE1} resembles the Einstein equations, except that the quantity $F_{ab} = F^{c}{}_{acb}$ and $F = F^{a}{}_{a}$ are not the Ricci tensor and scalar, but contractions of the curvature tensor. 

The full curvature tensor can be reconstructed once one solves for the contorsion tensor via the torsion condition of the modified theory. This condition is arrived at by requiring that the variation of the action with respect to the spin connection vanishes. Let us then rewrite the CS $3$-form via
\be
\Omega_{3} = \omega^{I}{}_{J} \wedge F^{J}{}_{I} - \frac{1}{3} \omega^{I}{}_{J} \wedge \omega^{J}{}_{K} \wedge \omega^{K}{}_{I}.
\ee
such that its variation with respect to $\omega_{KL}$ reduces to $F^{KL}$. The variation of the action with respect to $\omega_{KL}$ reduces to
\be
\delta S = -\frac{\kappa}{2} \int_{{\cal{V}}} D\left(\epsilon_{IJKL} e^{I} \wedge e^{J} \right) \delta \omega^{KL} - \frac{\alpha}{2} \int_{{\cal{V}}} d\vartheta \wedge F_{KL} \delta \omega^{KL},
\ee
where we have set $\beta = 0$ for simplicity and we have integrated the first term by parts. Cantcheff~\cite{Cantcheff:2008qn} has shown that $\delta S = =0$ is equivalent to the torsion constraint
\be
D \left( \epsilon_{IJKL} e^{I} \wedge e^{J} \right) = - \frac{\alpha}{\kappa} d\vartheta \wedge F_{KL}, 
\ee
which is nothing but the CS modified second Einstein-Cartan structure equation. This equation can also be rewritten in terms of the torsion tensor as
\be
\label{Torsion-condition}
\epsilon_{IJKL} \; T^{I} \wedge e^{J} = - \frac{\alpha}{2 \kappa} d\vartheta \wedge F_{KL}. 
\ee

The torsion condition of Eq.~\eqref{Torsion-condition} shows that in CS modified gravity the connection must be torsionfull. Solving for the torsion in Eq.~\eqref{Torsion-condition} is non-trivial, due to the implicit appearance of torsion itself through the curvature tensor on the right-hand side of this equation. However, if we replace this tensor by the Riemann ({\emph{torsionless}}) tensor, then we can solve for the torsion tensor exactly to find\footnote{This solution is the same as that found by~\cite{Cantcheff:2008qn} with $\alpha = 2$ and shortly after by~\cite{Alexander:2008wi} with $\alpha = \kappa$. We have further checked that Eq.~\eqref{torsion} is indeed a solution to the torsion constraint of Eq.~\eqref{Torsion-condition}}.
\be
\label{torsion}
T^{I} = - \frac{\alpha}{4 \kappa} \; \epsilon^{IJKL} \; v_{J} \; R_{KL},
\ee
where here we have used the first Bianchi identity $R_{[IJK]L} = 0$. Equation~\eqref{torsion} should be interpreted as an approximate solution to the torsion constraint, where we have neglected higher than second powers of $\theta$~\cite{Cantcheff:2008qn,Alexander:2008wi}. 

One can now relate this torsion tensor to the gravitational field of a compact binary in the PPN formalism~\cite{Alexander:2008wi}. One can show that this tensor is proportional to the product of $v^{a}$ with $\nabla \times V$ and $\vec{\nabla}{U}$, where $V^{i}$ and $U$ are the PPN vector and Newtonian potentials. In fact, there is no choice of $\theta$ for which all components of the torsion tensor vanish. Since this tensor affects the motion of point particles (in particular the frame-dragging effect~\cite{Alexander:2008wi}), one concludes that non-dynamical CS modified gravity generically leads to modified precession, irrespective of $\theta$. Care must be taken, however, since in spite of generically being non-zero,  the torsion tensor is after all proportional to the dual to the Riemann tensor. Thus, Earth-based experiments that search for non-vanishing torsion~\cite{Kostelecky:2003fs,Belyaev:2007fn,Kostelecky:2007kx} cannot measure this effect, since the Riemann tensor on Earth is prohibitively small. 

Are the first and second-order formulations of CS modified gravity equivalent? The modified field equations of Eq.~\eqref{1st-FE1} resemble the Einstein equations, except that the curvature tensor contains torsional pieces. When Eq.~\eqref{torsion} is used as the torsion, one can show that all these torsional pieces conspire to produce the C-tensor of Eq.~\eqref{eq:eom}~\cite{Cantcheff:2008qn}. One then finds that the first and second-order formulations of CS gravity are equivalent {\emph{if and only if}} the CS action in first-order form is defined in terms of the torsion-free curvature tensor (ie, in terms of the symmetric connection), such that Eq.~\eqref{torsion} is the exact solution to the torsional condition. 

The CS $3$-form, however, has been here introduced in terms of the generalized spin connection, for which the torsion condition is non-linear, depending explicitly on the curvature tensor. For such a torsion condition, the torsion tensor in Eq~\eqref{torsion} is still formally valid but only to linear order in $\theta$, and thus, the first-order formalism is still equivalent to the second-order one but only to ${\cal{O}}(\vartheta)$. The ${\cal{O}}(\vartheta)^{2}$ corrections to Eq.~\eqref{torsion} will  modify the field equations, and thus, force them to not be equivalent to those of the second-order formalism. Cantcheff~\cite{Cantcheff:2008qn} has further shown that line elements that are solutions in the second-order formulation of CS gravity ({\emph{eg.}}~the Schwarzschild metric, which leads to a vanishing C-tensor) are not necessarily solutions to the field equations of the first-order formulation if higher-order in $\theta$ terms are included.  

The inequivalence between the first- and second-order formalism thus depends on quadratic or higher powers of the CS scalar. The CS modified action considered in Sec.~\ref{Formulation}, however, only considers linear terms in $\theta$. In principle, there will be $\theta^{2}$ and higher-order terms in the action that one could have to include, since these will also generically break parity [eg.~$\vartheta^{2} (\pont)^{2}$]. A consistent comparison between first- and second-order formalisms thus requires that such terms be taken into account, if one wishes to define the CS action in first order form in terms of the torsionfull curvature tensor, instead of the Riemann tensor.

\subsection{Fermions and CS Modified Gravity }

One of the advantages of the first-order formalism is that it allows for the inclusion of fermions and bosons in the action. Let us then considering the inclusion of the following piece to the full action of Sec.~\ref{Formulation} 
\be
\label{Dirac-S}
S_{D} = \frac{\epsilon}{2} \int d^4x \sqrt{-g} \left(i \bar{\psi} \gamma^I e_I^{a} {\cal{D}}_{a} \psi + {\textrm{c.c.}} \right),
\ee
where ${\textrm{c.c.}}$ stands for complex conjugation, $\psi$ is a Dirac spinor, $\gamma^I$ are gamma matrices and $\epsilon$ is a coupling constant. Fermions are here represented by Dirac spinors, which are gauge field that live naturally in $SU(2)$. The tetrad and the spin connection of GR, on the other hand, are fields that live in $SO(1,3)$. Thus, without the machinery of fiber bundle theory one could not easily couple fermions to CS modified gravity. 

From the group structure of $\psi$, one can deduce how the generalized $SU(2)$ covariant derivative acts on Dirac spinors: ${\cal{D}}_a \psi := \partial_a \psi - (1/4) \omega^{IJ}{}_{a} \gamma_I \gamma_J \; \psi$, where we shall here follow the sign conventions of~\cite{Perez:2005pm}. Variation of the full action with respect to Dirac fermions then leads to the massless Dirac equation~\cite{Alexander:2008wi}:
\be
\gamma^{a} D_{a}^{(\Gamma)} \psi = \frac{1}{4} e^{a}_{M} C_{a}{}^{KL} \gamma^{M} \gamma_{K} \gamma_{L} \psi
\label{Dirac-Eq},
\ee
where notice that we have not included a mass term for the fermions for simplicity. We see then that the Dirac equation is modified in CS modified gravity by a source term that depends on the contorsion tensor. Equation~\eqref{Dirac-Eq} implies that  the CS effects might be enhanced in spacetime regions where the momentum of Dirac fermions  is large.

Before we can vary the full action with respect to the connection it is convenient to recast it in first-order form. Doing so, Eq.~\eqref{Dirac-S} becomes 
\be
S_{D} = \frac{\epsilon}{12} \int \epsilon_{IJKL} e^{I} \wedge e^{J} \wedge e^{K} \wedge \left(i \bar{\psi} \gamma^{L} {\cal{D}} \psi + {\textrm{c.c.}} \right).
\ee
Upon variation of the action with respect to the connection, Eq.~\eqref{Torsion-condition} becomes
\be
\epsilon_{IJKL} T^{I} \wedge e^{J} = - \frac{\alpha}{2 \kappa} d\vartheta \wedge F_{KL} + \frac{\epsilon}{4 \kappa} e_{K} \wedge e_{L} \wedge J_{(5)},
\ee
where $J_{5}^{L} := \bar{\psi} \gamma_{5} \gamma^{L} \psi$ is the fermion axial current, $e$ is the determinant of the tetrad field and we have used the identity
\be
\gamma^{I} \gamma^{[J} \gamma^{K]} = - i \epsilon^{IJKL} \gamma_{5} \gamma_{L} + 2 \eta^{I[J} \gamma^{K]}.
\ee
This equation can be solved for the torsion tensor if we once again replace the curvature tensor by the Riemann tensor on the right-hand side. Doing so, we find~\cite{Alexander:2008wi}
\be
T^{I} = - \frac{\alpha}{4 \kappa} \epsilon^{IJKL} v_{J} R_{KL} - \frac{\epsilon}{8 \kappa} \epsilon^{I}{}_{JKL} J_{5}^{L} e^{J} \wedge e^{K},
 \ee
which essentially is an approximate solution that neglects quadratic and higher powers of $\theta$. From this, the contorsion tensor can be calculated using Eq.~\eqref{cont-T} to find
\be
C_{IJ} = \frac{3 \alpha}{8 \kappa} v_{N} \epsilon_{[I}{}^{NML} R_{JK]ML} + \frac{\epsilon}{8 \kappa} \epsilon_{IJKL} J_{5}^{L} e^{K}.
\ee

Since the first- and second-order formalism are equivalent to leading order in $\theta$, one can compute the interacting action be reinserting the torsion solution in the full action. Doing so, one finds~\cite{Alexander:2008wi}
\be
S = \frac{3 \epsilon^{2}}{32 \kappa} \int_{{\cal{V}}} \tilde{\sigma} J_{5}^{a} J_{5 \;a}
+  \frac{\epsilon \alpha}{16 \kappa} \int_{{\cal{V}}} \tilde{\sigma} \left( 2 J_{5}^{a} v^{b} R_{ab} - J_{5}^{a} v_{a} R\right) 
\ee
where we have neglected other terms that are either higher-order in $\theta$ or in the gravitational coupling constant. The first term is the standard $4$-fermion interaction that arises in Einstein-Cartan theory coupled with fermions. The second term is a new CS contribution that depends on the embedding coordinate, as well as the Ricci tensor and scalar. This new term represents a $2$-fermion interaction and it is not suppressed by the gravitational coupling constant. We can then conclude that, at least to linear order, fermion current tend to enhance the CS correction, some of the implications of which shall be discussed in Sec.~\ref{Sec:AstroTests}.

\section{Astrophysical Tests}
\label{Sec:AstroTests}
All tests of CS modified gravity to date have been performed with astrophysical observations and concern the non-dynamical framework. After Alexander and Yunes~\cite{Alexander:2007vt,Alexander:2007zg} realized that the modified theory predicts an anomalous precession effect, Smith, {\emph{et.~al.~}}\cite{Smith:2007jm} tested the non-dynamical model with canonical CS scalar with LAGEOS~\cite{Ciufolini:2007wx,Ciufolini:2004rq,Ciufolini:2004gp} and Gravity Probe B~\cite{GPBwebsite,Will:2002ma} observations, placing the first, albeit weak, bound on the CS scalar. In view of these results, Konno, et.~al.~\cite{Konno:2008np} proposed that the CS correction could be used to explain the flat, rotation curves  of galaxies , which in turn could yield yet another constraint on the non-dynamical theory for non-canonical $\vartheta$. Recently, Yunes and Spergel~\cite{Yunes:2008ua} used double binary pulsar data to place a bound on the non-dynamical model with canonical CS scalar that is eleven orders of magnitude stronger than the Solar System one. 

The dynamical model remains untested today, mainly due to the difficulty in calculating observable quantities in a consistent way. The only possible avenue to perform such a test seems currently to be through gravitational wave observations~\cite{Alexander:2007kv}. Cosmological  tests of the modified theory will be discussed in the next section.  

\subsection{Solar System Tests}
\label{PPN-tests}

The non-dynamical modified theory has been so far only tested through Solar System, frame-dragging experiments. Smith, {\emph{et.~al.~}}\cite{Smith:2007jm} studied the anomalous precession inherent to the non-dynamical model with the conventions $\alpha = - \ell/3$ and $\beta = 1$ and we shall summarize these results here. With these conventions, $\vartheta$ has units of inverse length or mass and $m_{\textrm{CS}}$ is a characteristic mass scale defined in Eq.~\eqref{CS-mass}. When testing CS modified gravity, we shall in fact place bounds on this parameter, which can be trivially related to $\vartheta$ by the inversion of Eq.~\eqref{CS-mass}.

Only Solar System tests that sample the gravitomagnetic sector of the gravitational field can be used to test CS modified gravity in the non-dynamical model. As shown in Sec.~\ref{Sec:ApproxSol}, the non-dynamical modified theory possesses the same PPN parameters as GR, except for the gravitomagnetic potential. This implies, in particular, that the perihelion shift of Mercury or light deflection by the Sun cannot be used to constrain the modified theory. The only physical effect of the non-dynamical CS modification is the induction of anomalous precession effects. 

Precession is a generic term used to address the change in the rotation $3$-vector of some spinning object, {\emph{ie.}}~a non-vanishing $4$-gradient of the spin angular momentum.  Two types of precession can be distinguished: {\emph{torque-free}} and {\emph{torque-induced}}. The former corresponds to situations in which the spin angular momentum is not coaligned with the axial Killing vector. The latter, also known as {\emph{gyroscopic precession}}, occurs in situations where there is an additional torque (such as that of a gyroscope) that pushes on the spin angular momentum vector, forcing it to wobble.  Gyroscopic precession can be studied in a Newtonian framework, but relativity adds three additional corrections: {\emph{Thomas precession}}, an additional special relativistic correction due to the observer's non-inertial rotating frame; {\emph{de Sitter or geodetic precession}}, a GR effect that accounts for Schwarzschild-like deviations from flat spacetime; {\emph{Lense-Thirring precession}}, a GR correction due to the gravitomagnetic sector of the Kerr metric. 

The CS modification can correct several different types of precession, depending on the physical scenario under consideration. Torque-free precession can occur if one considers the far-field expansion of a non-dynamical CS spinning black hole metric, where the axis of rotation seems not to be co-aligned with the axial axis of symmetry. In this one-body scenario, in the absence of external torques, the spin angular momentum of the black hole precesses around the symmetry axis in a wobbling fashion. The evolution of the wobble angle requires the determination of the Killing axis, from which one can obtain the frequency of precession, via $\omega_{f} \approx J/I_{1}$ to Newtonian order, where $I_{1}$ is the moment of inertia about the symmetry axis~\cite{Jones:2001yg}. Since such an arrangement is asymmetric, one expects GW emission leading to spin-down and alignment. In the non-dynamical formalism, then, spinning black holes would tend to ``unspin'' themselves via interactions with the CS scalar, thus relaxing to the Schwarzschild solution. If so, observations of spinning black holes could be used to constrain the magnitude of the canonical scalar~\cite{tegpriv}. Such a possibility has not yet been studied in detail.

Torque-induced precession is also modified by the CS correction through the correction to the gravitomagnetic sector of the metric. Consider first the motion of a test body in the external field of a CS spinning source. The far-field solution for such a source was summarized in Sec.~\ref{Sec:ApproxSol} for extended bodies~\cite{Smith:2007jm}. In such a field, the orbital elements of a test body will experience Lense-Thirring precession~\cite{Iorio:1999ru}, which will be different in CS gravity relative to GR. Smith, {\emph{et.~al.~}}\cite{Smith:2007jm} studied the secular time variation of the longitude of the ascending node $\dot{\Omega}_{\textrm{orb}}$~\cite{Murray-Book} in the non-dynamical modified theory and found it to be given by
\be
\dot{\Omega}_{\textrm{orb}} = \Omega_{\textrm{GR}}  + \Omega_{\textrm{CS}} ,
\ee
where the GR Lense-Thirring drag is given by 
\be
\dot{\Omega}_{\textrm{GR}}  = \frac{2 G J}{a^{3} (1 - e^{2})},
\ee
with eccentricity $e$, the magnitude of the spin angular momentum of the central body $J$  and the CS correction $\dot{\Omega}_{\textrm{CS}}$ given by
\be
\label{Smith}
\dot{\Omega}_{\textrm{CS}} = \frac{15 a^{2}}{R^{2}} j_{2}(m_{\textrm{CS}} R) y_{1}(m_{\textrm{CS}} a),
\ee
with the semi-major axis $a$, Earth's radius $R$, and the spherical Bessel functions of the first and second kind $j_{\ell}(\cdot)$ and $y_{\ell}(\cdot)$.

The LAGEOS satellites have measured $\dot{\Omega}$ and found it in agreement with GR up to experimental error, which thus allows for a test of non-dynamical CS gravity~\cite{Smith:2007jm}. Figure~\ref{fig:OmegaCSdot} shows the ratio of the GR and CS predictions as a function of the characteristic CS mass~\cite{Smith:2007jm}, where the shaded region corresponds to a $1\sigma$ deviation from the experimentally measured value. The region where the CS prediction is in agreement with experiment is then all $m_{\textrm{CS}} \gtrsim 2.5 \times 10^{-3} \; {\textrm{km}}^{-1}$\footnote{The relation in Eq.~\eqref{Smith} is non-monotonic, and thus, many isolated islands exist for which the observed precession is consistent with the CS prediction. Nonetheless, it is always true that for the range of values quoted here the observed precession is always consistent with the CS prediction. The bound quoted by Smith, {\emph{et.~al.~}} are somewhat more stringent, thus including more island but also including regions of the parameter space that are inconsistent with observations~\cite{Richard}}. A $2\sigma$ or $3 \sigma$ constraint would increase the shaded region by roughly a factor of two or three, thus forcing $m_{\textrm{CS}} \gtrsim 2 \times 10^{-3} \; {\textrm{km}}^{-1}$ or $m_{\textrm{CS}} \gtrsim 1.5 \times 10^{-3} \; {\textrm{km}}^{-1}$. Taking the most conservative estimate translates into a bound for the CS scalar of $|\dot{\vartheta}| \leq 3000 \; (\kappa/\alpha) \; {\textrm{km}}$ with more than $99 \%$ confidence. 
\begin{figure}[ht]
\centerline{\epsfig{file=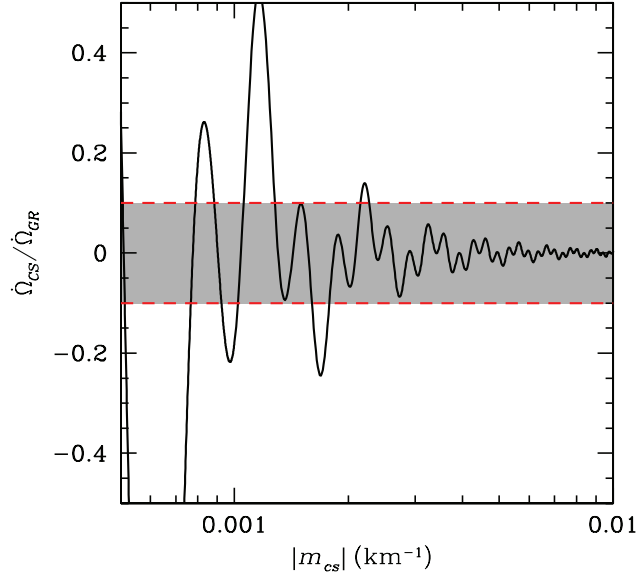,height=18pc,angle=0}}
\caption{This figure shows the ratio $\dot{\Omega}_{\rm CS}/\dot{\Omega}_{\rm GR}$ as a function of CS characteristic mass for the LAGEOS satellites, with semi-major axis $a \approx 12,000\ \mathrm{km}$.  The shaded region corresponds to a $10\%$ experimental error bar with a $1\sigma$ detection confidence~\cite{Ciufolini:2007wx,Ciufolini:2004rq,Ciufolini:2004gp}.}
\label{fig:OmegaCSdot}
\end{figure}

Another type of torque-induced precession is also affected by CS modified gravity, namely that experienced by a gyroscope. Consider then a gyroscope with spin angular momentum $S$ in circular orbit around the Earth. Neglecting geodetic precession, which is unaffected in the non-dynamical modified theory, the rate of change $S$ is given by
\be
\dot{S}^{i} = 2 \epsilon^{0ijk} B_{j} S_{k},
\ee
where $B^{i}$ is the gravitomagnetic field. The precessional angular velocity is then given by $\dot{\Phi} := |\dot{S}^{i}|/|S^{i}|$, which is corrected in CS by
\be
\label{Eq-LT}
\frac{\dot{\Phi}_{\textrm{CS}}}{\dot{\Phi}_{\textrm{GR}}} = 15 \frac{a^{2}}{R^{2}} j_{2}(m_{\textrm{CS}} R) \left[ y_{1}(m_{\textrm{CS}} a) + m_{\textrm{CS}} a y_{0}(m_{\textrm{CS}} a) \right],
\ee
where $\dot{\Phi}_{\textrm{GR}}$ is the GR prediction~\cite{Smith:2007jm}. In Eq.~\eqref{Eq-LT}, $R$ is the distance from the center of the Earth to the gyroscope (roughly $7000 {\textrm{km}}$ for Gravity Probe B)

Given an experimental verification of the Lense-Thirring gyromagnetic effect, one could then place a constraint on the CS scalar, as explained previously with the LAGEOS satellites. Gravity Probe B~\cite{GPBwebsite,Will:2002ma} was designed to measured precisely this effect to percent accuracy, but since its launch it has faced certain difficulties that might degrade its accuracy~\cite{Brunfiel}. Smith, {\emph{et.~al.~}}studied the possibility that Gravity Probe B could place a stronger constraint than the LAGEOS satellites, but this was found to not be the case~\cite{Smith:2007jm}. 

The estimates of Smith, {\emph{et.~al.~}}\cite{Smith:2007jm} presented above assume a spherical Earth, where its internal structure is neglected. The Earth, however is an oblate spheroid with layers of different density. These non-spherical corrections seem not to matter since the CS correction is affected by them in the same way as GR. Thus, the relative difference between the GR and CS effects remains roughly the same. Lastly, the estimates described above assume the Earth is not moving in the barycenter frame. In the point particle case, we saw that if their velocity is non-vanishing then the CS modification leads to a non-boundary correction to the gravitomagnetic field. These corrections remain to be studied further.

\subsection{Binary Pulsar Test}

Non-dynamical (and possible dynamical) CS modified gravity has been shown to modify only the gravitomagnetic sector of the metric, which does not influence most astrophysical processes. This is particularly true outside the Solar system, where stars inside some galaxy will possess randomly oriented velocities that will lead to a vanishing averaged CS correction. On cosmological scales, the CS effect is also mostly irrelevant since, for example, the equations of structure formation are not corrected, except for observations of the cosmic microwave background, which we shall discuss in the next chapter. 

Some astrophysical process, however, are CS modified. One example of this is the formation of accretion discs around protoplanetary systems, but although non-zero the CS correction would be difficult to measure because it would be greatly suppressed by the almost negligible compactness of such systems. Another, perhaps more interesting astrophysical scenario are double binary pulsars, such as PSR J$0737-3039$A/B~\cite{Burgay:2003jj}, which consist of two rapidly rotating neutron stars that are in orbit around each other. Neutron stars possess a mass of approximately $1.4 M_{\odot}$, while its radius is on the order of $10 \; {\textrm{km}}$, which implies their compactness is roughly $1/5$. Such a large compactness leads to strong gravitational fields that can be used to test GR to an unprecedented level~\cite{Hulse:1974eb}.

Consider then double binary pulsars and let us model their orbital evolution via a geodesic study of a compact object in the background of a rotating, homogeneous sphere. Under these assumptions, we can employ the gravitomagnetic field found by Smith, {\emph{et.~al.~}}\cite{Smith:2007jm}, where the motion is determined by the four-acceleration $\vec{a} = -4 \vec{v} \times \vec{B}$, $v$ is the velocity of one of the binary components and $\vec{B}$ is the gravitomagnetic field of the other [Eqs.~\eqref{BfullGR} and~\eqref{BfullCS}].  To leading order in $\dot\theta$~\footnote{Employing $\dot{\theta}$ as an expansion parameter is not formally valid, since this quantity can be dimensional. Corrections are actually proportional to ${\cal{O}}(\dot{\theta}/R)$ or ${\cal{O}}(\dot{\theta}/a)$, both of which can be shown to be much less than unity here.}, the CS correction to the gravitomagnetic field is 
\be
\vec{B}_{\CS} = 
\frac{c_{0}}{r} \cos[\xi(r)]  \left[ 
 \vec{{\cal{J}}}
-\tan\xi \left(\vec{{\cal{J}}} \times \hat{r}\right) 
- \left(\vec{{\cal{J}}} \cdot \hat{r}\right)  \hat{r} \right],
\ee
with $\xi(r) = 2 r \kappa/(\dot{\theta} \alpha)$,  $c_{0} = 15 \alpha \dot{\theta}/(4 \kappa R)  \sin[\xi(R)]$, $\hat{r} = \vec{r}/r$, $\vec{{\cal{J}}} = \vec{J}/R^{2}$ and $(\cdot,\times)$ the Euclidean dot and cross products. 

Let us now parameterize the trajectory with equatorial coordinates, where~\cite{satbook}
\ba
\hat{r} &=& \left[ \cos{u}, \cos{\iota} \sin{u}, \sin{\iota} \sin{u} \right],
\nonumber \\
\hat{t} &=& \left[ -\sin{u},\cos{\iota} \cos{u}, \sin{\iota} \cos{u} \right],
\nonumber \\ 
\hat{n} &=& \left[0,-\sin{\iota},\cos{i}\right],
\label{eq-coord}
\ea
are the radial, transverse and normal unit vectors relative to the comoving frame in the orbital plane. In Eq.~\eqref{eq-coord}, $\iota$ is the inclination angle, $w$ is the argument of perigee, $u = f + w$ with $f$ the true anomaly and $\Omega=0$ is the right ascension of the ascending node, where here this is co-aligned with the $\hat{x}$ vector~\cite{satbook}. 

The variation of the Keplerian orbital elements is determined by the perturbation equations, which can be obtained by projection of the geodesic acceleration onto the above triad. To leading order, the radial projection $a_{r} := \vec{a} \cdot \hat{r}$ is the only one CS modified, with~\cite{Yunes:2008ua} 
\be
a_{r}^{\CS} = -4 c_{0} \; \dot{u} \; {\cal{J}} \;  \left\{ \cos{\iota}  \cos\left[\xi(r)\right]  + \sin{\iota} \cos{u} \sin\left[\xi(r)\right]  \right\}.
\ee
The precession of the perigee is given by the perturbation equation $\dot{w} = - a_{r}/(n\; a \; e)  \cos{f}$, where the mean motion $n = \sqrt{M/a^{3}}$ and $e$ is the eccentricity. 

We shall consider next only the averaged rate of change of orbital elements, and in particular $<\dot{w}>$. Averaging over one orbital period~\cite{Yunes:2008ua} 
\be
<\dot{\omega}> := \int_{0}^{T} \frac{\dot{\omega}}{P} dt =  \int_{0}^{2 \pi} \dot{\omega} \frac{\left(1 - e^{2}\right)^{3/2}}{2 \pi \left(1 + e \cos{f} \right)^{2}} df,
\label{ave}
\ee
during which we assume the pericenter is constant and the motion is Keplerian, such that
\ba
\dot{u} &\sim& \dot{f} =  n \left(1 + e \cos{f} \right)^{2} \left(1 - e^{2}\right)^{-3/2},
\nonumber \\
r &=& a \left(1 - e^{2} \right) \left(1 + e \cos{f}\right)^{-1}.
\ea
This last assumption is justified since in the weak field the CS-corrected motion of test particles about any background remains CS unmodified.~\cite{Sopuerta:2009iy}. 

The CS correction to the averaged rate of change of the perigee is then, to linear order in the eccentricity~\cite{Yunes:2008ua},
\be
\left< \dot{w} \right>_{\CS} 
= \frac{15}{2a^{2} e} \frac{J}{R^{2}} \frac{\dot{\theta}}{R}  X \sin\left( \frac{2 \kappa R}{\alpha \dot{\theta}} \right) \sin\left(\frac{2 a \kappa}{\alpha \dot{\theta}}\right),
\label{dwdt_ave}
\ee
where $X := a \sin{\iota}$ is the projected semi-major axis. Note that the limit $e \to 0$ is meaningless because the orbital orientation is ill-defined for circular orbits. The scaling found here is consistent with the Solar System results presented in Eq.~\eqref{Smith}~\cite{Smith:2007jm}. The ratio of CS correction to the GR expectation is proportional to $a^{2} \dot{\theta}/R^{3}$, since $<\dot{\omega}>_{\GR} \sim J/a^{3}$. We see then that a Solar System test will naturally be weak since $a/R \sim 1$, while for binaries $a/R \sim {\cal{O}}(10^{4})$.

Formally speaking, the sole calculation of perigee precession is not sufficient to constrain an alternative theory with binary pulsar observations~\cite{lrr-2006-3}. In principle, at least two other quantities must be computed and measured in order to break the degeneracy in the determination of the individual component masses.  In CS gravity, however, other binary pulsar-relevant quantities will not be corrected, because $C_{00}$ identically vanishes for the canonical choice of CS scalar, or the correction becomes subleading, as in the case of the quadrupole gravitational wave emission formula. The only post-Keplerian parameters that is CS corrected to leading order is $\dot{\omega}$ in the non-dynamical theory with canonical CS scalar and all parameters can be obtained from~\cite{Kramer:2006nb}. 

Yunes and Spergel~\cite{Yunes:2008ua} used this calculation to place a strong bound on the non-dynamical framework. Using the relevant system parameters of PSR J$0737-3039$A/B found in~\cite{Kramer:2006nb}, the CS scalar was constrained to be 
\be
\dot{\theta} \lesssim 4 \times 10^{-9} \; {\textrm{km}}, \rightarrow m_{\CS} \gtrsim 100 \; {\textrm{meV}}
\ee
which is $10^{11}$ times stronger than current Solar System constraints. A similar analysis in the full, strong-field dynamical formalism is still lacking. In particular, the coupling of fermions to the CS term might be relevant for calculations involving neutron stars~\cite{Alexander:2008wi}.

\subsection{Galactic Rotation Curves}

The only galactic study that has been currently carried out is related to the flat-rotation curves of galaxies~\cite{Konno:2008np}. Consider a collection of stars in a spiral galaxy and measurements of their orbital velocity $v$ and their distance to the galaxy center $r$, through the shift of spectral lines. The plot of this velocity as a function of distance is the so-called {\emph{galaxy rotation curve}}, which according to Newtonian mechanics should obey a square-root fall-off $v \propto r^{-1/2}$. For a large number of such galaxies, with different luminosities, Rubin and Ford Jr.~\cite{1970ApJ...159..379R,Rubin:1980zd} have found that the galaxy rotation curve flattens with distance, which implies the existence of additional non-visible, or dark, matter. 

Konno, {\emph{et.~al.~}}\cite{Konno:2008np} have attempted to explain the flatness of rotation curves through non-dynamical CS modified gravity ($\alpha = - l \kappa$ and $\beta = 0$, where $l$ is a coupling constant) with a non-canonical CS scalar. We have already discussed the form of the metric obtained in the slow-rotation limit, when assuming $\vartheta \propto r \cos{\theta}$ [see eg.~Eq.~\eqref{Konno-scalar} and Eq.~\eqref{KMTsol}].  With this solution, Konno, {\emph{et.~al.~}}\cite{Konno:2008np} studied circular, equatorial geodesic motion, and found that the orbital angular velocity $v := r d\phi/dt  = -L f/(r E)$, where $E$ and $L$ and the conserved energy and angular momentum, which for large radius becomes
\be
v = \pm \sqrt{\frac{M}{r}} + \frac{C_{2}}{2} + {\cal{O}}(J)^{2},
\ee
and we recall that$C_{2}$ is a constant that depends on the spin angular momentum. This result is to be contrasted with the Kerr metric, for which $v = (M/r)^{1/2} - J/r^{2}$. 

The solution found by Konno, {\emph{et.~al.~}} is interesting in that it sheds light on some of the effects of non-dynamical CS modified gravity on certain observables with a non-conventional CS scalar, but before victory can be claimed over the rotation curves one must consider the solution more carefully. In doing so, one discovers that this solution possesses a few drawbacks that render it rather unphysical as a true spinning, BH solution. One of main problem is rooted in that the solution was found in the non-dynamical formulation, which as we have argued is quite contrived, arbitrary and probably not well-posed. 

One then hopes that an embedding of this solution in the dynamical framework can be found, but as we explain below this is impossible. The dynamical formulation requires that a well-defined scalar field couple to the field equations via a stress-energy tensor. The scalar fields studied in the non-dynamical formulation, however, all possess the feature of leading to infinite total energy in the scalar field. This issue cannot be bypassed by simply stating that the stress-energy contribution is second-order in the slow-rotation parameter, since the energy contribution is not large, but infinite. Thus, the solution discussed above is not a self-consistent one in the dynamical framework. 

Moreover, on closer inspection, Yunes and Pretorius~\cite{slow-rot} have found additional solutions in the non-dynamical formulation for other choices of CS scalar [see eg.~Eq.~\eqref{w-good}]. This family of solutions is better-behaved in that they do not contain any logarithmic divergences in the metric, but they lead to an orbital angular velocity
\be
v_{\phi} \sim \sqrt{\frac{M}{r}} -  \frac{M}{r^{2}},
\ee
in the far-field limit $M/r \ll 1$, which cannot explain the flat rotation curves.

At this juncture, one could make the argument that the above analysis is nothing but evidence that Nature somehow selects the solution found by Konno, {\emph{et.~al.~}}, but we shall here argue precisely the opposite. The non-dynamical framework does not suggest that either of the two solutions discussed here is more valid or less valid than the other. In fact, it is the immense freedom in the choice of CS scalar the leads to two different observables and points at an incompleteness of the framework. This observation, coupled with the overconstraining feature of the non-dynamical framework, casts some doubt as to validity of suggesting the CS correction as an explanation of the flatness in galactic rotation curves.

\subsection{Gravitational Wave Tests}

Several tests have recently been proposed of the modified theory with GWs. All such tests have so far concentrated on waves generated by binary systems in the early inspiral phase, where CS correction arises due to the propagation of the wave (instead of the mechanism). As we reviewed in Sec.~\ref{Sec:ApproxSol}, the main effect of the CS correction on the propagation of GWs is an amplitude birefringence, characterized by the parameter $\zeta$ defined in Eq.~\eqref{zeta}. Thus, if a GW detection can constrain the magnitude of $\zeta$, one can derive a bound on the CS scalar. 

Alexander, {\emph{et.~al.~}}\cite{Alexander:2007kv} have proposed a GW test of non-dynamical CS modified with a generic CS scalar through the space-borne GW detector LISA~\cite{Bender:1998,Danzmann:1997hm,Danzmann:2003tv,Sumner:2004,Merkowitz:2006yh}. The sources in mind are supermassive black hole binaries located at redshifts $z < 30$. In principle, in order to determine how good of a constraint LISA could place on CS modified gravity, one would have to carry out a full covariance (Fisher) matrix analysis, including all harmonics in the signal amplitude, since the CS correction affects precisely this amplitude. 

One can obtain an order-of-magnitude estimate, however,  by making the following assumptions. First, let there be a GW detection in two GW detectors, such that one can reconstruct both the right- and left-polarized amplitudes. Second, let us model the noise as white, with one-sided spectral noise density $S_{0}$. Third, focus attention on the Fisher matrix $\Gamma_{ij}$ related to the parameters that affect the amplitude of the GW signal, neglecting the phase parameters, namely~\cite{Finn:1992wt,Finn:1992xs}
\be
\Gamma_{ij} = \sum_{k=\text{R,L}}\frac{2}{S_{0}} \int_{t_i}^{t_f}
\Re\left(\frac{\partial{m_k}}{\partial
    \lambda^i}\right)\,\Re\left(\frac{\partial{m_k}}{\partial \lambda^j}\right)\, dt,
\ee
where the observation period is $(t_{i},t_{f})$, $\lambda^{i} = ({\cal{D}},\chi,\zeta)$ is a vector of parameters that affect the amplitude (the natural logarithm of the luminosity distance to the source ${\cal{D}}=\ln d_{L}$ and the cosine inclination angle $\chi = \cos{\iota}$) and $m_{\textrm{R,L}}$ is the scalar detector response. The ensemble average co-variance is then simply $\nu_{ij} = \left(\Gamma^{-1}\right)_{ij}$, where the diagonal components are a measure of the accuracy in the determination of the $ii$ parameter.

For a GW signal detected on plane ($\chi =0$), the accuracy in the determination of $\zeta$ is given by~\cite{Alexander:2007kv}
\be
\label{nuzetazeta}
\nu_{\zeta \zeta} = \frac{({\cal{M}} H_{0})^{2}}{4 \rho^{2}} \frac{{\cal{I}}^{2}}{{\cal{I}} {\cal{K}} - {\cal{J}}}
\ee
where ${\cal{M}} = m_{1} m_{2}/(m_{1} + m_{2})^{2}$ is the chirp mass, with binary mass components $m_{1,2}$, and $\rho^{2}$ is the squared signal-to-noise ratio
\be
\rho^{2} = \frac{1}{S_{0}} \int_{t_{i}}^{t_{f}} dt \left(A_{\textrm{R}}^{2} + A_{\textrm{L}}^{2} \right),
\ee
with $A_{\textrm{R,L}}$ the GW amplitudes. In Eq.~\eqref{nuzetazeta}, the noise moments are defined via
\ba
\mathcal{I} &:=& \int_{t_i}^{t_f} \left(\frac{k_0(t)\mathcal{M}}{2}\right)^{4/3}\frac{dt}{S_0} = -\frac{2^{1/3}\mathcal{M}}{S_0}\frac{5}{64}\left.\left(k\mathcal{M}\right)^{-4/3}\right|_{k_{\min}}^{k_{\max}}
\\
\mathcal{J} &:=& \int_{t_i}^{t_f} \left(\frac{k_0(t)\mathcal{M}}{2}\right)^{7/3}\frac{dt}{S_0} = -\frac{2^{1/3}\mathcal{M}}{S_0}\frac{5}{32}\left.\left(k\mathcal{M}\right)^{-1/3}\right|_{k_{\min}}^{k_{\max}}\\
\mathcal{K} &:=& \int_{t_i}^{t_f} \left(\frac{k_0(t)\mathcal{M}}{2}\right)^{10/3}\frac{dt}{S_0}= \frac{2^{1/3}\mathcal{M}}{S_0}\frac{5}{128}\left.\left(k\mathcal{M}\right)^{2/3}\right|_{k_{\min}}^{k_{\max}}.
\ea

For a binary black hole system with total mass $M = m_{1} + m_{2}=10^6\,M_{\odot}\left(1+z\right)^{-1}$ at redshift $z$, a GW one year before coalescence has a wavelength in the range $\lambda = (10^{2},10^{4}) c\,\text{Hz}^{-1}$. For such a system~\cite{Alexander:2007kv} 
\ba
\rho &=&  \frac{10^5h_{100}}{1+z-\sqrt{1+z}} \left(\frac{10^{-40}\,\text{Hz}^{-1}}{S_0}\right)^{1/2}
\\
\nu_{\xi\xi}&=& 3.1\times10^{-40}\left(\frac{S_0}{10^{-40}\,\text{Hz}^{-1}}\right)\left(1+z-\sqrt{1+z}\right)^2,
\ea
where $h_{100} =  H_{0}/100 \, {\textrm{s}} \, {\textrm{Mpc}}/{\textrm{km}}$. Such a result implies a $1\sigma$ upper bound on $\xi$ of order $10^{-19}$ for a LISA GW detection at $z=15$. With the canonical choice of $\vartheta$, this translates to a constraint on the CS scalar of approximately $\alpha\dot{\vartheta}/\kappa < 10^{-2} \, {\textrm{km}}$. Not only is the possible constraint on $\dot{\vartheta}$ five orders of magnitude better than with Solar System tests, a GW detection also constrains a different sector of the modified theory, since it samples the temporal evolution of the CS scalar, instead of its local value. This is because a GW detection really constrains the evolution of the CS scalar from the time of emission of the GW to its detection on Earth. 

Perhaps a more interesting test of CS modified gravity can be performed using GWs emitted by extreme-mass ratio inspirals or binary 
black hole mergers~\cite{Sopuerta:2009iy}. These systems sample the strong-gravitational regime of spacetime, in which the CS 
correction is enhanced, as shown by the CS modified Kerr solution~\cite{slow-rot}. The generation of GWs would then also be CS 
modified, not only due to modifications in the trajectories due to corrections to the background metric, but also due to the fact that the CS 
scalar must carry energy-momentum away from the system. Even ignoring the latter, Sopuerta and Yunes~\cite{Sopuerta:2009iy} have 
shown that the background modifications lead to extreme- and intermediate-mass ratio inspirals, whose waveforms are sufficiently 
distinct from their GR counterpart to allow for a test of the radiative sector of the dynamical theory over a few-month integration period. A more detailed data analysis study is currently underway to determine the accuracy to which the theory can be tested.

\section{Chern-Simons Cosmology}
\label{SEC:CScosmology}

After the quintessence model was proposed to account for the acceleration seen from Type Ia supernovae~\cite{Riess:2004nr}, Carroll proposed that the quintessence field can generically couple to parity violating terms in the electromagnetic sector~\cite{Carroll:1990vb,Carroll:1998zi}.  This was used as a constraint on the quintessence dark energy models since such parity violating, ``birefringent'' couplings are generic. Soon after, the cosmological study of CS modified gravity was first proposed by Lue, Wang and Kamionkowski~\cite{Lue:1998mq} as a way to search for parity-violating effects from the GW sector of the CMB polarization spectrum. 
This was so because, in CS modified gravity, an asymmetry of left and right-handed GWs leads to an anomalous cross correlation in the GW power spectrum. Such a pioneering study was then followed up by Saito, et.~al.~\cite{Saito:2007kt}, who improved on their calculation.

The possibility of parity violation in the gravitational sector led Alexander, Peskin and Sheikh-Jabbari (APS)~\cite{Alexander:2004us,Alexander:2007qe} to propose an explanation for the observed matter-antimatter asymmetry in the universe.  The lepton asymmetry is generated from one simple and generic ingredient: the CS correction to the action, which leads to an asymmetric left/right production of GWs, and due to the gravitational ABJ anomaly~\cite{1969PhRv..177.2426A,Bell:1969ts}, chiral leptons are produced.  APS were able to account for a correct amount of lepton asymmetry using the bounds placed on the GW power spectrum amplitude and the scale of inflation. In what follows, we shall review the inflationary production of GWs with the CS term present.  We show how the GW power spectrum is modified and apply the GW solutions to the inflationary leptogenesis mechanism of Alexander, Peskin and Sheikh-Jabbari.  Following this, we will review the LWK analysis of constraining parity violation in the CMB.

\subsection{Inflation and the Power Spectrum}
\label{Sec:Inflation}

In this chapter we shall consider CS modified gravity in the context of inflation.
The idea of inflation is quite simple: a period $(t_{i},t_{f})$ in which the scale factor $a(t) \sim e^{H(t_{f} - t_{i})}$ grows exponentially.  During such an epoch, the FRW equations govern the dynamics of the scale factor $a(t)$ via
\be 
\left(\frac{\dot{a}}{a}\right)^{2} = \frac{8\pi}{3}G \rho + \frac{k}{a^{2}}  \qquad 
{\textrm{and}} \qquad 
 \frac{2{\ddot{a}}}{a} + \frac{{\dot{a}}^{2}}{a^{2}} + \frac{k}{a^{2}} =-8\pi G \rho, 
 \ee
where $H:= \frac{{\dot{a}}}{a}$ is the Hubble parameter.  We see that inflation, $a(t) \sim e^{tH}$, occurs when there is a fluid of negative pressure
$p =-\rho$.

In scalar field theories, this requirement can be obtained if the scalar field is slowly rolling down its false vacuum potential, often referred to as {\emph{slow-roll}} inflation.   In what follows, we shall assume slow-roll inflation (see eg.~\cite{Mukhanov:1990me} for further details) and we shall study  the power spectrum of CS birefringent GWs during the inflationary epoch.  We will then rederive the consistency relation between the scalar to tensor ratio and the slow-roll index of inflation following Ref.~\cite{Alexander:2004us}.

\subsubsection{Gravitational Waves and the Effective Potential}

Consider inflation driven by a pseudoscalar field $\phi$ with a standard kinetic term and a potential $V(\phi,\lambda)$. 
Consider also the Einstein-Hilbert action coupled to the gravitational CS
term, as given by Eq.~\eqref{CS-action} with the choices $\alpha = 2 \kappa$, $\beta = 0$ 
and $f(\phi)$ a functional of the inflation, which is identified with the CS scalar.

Let us now linearize the action in an FRW background, as explained in Sec.~\ref{Sec:ApproxSol}.
The field equations for the GW become essentially Eq.~\eqref{GW-FE}
and, concentrating on plane-wave solutions, we find the evolution equation for the phase
in a left/right-basis [Eq.~\eqref{eq:disp}]. Let us now introduce the quantity $z_s$ via
\begin{equation}
\label{defz}
z_s \left(\eta, {\mathbf k}\right)\equiv a\left (\eta \right)
\sqrt{1-\lambda ^sk\frac{f'}{a^2}}\,
\end{equation}
and the new amplitude $\mu _{\mathbf k}^s(\eta )$ defined by $\mu
_{\mathbf k}^s\equiv z_s h_{\mathbf k}^s$, where $s = \{R,L\}$, $\lambda_{R,L} = \pm$ 
and the subscript {\bf{k}} reminds us of the wavenumber dependance. The evolution 
equation for the phase then becomes a parametric oscillator equation for $\mu_{\mathbf k}^s$:
\begin{equation}
\label{eq:motionmu}
\left(\mu _{\mathbf k}^s\right)''
+\left(k^2-\frac{z_s''}{z_s}\right)\mu _{\mathbf k}^s=0\, .
\end{equation}
where primes stand for derivatives with respect to conformal time $\eta$. 
The effective potential $z_s''/z_s$ depends on time, on polarization but also on wavenumber, 
which distinguishes this CS effect from the standard GR case, where the effective
potential depends on conformal time only. 


Let us study the effective potential in more detail. 
Using Eqs.~\eqref{defz} and~\eqref{eq:motionmu}, this potential is given exactly by 
\begin{equation}
\label{poteff}
\frac{z_s''}{z_s} = \frac{a''}{a}-{\cal H}\lambda ^sk
\frac{\left(f'/a^2\right)'}{1-\lambda ^sk\left(f'/a^2\right)}
-\frac{\lambda ^sk}{2}\frac{\left(f'/a^2\right)''}{1-\lambda
^sk\left(f'/a^2\right)} 
 -\frac{\left(\lambda ^sk\right)^2}{4}
\frac{\left[\left(f'/a^2\right)'\right]^2}{\left[1-\lambda
^sk\left(f'/a^2\right)\right]^2}\, .
\end{equation}
For convenience, we choose specify the CS coupling functional $f[\phi]$ via~\cite{Alexander:2004us,Alexander:2007qe}
\begin{equation}
f=\frac{{\cal N}}{16\pi ^2M_{_{\rm Pl}}^2}\frac{ \phi }{M_{_{\rm Pl}}}\, .
\label{f-param}
\end{equation}
where $M_{_{\rm Pl}}\equiv m_{_{\rm Pl}}/\sqrt{8\pi }$ is the reduced
Planck mass and ${\cal N}$ a number that can be related to the string scale. In
terms of the slow-roll parameters $\epsilon \equiv -\dot{H}/H^2$,
$\delta \equiv -\ddot{\phi }/(H\dot{\phi })$ and $\xi \equiv
(\dot{\epsilon }-\dot{\delta })/H$ (an overhead dot means a derivative with
respect to cosmic time), we have at leading order in these 
parameters (see also Ref.~\cite{Martin:1999wa}) 
\be
a(\eta ) \sim (-\eta )^{-1-\epsilon }\, ,
\quad 
\phi ' \simeq -M_{_{\rm Pl}}
{\cal H}\sqrt{2\epsilon}\, .
\ee
From these expressions, one deduces that to leading order
\begin{equation} \frac{f'}{a^2}=-\frac{{\cal N}}{16\pi
^2M_{_{\rm Pl}}^2} \frac{\cal H}{a^2}\sqrt{2\epsilon}\simeq
\frac{{\cal N}}{16\pi ^2}\left(\frac{H_{_{\rm
inf}}}{M_{_{\rm Pl}}}\right)^2 \sqrt{2\epsilon }\eta
\end{equation}
because ${\cal H}\simeq -(1+\epsilon)/\eta $. 


The equation of motion is still in a rather involved form, 
but it can be simplified further by introducing the characteristic scale $k_{\rm C}$ 
and $\Theta$ parameter~\cite{Alexander:2004us,Alexander:2007qe}
\begin{equation}
\label{deftheta}
k_{_{\rm C}} \equiv  k\frac{{\cal N}}{32\pi ^2} \left(\frac{H_{_{\rm
inf}}}{M_{_{\rm Pl}}}\right)^2 \sqrt{2\epsilon }=k\frac{\Theta }{16}\, ,
\qquad
\Theta \equiv \frac{{\cal N}}{2\pi ^2}
\left(\frac{H}{M_{_{\rm Pl}}}\right)^2\sqrt{2\epsilon}\, .
\end{equation}
Let us further introduce the variable $x\equiv\Theta k\eta /8<0$, such that Eq.~\eqref{eq:motionmu} takes the form
\begin{equation}
\frac{{\rm d}^2\mu }{{\rm d}x^2}+\left[\frac{64}{\Theta ^2}
-f_{s}(x)\right]\mu =0\, ,
\end{equation} 
with 
\begin{equation}
f_{s}(x) = \frac{2+3\epsilon }{x^2}+\frac{\lambda_{s}}{x(1 - \lambda_{s} x)}-\frac14 
\frac{1}{(1- \lambda_{s}x)^2}\, ,
\end{equation}
Using slow-roll perturbation theory and the functional relations for $f(\phi)$, the equation of motion can be 
maximally simplified into   
\begin{equation}
\frac{{\rm d}^2 \mu _{\mathbf k}^{_{\rm L}}}{{\rm d}\tau ^2}
+\left[-\frac14+\frac{i\Theta }{16 \tau }+
\frac{1}{4\tau ^2}\right]\mu _{\mathbf k}^{_{\rm L}}=0\, .
\label{Whit1}
\end{equation} 
where we have defined $\tau \equiv 16i(1+x)/\Theta $. We recognize Eq.~\eqref{Whit1}
as the Whittaker equation [see eg.~Eq.~(9.220.1) of
Ref.~\cite{Grad}], whose corresponding, normalized solution is
\begin{equation}
\mu _{\mathbf k}^{_{\rm L}}=-\frac{4\sqrt{\pi }\ell _{_{\rm Pl}}}{\sqrt{2k}}
{\rm e}^{ik\eta _{_{\rm i}}}{\rm e}^{-\pi \Theta /32}
{\rm W}_{i\Theta/16, 0}\left[\frac{16i(1+x)}{\Theta }\right]\, ,
\end{equation}
where ${\rm W}_{\kappa ,\mu }(z)$ is the Whittaker function.  

\subsubsection{CS Corrected Inflationary Gravitational Waves on Large Scales}

Let us now concentrate on super-horizon scales, ie.~$x\sim0$, such that we can make contact with CMB observables.
The effective potentials can then be approximated via
\be
\label{approxpot}
f_{s}(x) \simeq  \frac{2+3\epsilon }{x^2}+\frac{\lambda_s}{x}-\frac14\,,
\ee
where the first term gives the standard slow-roll term behavior, while the second term
represents the birefringent CS correction. With this potential, the equation of motion becomes
\begin{equation}
\frac{{\rm d}^2\mu _{\mathbf k}^s}{{\rm d}x ^2}
+\left[\frac{64}{\Theta ^2}+\frac14-\frac{\lambda_s}{x}
-\frac{2+3\epsilon }{x^2}\right]\mu _{\mathbf k}^s=0\, .
\label{whit2}
\end{equation}
Using results from Ref.~\cite{Martin:2000ei}, we define
\begin{equation}
y \equiv i \sqrt{1+\frac{256}{\Theta ^2}}x\, ,\quad \kappa \equiv
\frac{i\lambda ^s}{\sqrt{1+256/\Theta ^2}}, \quad \xi \equiv
\frac32+\epsilon \,.
\end{equation}
and simplify the equation of motion in the limit $256/\Theta ^2\gg 1$, which corresponds
to large scale behavior, such that Eq.~\eqref{whit2} becomes
\begin{equation}
\frac{{\rm d}^2\mu _{\mathbf k}^s}{{\rm d}y^2}
+\left[-\frac14+\frac{\kappa}{y }+\left(\frac14-\xi ^2\right)
\frac{1}{y^2}\right]\mu _{\mathbf k}^s=0\,.
\end{equation} 
The exact general solution to this equation is given in terms of Whittaker
functions
\begin{equation}
\mu _{\mathbf k}^s(\eta )=C_1^s(k)W_{\kappa, \xi}\left(y\right)
+C_2^s(k)W_{-\kappa, \xi}\left(-y\right)\, ,
\end{equation}
where $C_1^s(k)$ and $C_2^s(k)$ are two constants of integration.

Initial conditions are determined in the divergence-free region $-1<x\ll 0$, where
we assume $\mu^{s}_{\bf{k}}$ has a plane wave behavior~\cite{Martin:2000ei} 
\begin{equation}
C_1^s(k) = -\frac{4\sqrt{\pi }\ell _{_{\rm Pl}}}{\sqrt{2k}} {\rm
e}^{iq\eta _{\rm i}}\exp\left(-\frac{\lambda ^s\pi
\Theta}{32}\right)\, , 
\qquad
C_2^s(k) = 0\, ,
\end{equation}
where $\ell_{_{\rm{Pl}}}$ is the Planck length and we have used that $W_{\kappa, \xi}(y) \sim {\rm e}^{-y/2} y^{\kappa }$ as $y \to +\infty$.
This assumption is equivalent to
requiring that non-linear phenomena occurring near the divergence of the effective potential 
does not affect the initial conditions in the region $x>-1$. Indeed, such an assumption is also
made in inflationary cosmology, where the vacuum is assumed to be the
correct initial state, in spite of the fact that modes of astrophysical interest today originate from the
trans-Planckian region~\cite{Martin:2000xs}. A possible
weakness of the above comparison is that, in the case of the
trans-Planckian problem of inflation~\cite{Martin:2000xs,Lemoine:2001ar}, one can show
that (under certain conditions) the final result can be robust to changes in the short distance
physics~\cite{Martin:2000xs,Lemoine:2001ar}. In the present context, however, it
is more difficult to imagine that the non-linearities will not affect
the initial conditions. On the other hand, in the absence of a
second-order calculations and as a first approach to the problem, this
seems to be quite reasonable.

\subsubsection{Power Spectrum and Tensor-to-Scalar Ratio}

The power spectrum can be calculated either as the two-point correlation
function in the vacuum state or as a classical spatial average. We
take here the latter view, since a fully consistent quantum formulation 
of the present theory is not yet available. The power spectrum is then defined as
\begin{equation}
\langle h_{ij}\left(\eta ,{\mathbf x}\right)
h^{ij}\left(\eta ,{\mathbf x}\right)\rangle 
=\frac{1}{V}\int {\rm d}{\mathbf x}\, h_{ij}\left(\eta ,{\mathbf x}\right)
h^{ij}\left(\eta ,{\mathbf x}\right)\, ,
\end{equation}
with $V=\int {\rm d}{\mathbf x}$ is the total volume. Using the
properties of the polarization tensor, straightforward calculations
show that
\begin{equation}
\langle h_{ij}\left(\eta ,{\mathbf x}\right)
h^{ij}\left(\eta ,{\mathbf x}\right)\rangle 
=\frac{1}{\pi ^2}\sum _{s={\rm L},{\rm R}}\int _0^{+\infty }
\frac{{\rm d}k}{k}k^3\left\vert h_{\mathbf k}^s\right \vert ^2\, ,
\end{equation}
from which we deduce the power spectrum
\begin{equation}
\label{spec}
k^3P_h^s(k)=\frac{k^3}{\pi ^2}\left\vert \frac{\mu _{\mathbf
k}^s}{a(\eta )\sqrt{1-\lambda ^skf'/a^2}}\right \vert ^2\, .
\end{equation}
The power spectrum is usually proportional to $2k^3/\pi ^2$, where here the 
factor of $2$ is missing because we have not summed over polarizations.

The CS corrected power spectrum could be calculated exactly in terms of the Whittaker
function, but only the large scale behavior is needed so, in this regime, one has 
\begin{equation}
k^3P_h^s=\frac{16}{\pi}\frac{\ell _{_{\rm Pl}}^2}{a_0^2}
\frac{k^{-2\epsilon }}{2^{2\xi }} \frac{\Gamma ^2\left(2\xi
\right)}{\vert \Gamma \left(1/2+\xi -i\lambda ^s\Theta /16\right)\vert
^2}{\rm e}^{-\lambda ^s\pi \Theta /16}\, .
\label{powerspec1}
\end{equation}
where $\Gamma[\cdot]$ is the Gamma function and $\a_{0}$ is the value of the scale factor today.
Eq.~\eqref{powerspec1} can be expanded to first order in the slow-roll parameter to obtain 
\begin{equation}
k^3P_h^s(k) = \frac{16H_{_{\rm inf}}^2}{\pi m_{_{\rm Pl}}^2}
\frac12 {\cal A}^s\left(\Theta \right)
\left[1-2\left(C+1\right)\epsilon -2\epsilon \ln
\frac{k}{k_*}-\epsilon {\cal B}(\Theta )\right] \, ,
\label{Tspec}
\end{equation}
with, 
\begin{equation}
{\cal A}_{s} \equiv 1- \lambda_{s} \frac{\pi }{16}\Theta + \left(\frac{\pi
^2}{384}-\frac{1}{256}\right)\Theta ^2 +{\cal O}\left(\Theta
^3\right)\, ,\qquad
{\cal{B}} \equiv 4 \Psi - 2 \Psi(2),
\end{equation}
where $\Psi$ is related to the derivative of the Gamma function via $\Psi =\Gamma'/\Gamma$.
The amplitude of the CS corrected right-polarization state is
reduced while the one of the left-polarization state is
enhanced. Moreover, at leading order in the slow-roll parameter, the spectral index
remains unmodified, since $ n_{_{\rm T}}^s={\rm d}\ln
\left(k^3P_h^s\right)/{\rm d}\ln k=-2\epsilon $ for each polarization
state.

Let us now compute how the tensor to scalar ratio $T/S$ in the modified theory. 
In CS theory, the scalar power spectrum is not modified (see also Ref.~\cite{Choi:1999zy}) and
reads~\cite{Martin:1999wa} 
\begin{equation}
k^3P_{\zeta }=\frac{H^2_{_{\rm inf}}}{\pi m_{_{\rm Pl}}^2\epsilon }
\left[1-2\epsilon -2C(2\epsilon -\delta )-2(2\epsilon -\delta ) \ln
\frac{k}{k_*}\right]\,,
\end{equation}
while the tensor power spectrum is given by Eq.~\eqref{Tspec}.
The T/S ratio is then given by
\begin{eqnarray}
\frac{T}{S} &\equiv & \frac{1}{\left(k^3P_{\zeta }\right)}\left(\sum
_{s={\rm L},{\rm R}} k^3P_h^s\right)\Biggl\vert _{k=k_*} 
=
16\epsilon \times \frac12 \left[{\cal A}^{\rm L}\left(\Theta
\right)+{\cal A}^{\rm R} \left(\Theta \right)\right]
\\
&\simeq & 16\epsilon \times \left[1+\left(\frac{\pi
^2}{384}-\frac{1}{256}\right)\Theta ^2\right]\,,
\end{eqnarray}
where we see that the linear corrections in $\Theta $ has canceled out, and one
is left with a second-order correction only. Alternatively, we can express the above
result as the fraction CS correction
\begin{equation}
\label{cccheck}
\frac{\left(T/S\right)_{\Theta \neq 0}}{\left(T/S\right)_{\Theta =0}}
\simeq 1+0.022 \times \Theta ^2\,,
\end{equation}
from which it is clear that the CS correction is not
observable, since one has assumed here that $\Theta \lsim 10^{-5}$ 
(ie.~for the divergence of the effective potential to be in the trans-Planckian region).

The super-Hubble power spectrum exhibits two interesting regimes: a linear
and non-linear one.  The non-linear regime occurs when $k\eta \sim \Theta
^{-1}$, because the effective potential controlling the evolution of
the linear perturbations diverges and linear cosmological perturbation theory
becomes invalid. This divergence occurs for all modes (ie.~for all comoving wavenumber $k$)
but at different times. The full non-linear regime has not yet been investigated.

The linear regime is compatible with the stringy embedding of inflationary
baryogenesis~\cite{Alexander:2004xd}, part of which we discussed in Sec.~\ref{String-Theory and Chern Simons Gravity}.
In this context, $\Theta$ is enhanced, possibly leading to resonant frequencies that could be associated with the
observed baryon asymmetry. Since $\Theta$ is completely determined by the string scale and string coupling 
in a model-independent fashion, one obtains a direct link between stringy quantities and CMB anisotropies:
\begin{equation}
\frac{\left(T/S\right)_{\Theta \neq 0}}{\left(T/S\right)_{\Theta =0}}
\simeq 1+\frac{0.022}{4}\left(\frac{H_{_{\rm inf}}}{M_{\rm
10}}\right)^4 g_{\rm s}\epsilon \,,
\end{equation}
where we have used that
\begin{equation} 
{\cal N}= \pi ^2\sqrt{\frac{g_{\rm s}}{2}} \left(\frac{M_{_{\rm
Pl}}}{M_{_{\rm 10}}}\right)^2\, ,
\end{equation}
$M_{_{\rm 10}}$ is the ten-dimensional fundamental scale and
$g_{\rm s}$ is the string coupling. For reasonable values of string coupling (ie.~weak) and the string scale
set to $10^{16}{\rm GeV}$,  $\Theta \sim 10^{-2}$, but the stringy embedding admits much larger values of $\Theta$, 
forcing the analysis into the non-linear regime.

Large values of $\Theta$ (eg.~$\Theta \gsim 10^{-5}$) require a non-linear calculation, 
through which one could hope to obtain a significant modification to $T/S$ that might lead 
to an observable CMB signature. Unfortunately, it is precisely in this observable regime where
technical difficulties have prevented a full analysis of the cosmological perturbations. 

\subsection{Parity Violation in the CMB}
One of the major issues in the standard model of particle physics is the origin of parity violation in the weak interactions.   While we know that the other gauge interactions respect parity, it may be the case that the there is a definite handedness on cosmological scales.  The polarization pattern in the CMB fluctuations can leave an imprint of parity violation in the early universe through a positive measurement of cross correlation functions that are not parity invariant. 

The measurement of parity violation from CMB polarization was first discussed by Lue, Wang and Kamionkowski~\cite{Lue:1998mq}.  They realized that the presence of the CS term naturally leads to a rotation of the plane of polarization as a CMB photon travels to the observer.   It was later realized by Alexander~\cite{Alexander:2006mt} that gravitational backreaction of parity violating modes can lead to loss of power for parity-odd spherical harmonics, which lacking a systematic explanation, is observed in the CMB for low multipole moments.  

Such considerations can be understood by studying the polarization state of light as described through the Stokes parameters. Let us consider a classical electromagnetic plane-wave with electric field given by the following components:
\begin{eqnarray}
	\label{eq:2.2}
		E _{1}(t)=a_{1}\sin(\omega t- \epsilon _1)	\quad \text{and} \quad 	E _{2}(t)=a_{2}\sin(\omega t- \epsilon _2)		
\end{eqnarray} 
where we assume, for simplicity, that the wave is nearly monochromatic with frequency $\omega$, such that $a_1$, $a_2$,  $\epsilon _1$, and  $\epsilon _2$ only vary on time scales long
compared to $\omega^{-1}$. The Stokes parameters in the linear polarization basis are then defined as
\begin{eqnarray}
		I &\equiv & \left\langle(a_{1})^2 + (a_{2})^2\right\rangle, 
		\qquad
		Q \equiv  \left\langle(a_{1})^2 - (a_{2})^2\right\rangle,\\
		U &\equiv & \left\langle2a_{1}a_{2}\,\cos\delta\right\rangle,\qquad 
		V \equiv  \left\langle2a_{1}a_{2}\,\sin\delta\right\rangle,
\end{eqnarray} 
where  $\delta \equiv \epsilon _2-\epsilon _1$ and the brackets signify a time
average over a time long compared to $\omega^{-1}$. The $I$ parameter measures the intensity of the radiation, while the parameters $Q$, $U$, and $V$ each carry information about the polarization of the radiation. Unpolarized radiation (so-called \textit{natural light}) is described by $Q = U = V = 0$. The linear polarization of the radiation is encoded in $Q$ and $U$, while the parameter $V$ 
is a measure of elliptical polarization with the special case of circular polarization ocurring when $a_1 = a_2$ and $\delta = \pm \pi/2$. From here on we will simply refer to $V$ as the measure of circular polarization, which is technically correct if $Q=0$.

While $I$ and $V$ are coordinate independent, $Q$ and $U$ depend on the orientation of the coordinate system used on the plane orthogonal to the light's direction of propagation. Under a rotation of the coordinate system by an angle $\phi$, the parameters $Q$ and $U$ transform according to
\begin{equation}
\nonumber Q' = Q\, \cos(2\phi )+ U \, \sin(2\phi ), \qquad {\textrm{and}} \qquad
\nonumber U' = -Q\, \sin (2\phi ) + U\,\cos(2\phi ),
\end{equation}
while the angle defined by
\begin{equation}
\nonumber \Phi = \half \arctan \bigg( \frac{U}{Q} \bigg),
\end{equation}
goes to $\Phi - \phi$ following a rotation by the angle $\phi$. Therefore, $Q$ and $U$ only define an orientation of the coordinate system and not a particular direction in the plane: after a rotation by $\pi$ they are left unchanged. 

Physically, such transformations are simply a manifestation of the oscillatory behavior of the electric field, which indicate that $Q$ and $U$ are part of a second-rank symmetric trace-free tensor $P_{ij}$, i.e. a spin-2 field in the plane orthogonal to the direction of propagation. Such a tensor can be represented as
\begin{equation}
\label{eq:2.7}
 P_{ij} 
=\left(
\begin{array}{cc}
P & 0  \\
0 & -P \\  \end{array}
\right), 
\end{equation}
in an orthonormal eigenbasis, where $P = (Q^2+U^2)^{1/2}$ is ususally called the magnitude of linear polarization.
For example, in two-dimensional, spherical polar coordinates $(\theta,\phi)$, the metric is $\Omega_{ij} = {\textrm{diag}}(1,\sin^{2}\theta)$ and the polarization tensor is
\be {\cal{P}}_{ij}(\hat{n}) =\left(
\begin{array}{cc}
Q(\hat{n})  & - U(\hat{n}) \sin\theta  \\
-U(\hat{n})\sin\theta & -Q(\hat{n})\sin^{2} {\theta} \\  \end{array}
\right), 
\end{equation}
where we recall that $(i,j)$ run over the angular sector only.

The temperature pattern on the CMB can be expanded in a complete orthornmal set of spherical harmonics:
\begin{equation}
     {T(\hatn) \over T_0}=1+\sum_{l=1}^\infty\sum_{m=-l}^l
     a^{\rm T}_{(lm)}\,Y_{(lm)}(\hatn)
\label{Texpansion}
\end{equation}
where
\begin{equation}
 a^{\rm T}_{(lm)}={1\over T_0}\int d\hatn\,T(\hatn) Y_{(lm)}^*(\hatn)
\label{temperaturemoments}
\end{equation}
are the coefficients of the spherical harmonic decomposition of the temperature/polarization map and $T_0$ is the
mean CMB temperature. Likewise, we can also expand the polarization tensor in terms of a complete set of
orthonormal basis functions for symmetric trace-free $2\times2$ tensors on the
2-sphere, 
\begin{equation}
     {{\cal P}_{ij}(\hatn)\over T_0} = \sum_{l=2}^\infty\sum_{m=-l}^l
     \left[ a_{(lm)}^{\rm G}
     Y_{(lm)ij}^{\rm G}(\hatn) + a_{(lm)}^{\rm C} Y_{(lm)ij}^{\rm C}
     (\hatn) \right],
\label{Pexpansion}
\end{equation}
where the expansion coefficients are given by
\begin{equation}
     a_{(lm)}^{\rm G}={1\over T_0}\int \, d\hatn {\cal P}_{ij}(\hatn)
                             Y_{(lm)}^{{\rm G} \,ij\, *}(\hatn), \qquad\qquad
     a_{(lm)}^{\rm C}={1\over T_0}\int d\hatn\, {\cal P}_{ij}(\hatn)
                                      Y_{(lm)}^{{\rm C} \, ij\, *}(\hatn),
\label{defmoments}
\end{equation}

The basis functions 
$Y_{(lm)ij}^{\rm G}(\hatn)$ and $Y_{(lm)ij}^{\rm C}(\hatn)$
are given in terms of
covariant derivatives of spherical harmonics by
\begin{equation}
     Y_{(lm)ij}^{\rm G} = N_l
     \left( Y_{(lm):ij} - {1\over2} \Omega_{ij} Y_{(lm):k}{}^k \right),
\label{Yplusdefn}
\end{equation}
and
\begin{equation}
     Y_{(lm)ij}^{\rm C} = { N_l \over 2}    
     \left(\vphantom{1\over 2} 
       Y_{(lm):ik} \epsilon^k{}_j +Y_{(lm):jk} \epsilon^k{}_i \right),
\label{Ytimesdefn}
\end{equation}
where $\epsilon_{ij}$ is the completely antisymmetric tensor on the $2$-sphere,
a colon in an index list stands for covariant differentiation on the $2$-sphere,
and $N_l^{2} \equiv 2 (l-2)!/(l+2)!$ is a normalization factor.
Since the $Y_{(lm)}$'s provide a complete basis for scalar
functions on the sphere, the $Y_{(lm)ij}^{\rm G}$ and
$Y_{(lm)ij}^{\rm C}$ tensors provide a complete basis for gradient-type (G)
and curl-type (C) STF tensors, respectively.  This G/C decomposition is
also known as the scalar/pseudo-scalar decomposition, which is similar to the 
tensor spherical harmonic decomposition of Sec.~\ref{Sec:ApproxSol}.

Integration by parts
transforms Eqs.~(\ref{defmoments})
into integrals over scalar spherical
harmonics and derivatives of the polarization tensor:
\begin{equation}
     a_{(lm)}^{\rm G} = {N_l\over T_0}
     \int d\hatn \, Y_{(lm)}^*(\hatn)\,
     {\cal P}_{ij}{}^{:ij}(\hatn),
\label{Gmomentseasy}
\end{equation}
\begin{equation}
     a_{(lm)}^{\rm C} = {N_l\over T_0} 
     \int d\hatn \, Y_{(lm)}^*(\hatn)\,
     {\cal P}_{ij}{}^{:ik}(\hatn) \epsilon_k{}^j,
\label{Cmomentseasy}
\end{equation}
where the second equation uses the fact that
$\epsilon^{ij}{}_{:k}=0$. Given that $T$ and ${\cal P}_{ij}$ are real, all of
the multipoles must obey the reality condition
$a_{(lm)}^{\rm X\,*} = (-1)^m a_{(l,-m)}^{\rm X}$,
where ${\rm X}= \{{\rm T,G,C}\}$.
The spherical harmonics $Y_{(lm)}$ and $Y_{(lm)ij}^{\rm G}$ have parity $(-1)^l$, but the 
tensor harmonics $Y_{(lm)ij}^{\rm C}$ have parity $(-1)^{l+1}$.

The two-point statistics of the temperature/polarization map is then given via
\be 
C_l^{XX'} \equiv 
\VEV{a_{(lm)}^X (a_{(lm)}^{X'})^*},
\ee
where the averaging is over all $2l+1$ values of $m$ and over many realizations of the sky. 
This two-point statistic is thus completely specified by the six ($TT$, $GG$, $CC$, $TG$, $TC$, and $GC$) 
sets of multipole moments. If the temperature/polarization distribution is parity
invariant, then $C_l^{TC}$ and $C_l^{GC}$ must vanish due to the symmetry properties of the
G/C tensor spherical harmonics under parity transformations.

Parity conservation, however, is a {\emph{theoretical bias}}.  
Lue, Wang and Kamionkowski~\cite{Lue:1998mq} provided the first time physical scenario where $C_l^{\rm TC}=C_l^{\rm GC} \neq 0$
due to parity violation in the GW power spectrum of the CMB\footnote{Lue et. al noticed that the CS term violates both parity and time  reflections.  Thus, since gravity is insensitive to charge, CPT is conserved.}.  This physical scenario consisted of GWs 
sourced by the CS interaction term in Eq.~\eqref{CS-action} with $(\alpha,\beta)=(1,0)$ and $\vartheta = f(\phi)$ 
some polynomial function of the inflaton field $\phi$. As we have discussed, CS modified gravity leads to amplitude
birefringence in GW propagation, which in turn leads to an excess of left- over right-cicularly polarized GWs that
lead to a non-vanishing $C^{TC}_{l}$~\cite{Lue:1998mq}. 

In order to understand this, consider GWs during the inflationary epoch. These waves stretch and become classical at
 wavelengths on the order of $\lambda \sim 1/\mu$, where $\mu \sim 1/f'$ is some CS energy scale, until they eventually 
freeze as they become comparable to the Hubble radius. When the waves exit this radius, the fraction of the accumulated 
discrepancy between left- and right-polarized GWs can be estimated through the index 
\be 
\epsilon \sim (M_{p}/\mu) (H/M_{P})^{3} (\dot{\phi}/H^{2})^{2} , \label{parityv}
\ee
where $H$ is the Hubble scale and $f^{''} \sim 1/\mu^{2}$. The factor $H^{2}/\dot{\phi} \sim 10^{-5}$ is associated with the amplitude of 
scalar density perturbations, while $H/M_{P}<10^{-6}$ is related to the amplitude of tensor perturbations~\cite{Lue:1998mq}.

Since long wavelength GWs produce temperature anisotropies of curl type, an excess of left- over right-polarized GWs produces 
a nonzero $C^{TC}_{l}$.  
This is because the multipole coeficients $a^{T,C}_{(lm)}$ will be non-vanishing (right) for circularly polarized GW
\begin{eqnarray}
     a_{(lm)}^T &=& 
     \begin{cases} 
     (\delta_{m,2}+ \delta_{m,-2}) A_l^T(k)
     & {\textrm{even}} \; $l$ \quad ($+$), \cr
     -i (\delta_{m,2}-\delta_{m,-2}) A_l^T(k) 
     & {\textrm{odd}} \; $l$ \quad (\times),\cr 
     \end{cases}
\label{eq:Tmoments}
\\
    a_{(lm)}^C &=& \begin{cases} (\delta_{m,2}+ \delta_{m,-2}) A_l^C(k)
     & {\textrm{even}} \; $l$ \quad (\times), \cr
     -i (\delta_{m,2}-\delta_{m,-2}) A_l^C(k) & {\textrm{odd}} \; $l$ \quad ($+$), \cr 
     \end{cases}
\label{eq:Cmoments}
\end{eqnarray}
where $A_{l}^{T,C}$ are temperature brightness functions and $(+,\times)$ stand for plus or cross, linear GW polarizations
(see eg.~\cite{Kamionkowski:1996ks}).   Likewise the multipole coefficients for the gradient component of the CMB polarization are  similar, with the replacement of $A^{T,C}_{l}$ for polarization brightness functions. For a left, circularly polarized GW, the sign of the even-$l$ moments is reversed. The above equations allow one to understand why parity is not violated with linearly polarized GWs. For example, let us assume that only $+$ modes are present, then $C_{l}^{TC}$ by construction. However, a right or left, circularly polarized GW possesses both $+$ and $x$ modes, and thus the cross-correction is non-vanishing:
\be 
C_l^{TC}=2(2l+1)^{-1} A_l^T(k) A_l^C(k)
 \ee

Lue, Wang and Kamionkowski~\cite{Lue:1998mq} conclude that the parameter $\epsilon$ in Eq.~\eqref{parityv} could in principle be measured by a post-Planck experiment with a sensitivity of $35 \mu K$, a result that was later confirmed by the more detailed study of Saito, et.~al.~\cite{Saito:2007kt}.
Such results have aroused interest in the polarization detection community, pushing them to improve their detection sensitivities to measure CS-inspired, CMB parity violation.  For example, Keating et al \cite{Miller:2009pt} have considered non-vanishing parity violating correlations induced by a class of telescope-beam systematics, which can mimic the birefringence effect.  Furthermore, other authors have generalized the parity violation cross-correlations to CPT violating cross correlations in the photon sector \cite{Cabella:2007br,Xia:2007qs,Feng:2006dp,Kamionkowski:2008fp,Dunkley:2008ie}.

\subsection{Leptogenesis and the Baryon Asymmetry}
Collider experiments have established a symmetry between matter and anti-matter, confirming the prediction of baryon number conservation in the Standard Model (SM) of elementary particle physics.  
In the visible Universe, however, there is an excess of matter over antimatter 
supported by the recent CMB determinations of the cosmological parameters, 
in particular by the WMAP experiment~\cite{Dunkley:2008ie}.
In view of this, one of the major puzzles in cosmology and particle physics is to understand why and how the matter 
asymmetry was generated during the course of the evolution of the Universe starting from a symmetric ``soup'' of matter 
and antimatter soon after the Big Bang. In this section we will show that  if CS gravity is active during the inflationary epoch,
a novel mechanism of leptogenesis is generic.   

Quantitatively, the baryon asymmetry can be expressed in terms of the ratio
of baryon density excess to photon density excess~\cite{Spergel:2003cb} 
\be\label{baryond}
\frac{n_B}{n_\gamma} = (6.5\pm 0.4)\times 10^{-10}\ ,
\ee
where $n_B= n_b- n_{\bar b}$ is the difference in number density of baryons and antibaryons 
and $n_\gamma$ is the number density of photons. This ratio is time independent, as
the evolution of the $n_b$ and $n_\gamma$ with the cosmic Hubble expansion are identical.
Such a large baryon excess cannot be explained within the SM~\cite{Huet:1994jb}, because 
baryon number violating interactions are here loop suppressed. The only SM source of CP violation 
in the hadronic sector is in the Dirac phase of the CKM mixing matrix, which is not enough to explain
the asymmetry of Eq.~\eqref{baryond}.

One can map the problem of baryon asymmetry to that of lepton
asymmetry. This is because the SM weak interactions contain processes,
mediated by {\it sphalerons} ($SU(2)$ instantons), which interconvert baryons and leptons and are
thermally activated at temperatures greater than 1TeV.  Thus, a baryon asymmetry can be generated 
through the generation of net lepton number at high temperature
through out-of-equilibrium and CP-asymmetric processes \cite{Kuzmin:1985mm,Fukugita:1986hr}. 
Such scenarios are commonly referred to as {\it leptogenesis}.

Leptogenesis is a valid route to explain the baryon asymmetry, provided 
one can construct a model that fulfills the so-called Sakharov conditions for leptons~\cite{Sakharov:1967dj}:
\begin{enumerate}
\item Baryon number violating interactions should be present.
\item Charge and parity (CP) should be violated. 
\item CP and baryon number violating interactions should be active when the Universe is out of thermal equilibrium.
\end{enumerate}
These three requirements constitute model-independent, necessary conditions to generate a
baryon asymmetry dynamically from symmetric initial conditions. 
Within the SM, however, there are no such leptogenesis models, and hence one is forced to associate the 
observed baryon asymmetry to physics beyond the SM.

\subsubsection{Outline of the mechanism}

A mechanism for the creation of baryon asymmetry, associated with gravitational inflationary fluctuations, 
was presented by Alexander, Peskin and Sheikh-Jabbari (APS)~\cite{Alexander:2004us}.  
The key to this mechanism is CS modified gravity, where we shall here follow APS and set 
$(\alpha,\beta) = (8 \kappa,0)$,  $\theta = f(\phi)$ and associate $\phi$ with the inflaton field. 
Referring back to Sec.~\ref{parity-CS}, the function $f(\phi)$ must be odd in $\phi$, thus implying 
that $f(\phi)$ is odd under P and CP transformations, and allowing the parameterization of Eq.~\eqref{f-param}.

Let us first spell out how the three Sakharov conditions are realized in the APS model of
baryon asymmetry, so-called {\it gravi-leptogenesis}:
\begin{enumerate}
\item{\bf{Lepton number violation}} is generated here via the triangle anomaly, discussed in 
Sec.~\ref{Particle-Physics and Anomalies}. In the SM, the lepton number current (and
hence the total fermion number density), has a gravitational anomaly~\cite{AlvarezGaume:1983ig}:
\be\label{Jlepton}
     \partial_a J^a_\ell  =  \frac{N}{16\pi^2}   \pont
\ee
where the lepton number current is given by $J^a_\ell =   \sum_{i=L,R} \bar \ell_i\gamma^a \ell_i + \bar \nu_i \gamma^a \nu_i$,
$N=N_L-N_R$ equals three in the SM, $\gamma^{a}$ are Dirac gamma matrices, and $\ell$, $\nu$ denote lepton and neutrino species respectively. In the SM, the anomaly is thus a consequence of an imbalance between left- and right-handed leptons.
\item{\bf{CP violation}} manifests itself naturally in CS modified gravity. Here, lepton number is generated due to a non-vanishing 
vacuum expectation value of the Pontryagin density in the evolution equation of the CS scalar, which is associated with the inflaton. 
In turn, this density is generically non-vanishing during inflation due to GW perturbations.
\item{\bf{Out-of-equilibrium}} conditions arise due to the (exponential)
growth of the background spacetime. Such a growth leads to lepton number production that is naturally out of
equilibrium.
\end{enumerate}

The APS leptogenesis model can be naturally realized if the inflaton field is associated with a complex
modulus field. In such a case, one must guarantee that the inflaton potential is sufficiently flat, such that the slow-roll
conditions are satisfied. The simplest model of this kind is that of single-field
inflation and a pseudo-scalar $\phi$ as the inflaton, known as natural inflation. Such a model 
can be expanded to include multiple axions, as in N-inflation models \cite{Dimopoulos:2005ac}.
Such models fit into extensions of the SM and in string-inspired 
inflationary models \footnote{For a concise review of string-inspired inflation see \cite{McAllister:2007bg}} .

In the remainder of this chapter, we shall use particle physics notation and conventions, 
where $h = c = 1$ and $M_\Pl = (8 \pi G)^{-1/2} \sim 2.44 \times 10^{18}$ GeV is the reduced Planck mass. 
In particular, we shall follow~\cite{LL} and choose $\alpha = 8 \kappa$, $\beta = 0$, and $\theta = f(\phi)$, although
we shall often work with the dimensionless functional $F(\phi) := f(\phi)/M_{\Pl}^{2}$. 

\subsubsection{Gravitational Wave Evolution During Inflation}

Although GW solutions have already been discussed in Sec.~\ref{Sec:ApproxSol}, 
it is instructive to present the expansion of the Lagrangian to second order in the metric 
perturbation. In the TT gauge, assuming a GW perturbed FRW metric and using a left/right basis for the 
GW polarization, (see Eq.~\eqref{FRW-metric}):  
\be\label{RRdual}
\begin{split}
{\cal L}&= -({h_L} \Box {h_R} + {h_R} \Box {h_L})\cr
& +{16i F(\phi)}\biggl[\left(\frac{\partial^2}{\partial z^2}{h_R}
\frac{\partial^2}{\partial t\partial z}{h_L} -
\frac{\partial^2}{\partial z^2}{h_L}
\frac{\partial^2}{\partial t\partial z}{h_R} \right)\cr
& + a^2 \left(\frac{\partial^2}{\partial t^2}{h_R}
\frac{\partial^2}{\partial t\partial z}{h_L} -
\frac{\partial^2}{\partial t^2}{h_L}
\frac{\partial^2}{\partial t\partial z}{h_R} \right) \cr
&  + Ha^2\left(
\frac{\partial}{\partial t}{h_R}\frac{\partial^2}{\partial t\partial z}{h_L}
-\frac{\partial}{\partial t}{h_L}\frac{\partial^2}{\partial t\partial
z}{h_R}\right)\biggr] +{\cal O}(h^4)
\end{split}
\ee
where $t$ stands for cosmic time $dt = a(\eta) d\eta$ and
$\Box= \partial_{t}^2 +3H \partial_{t} - \partial_{z}^2/a^{2}$.
As is clear from Eq.~\eqref{RRdual}, if $h_L$ and $h_R$ have the same {\it dispersion relation}, 
$\pont$ vanishes, while otherwise ``cosmological birefringence'' is induced.

From this Lagrangian, the equations of motion for $h_L$ and $h_R$ become
\be\label{LReqs}
  \ybox\, h_L = - 2i \frac{\Theta}{ a} {\dot h}^\prime_L \ , \qquad
  \ybox\, h_R = + 2i \frac{\Theta}{a} {\dot h}^\prime_R \ ,
\ee
where dots denote time derivatives, and primes denote differentiation of $F$
with respect to $\phi$. The quantity $\Theta \sim 2F'H\dot\phi/M^2_{Pl}$, which with 
the functional form of $F(\phi)$ becomes $\Theta = \sqrt{2\epsilon}/(2\pi^2) H^{2}/M_{Pl}^{2} {\cal N}$, 
where $\epsilon = \dot\phi^2/2 \; (H M_\Pl)^{-2}$ is the slow-roll parameter of inflation~\cite{LL}.
We have here used the fact that the inflaton is purely time-dependent and we have neglected terms 
proportional to $\ddot\phi$ by the slow-roll conditions. 

Gravitational birefringence is present in the solution to the equations of motion.
Let us focus on the positive frequency component of the evolution of $h_L$ 
and adopt a basis in which $h_{L}$ depends on $(t,z)$ only. 
In terms of conformal time, then, the evolution equation for $h_L$ becomes
\be\label{hLformula}
  {d^2\over d\eta^2} h_L - 2 {1\over \eta} {d\over d\eta} h_L
- {d^2\over dz^2} h_L
          =  -2i\Theta {d^2\over d\eta dz} h_L\ ,
\ee
which is a special case of Eq.~\eqref{GW-FE}. If we ignore $\Theta$ for the moment and let $h_L \sim e^{ikz}$, 
this becomes the equation of a spherical Bessel function, for which the positive frequency solution is
\be\label{Bessel}
h_L^{+}(k,\eta) =  e^{+ i k(\eta+z)} (1 - i k\eta)  \ .
\ee

Let us now peel-off the asymptotic behavior of the solution with non-zero $\Theta$. Let then 
\be\label{gdef}
     h_L = e^{ikz} \cdot (-ik\eta)
                        e^{k\Theta\eta}  g(\eta)
\ee
where  $g(\eta)$ is a Coulomb wave function that satisfies
\be\label{Coulomb}
   {d^2\over d\eta^2} g + \left[ k^2 (1-\Theta^2) - {2\over \eta^2}
                 - { 2 k \Theta\over \eta} \right]\, g\ =\ 0,
\ee
which in turn is the equation of a Schr\"odinger particle with $\ell = 1$ in a
weak Coulomb potential. For $h_L$, the Coulomb term is repulsive, while 
for $h_R$ this potential is attractive. Such a fact leads to attenuation of $h_L$ 
and amplification of $h_R$ in the early universe, which is equivalent to the 
exponential enhancement/suppression effect discussed in Sec.~\ref{Sec:ApproxSol}. 

As we shall see, generation of baryon asymmetry is dominated
by modes at short distances (sub-horizon modes) and at early times. This
corresponds to the limit $k\eta \gg 1$.
In this region, we can ignore the potential
terms in Eq.~\eqref{Coulomb} and take the solution to be approximately
a plane wave.  More explicitly,
\be\label{findg}
    g(\eta) =   \exp[ ik(1-\Theta^2)^{1/2} \eta (1 + \alpha(\eta))] \ ,
\ee
where $\alpha(\eta) \sim \log \eta/\eta$.

\subsubsection{The Expectation Value of the Pontryagin Density}

Let us now compute use Eq.~\eqref{gdef} to compute
the expectation value of $\pont$ in an inflationary spacetime.
This expectation value is dominated by the sub-horizon, {\it quantum} part of the
GW evolution. Hence, to compute $\langle \pont \rangle$ we only need the
two point (Green's) function $\langle h_L h_R\rangle$:
\be\label{Gdefin}
   G(x,t;x',t') = \VEV{h_L(x,t) h_R(x',t')} 
                = \int \, {d^3 k\over (2\pi)^3} e^{ik \cdot (x-x')} G_k(\eta,\eta')\ .
\ee
For $k$ parallel to $z$, the Fourier component $G_k$ satisfies
Eq.~\eqref{hLformula} with a delta-function source
\be\label{Gequation}
  \left[{d^2\over d\eta^2} - 2 ({1\over \eta}+k\Theta) {d\over d\eta}
+ k^2 \right]G_k(\eta,\eta')  =
          i  {(H\eta)^2\over M_\Pl^2} \delta(\eta - \eta') .
\ee

Let us first consider the case where $\Theta = 0$. The solution to Eq.~\eqref{Gequation} is
\be\label{Gvalue}
   G_{k0}(\eta, \eta') = \left\{\begin{array}{cc}
           \aleph   \; h_L^{+}(k,\eta) h_R^{-}(-k,\eta')
                   &   \quad \eta < \eta' \cr   \ \ \ &\ \ \\
           \aleph    \; h_L^{-}(k,\eta) h_R^{+}(-k,\eta')
                   &   \quad \eta' < \eta\ ,
\end{array}\right.
\ee
where $\aleph \equiv (H^{2}/M_{Pl}^{2}) k^{-3}/2$, $h_L^{-}$ is the complex conjugate of Eq.~\eqref{Bessel}, and
$h_R^{+}$, $h_R^{-}$ are the corresponding solutions of the $h_R$ equation.
Since here $\Theta = 0$, $h_{R}^{+,-} = h_{L}^{+,-}$.
The leading-order effect of $\Theta \neq 0$ is to introduce
an exponential dependence from Eq.~\eqref{gdef}. The generic solution to Eq.~\eqref{Gequation} is then
\be\label{myGk}
      G_k =
e^{-k\Theta\eta}  e^{+k\Theta \eta'}  G_{k0}
\ee
for both $\eta >\eta'$ and $\eta <\eta'$.

The Green's function in Eq.~\eqref{myGk}
can now be used to contract $h_L$ and $h_R$ and
evaluate the quantum expectation value of $\pont$.  The result is
\be\label{RRdualval}
  \VEV{\pont} =  {16\over a}\, \int \, {d^3 k\over (2\pi)^3}\
  {H^2\over 2 k^3 M_\Pl^2}
                      (k\eta)^2 \cdot k^4 \Theta  \ee
where we have neglected subleading corrections in  $k\eta \gg 1$.
This expectation value is nonzero because inflation is producing a CP asymmetry out of equilibrium. 
The above result and computations seem to depend on the choice of vacuum state and the form 
of the Green's function, but in fact this is not the case as one can verify by recalculating Eq.~\eqref{RRdualval}
using a different method, such as fermion level-crossing~\cite{Gibbons:1993hg}. 

\subsubsection{Lepton and Photon number density}

The leptogenesis model can be completed by computing the net lepton/anti-lepton asymmetry generated
during inflation. Inserting Eq.~\eqref{RRdualval} into Eq.~\eqref{Jlepton} and
integrating over the time period of inflation, we obtain
\be\label{netlept}
  n_{L} = \int^{H^{-1}}_0 d\eta\ \int \, {d^3 k\over (2\pi)^3}\
       {1\over 16\pi^2}\, {8 H^2  k^3 \eta^2  \Theta \over M_\Pl^2}  \
       ,
\ee
where $n_{L}$ is the lepton number density.
The $k$-integral runs over all of momentum space, up to the
scale $\mu = 1/F'$ at which the effective Lagrangian description breaks down, 
but it is dominated by modes at very short distances compared
to the horizon scale. The $\eta$-integral is dominated by modes at large $\eta$ and it represents
a compromise between two competing inflationary processes: exponential expansion 
to large $k$ and exponential dilution of the generated lepton number
through expansion. The dominant contribution to lepton generation then arises from modes in the 
region $k\eta \gg 1$.

Equation~\eqref{netlept} can be integrated exactly to return
\be\label{nval}
      n_{L}  =  {1\over 72\pi^4} \left({ H\over
            M_\Pl}\right)^2 \Theta H^3  \left(\frac{\mu}{H} \right)^6.
\ee 
The factor $(H/M_\Pl)^2$ is the magnitude of the GW power
spectrum. We should stress that the usual GW power
spectrum comes from the super-horizon modes, while the main
contribution to $n$ here comes from sub-horizon modes. The factor of
$\Theta$ is  a measure of the effective CP violation caused 
by birefringent GWs, while $H^3$ is the inverse horizon size at inflation, 
giving $n$ appropriate units. The factor of $({\mu/H})^6$ is an enhancement 
over one's first guess due to the use of strongly quantum, short distance fluctuations to
generate $\pont$, rather than the super-horizon modes which effectively behave classically.

The significance of the lepton number density can be understood by comparing it 
to the entropy density of the Universe just after reheating,
or equivalently to the photon number density. 
Recall that almost all of the entropy of the Universe
is generated during the reheating time and it is carried by the
massless degrees of freedom, ie.~photons. 
Let us then employ one of the simplest (and at the same time most naive) reheating
model: instant reheating. We shall assume that all of
the energy of the inflationary phase $\rho= 3 H^2 M_{Pl}^2$ is converted into
the heat of a  gas of massless particles $\rho = \pi^2 g_* T^4/30$ instantaneously. 
The entropy of this gas is $s =  2\pi^2 g_* T^3/45$, where $T$ is the reheating temperature, $g_*$ is the effective
number of massless degrees of freedom, $s=1.8g_*\ n_\gamma$ \cite{LL}, and we have assumed 
an adiabatic post-reheating evolution. With these assumptions we obtain the photon number density
\[
n_\gamma =  1.28 g_*^{-3/4} (H M_{Pl})^{3/2}.\]

Recalling that the ratio of the present baryon number to the lepton number originally generated in
leptogenesis is approximately $n_B/n_L = 4/11$ \cite{Kuzmin:1985mm}, in this model we obtain
\be\label{novers}
    \frac{n_B}{n_\gamma} = 4.05\times 10^{-5} g_*^{3/4} \left(\frac{H}{M_{Pl}}\right)^{7/2}
                       \ \Theta   \left(\frac{\mu}{H}\right)^6 ,
\ee
A less naive approach could be to follow the dilution of $n_{L}$ and $\rho$ with
the expansion of the universe to the end of reheating.  The final result is
the same (see, however, \cite{Linde} for a comment on this point).
With the adiabatic expansion assumption, Eq.~\eqref{novers}
can be compared directly to the present value of $n_B/n_\gamma$ given in Eq.~\eqref{baryond}.

\subsubsection{Is Gravi-Leptogenesis a Viable Model?}

In order to answer this question, one must numerically estimate  
$n_B/n_\gamma$. Substituting for $\Theta$, Eq.~\eqref{novers} becomes 
\be\label{n/n-final}
\frac{n_B}{n_\gamma} \simeq 2.9\times 10^{-6} g_*^{3/4}\sqrt{\epsilon}\ {\cal N} \left(\frac{H}{M_{Pl}}\right)^{-1/2}
\left(\frac{\mu}{M_{Pl}}\right)^6  .
\ee
Clearly, this ratio depends on five dimensionless parameters:
$g_*,\ H/M_{Pl},\ \mu/M_{Pl}$ and the slow-roll parameter $\epsilon$ and ${\cal N}$.

Some of these parameters are already constrained or given by theoretical considerations. 
For example, within the usual supersymmetric particle physics models, $g_* \sim 100$ is a reasonable choice.
Moreover, WMAP data, through the density perturbation ratio $\delta\rho/\rho$ (for a single field inflation),
leads to an upper bound on $H/M_{Pl}$ ratio~\cite{Dunkley:2008ie} of 
${H}/{M_{Pl}}\lesssim 10^{-4}$ and $\epsilon \lesssim 0.01$, or $H\lesssim 10^{14}\ GeV$.
Finally, the factor ${\cal N} \simeq 10^2-10^{10}$ is inferred from the string theory compactification and is proportional to
the square of the four dimensional $M_{Pl}$ to the ten dimensional (fundamental) Planck mass \cite{Alexander:2004xd} (see also the Appendix in~\cite{Alexander:2004xd}). If the gravi-leptogenesis model is to be viable, we then have that  
\be\label{theta-estimate}
\Theta\simeq 10^{-8}-10^{0},
\ee
assuming that $H$ saturates its current bound.

The physically viable range for the parameter $\mu$ depends on the details of the underlying
particle physics model on which our gravi-leptogenesis is based. For example, within the SM plus
three heavy right-handed neutrinos (and the seesaw mechanism), $\mu$ could be of the order of the 
right-handed neutrino mass, in which case $\mu \lesssim 10^{12} \; {\textrm{GeV}}$.  If we do not
restrict the analysis to the seesaw mechanism, $\mu$ could be larger, up to perhaps approximately 
the energy scale at which the effective field theory analysis breaks down. Therefore, depending on the details of the model
\be\label{mu-estimate}
10^{-6}  \lesssim \frac{\mu}{M_{Pl}} \lesssim 10^{-2}-10^{-1}.
\ee

Combining all these considerations, one finds that the gravi-leptogenesis model predicts 
\be\label{n-n-estimate}
    \frac{n_B}{n_\gamma} \simeq 10^{+5}
                       \ \Theta \  \left(\frac{\mu}{M_{Pl}}\right)^6 \ .
\ee
Gravi-leptogenesis thus possesses a large parameter space to explain the observed baryon asymmetry of the Universe, 
but its final prediction strongly depends on the ultraviolet cutoff of the effective theory~\cite{Lyth:2005jf,Fischler:2007tj}. 
In turn, this cutoff dependence arises because lepton number is generated by graviton fluctuations, 
which we know are non-renormalizable in GR. This issue and that of initial conditions are currently being investigated.

\section{Outlook}

We have provided a comprehensive review of CS modified gravity from a particle physics, 
astrophysical, and cosmological perspective. From the particle physics point of view, 
the presence of the CS term leads to a gravitational anomaly-cancellation mechanism in the standard model.  
In cosmology, the chiral anomaly works together with inflation to amplify the production of leptons, 
leading to a viable model of leptogenesis.  The parity violation present in the Pontryagin density 
leads to direct modifications to GW generation and propagation.  

While all of these avenues are promising directions of research, there is still much to understand.
One important issue that needs to be addressed in more detail is that of the coupling strength
and the potential of the CS field. In string theory, the perturbative evaluation of this coupling suggests 
a Planck suppression, while in LQG it is related to a quantum (Immirizi) ambiguity. 
The potential of the CS field, on the other hand, cannot be evaluated with current mathematical methods.
Thus, new {\emph{non-perturbative}} techniques need to be developed to fully evaluate these terms. 
In lack of a concrete, accurate estimate for these terms, we have taken an agnostic view in
this review: to study the observational consequences of the CS term and its possible constrains as a function
of the coupling strength. Such a view allows the possibility to search for CS effect, a detection of which 
would provide theorists with a strong motivation to develop the necessary tools to evaluate the 
CS coupling strength and the exact potential of the CS dynamical field. 

Another issue that remains open is that of backreation and propagation effects of fermions in the modified theory.   Massive sterile neutrinos
may be a viable candidate for dark matter.  In spiral galaxies, dark matter is expected to be at the center of the galaxy where a supermassive black hole might reside.
 In the context of CS gravity, the trajectory of
sterile neutrinos would be controlled by the modified Dirac equation~\cite{Alexander:2008wi}, which could now be studied in the new 
CS-modified, slow-rotating background of Yunes and Pretorius~\cite{slow-rot}. A modification in the behavior of fermions
in highly-dense, strongly-gravitating systems could change the rate of black hole formation, such as in the the symbiotic supermassive formation mechanism \cite{Richter:2006fa}.

A further avenue of future research that is worth exploring is that of an exact solution that represents the exterior gravitational 
field of rapidly-rotating black holes and neutron stars in Dynamical CS modified gravity.  
The slow-rotating solution found by Yunes and Pretorius~\cite{slow-rot} has provided some hints as to what type of modification is introduced
to the metric by the dynamical theory, but an exact rapidly rotating solution is still missing.   
Such a solution would allow for the study of extreme-mass ratio inspirals and
their associated GWs in the modified theory, which could lead to interesting GW tests via
observations with Advanced LIGO or LISA. Moreover, such a solution could also be used to study modifications to the location 
of the innermost stable circular orbit and the event horizon, which could have implications in X-ray astrophysics. 

A final avenue worth exploring is that of the connection of the dynamical formulation and GW theory.
The former could serve as a platform upon which to test well-defined and well-motivated alternative theories 
of gravity with GW observations. An excellent source for such waves are binary systems, the detection of which
requires accurate templates. GW tests of dynamical CS modified gravity would thus require the construction 
of such templates during both the inspiral and merger. Inspiral templates could be built through the post-Newtonian 
approximation, while the merger phase will probably have to be modeled numerically.
Such studies would then allow GW observatories to explore, for the first time, the non-linear quantum nature of spacetime.

\bibliographystyle{elsarticle-num}
\bibliography{phyjabb,bibliography}

\begin{thebibliography}{100}
\expandafter\ifx\csname url\endcsname\relax
  \def\url#1{\texttt{#1}}\fi
\expandafter\ifx\csname urlprefix\endcsname\relax\def\urlprefix{URL }\fi

\bibitem{Dunkley:2008ie}
J.~Dunkley, et~al., {Five-Year Wilkinson Microwave Anisotropy Probe (WMAP)
  Observations: Likelihoods and Parameters from the WMAP data}\href
  {http://arxiv.org/abs/0803.0586} {\path{arXiv:0803.0586}}.

\bibitem{Wittman:2000tc}
D.~M. Wittman, J.~A. Tyson, D.~Kirkman, I.~Dell'Antonio, G.~Bernstein,
  {Detection of weak gravitational lensing distortions of distant galaxies by
  cosmic dark matter at large scales}, Nature 405 (2000) 143--149.
\newblock \href {http://arxiv.org/abs/astro-ph/0003014}
  {\path{arXiv:astro-ph/0003014}}, \href {http://dx.doi.org/10.1038/35012001}
  {\path{doi:10.1038/35012001}}.

\bibitem{Markevitch:1997ic}
M.~Markevitch, W.~R. Forman, C.~L. Sarazin, A.~Vikhlinin, {The Temperature
  Structure of 30 Nearby Clusters Observed with ASCA. Similarity of Temperature
  Profiles}, Astrophys. J. 503 (1998) 77.
\newblock \href {http://arxiv.org/abs/astro-ph/9711289}
  {\path{arXiv:astro-ph/9711289}}, \href {http://dx.doi.org/10.1086/305976}
  {\path{doi:10.1086/305976}}.

\bibitem{Riess:2004nr}
A.~G. Riess, et~al., {Type Ia Supernova Discoveries at z>1 From the Hubble
  Space Telescope: Evidence for Past Deceleration and Constraints on Dark
  Energy Evolution}, Astrophys. J. 607 (2004) 665--687.
\newblock \href {http://arxiv.org/abs/astro-ph/0402512}
  {\path{arXiv:astro-ph/0402512}}, \href {http://dx.doi.org/10.1086/383612}
  {\path{doi:10.1086/383612}}.

\bibitem{Svrcek:2006yi}
P.~Svrcek, E.~Witten, {Axions in string theory}, JHEP 06 (2006) 051.
\newblock \href {http://arxiv.org/abs/hep-th/0605206}
  {\path{arXiv:hep-th/0605206}}.

\bibitem{Alexander:2008wi}
S.~Alexander, N.~Yunes, {Chern-Simons Modified Gravity as a Torsion Theory and
  its Interaction with Fermions}\href {http://arxiv.org/abs/0804.1797}
  {\path{arXiv:0804.1797}}.

\bibitem{Alexander:2007vt}
S.~Alexander, N.~Yunes, {Parametrized Post-Newtonian Expansion of Chern-Simons
  Gravity}, Phys. Rev. D75 (2007) 124022.
\newblock \href {http://arxiv.org/abs/arXiv:0704.0299 [hep-th]}
  {\path{arXiv:arXiv:0704.0299 [hep-th]}}.

\bibitem{Sun:2006sd}
C.-Y. Sun, D.-H. Zhang, {A cyclic cosmological model in the landscape
  scenario}\href {http://arxiv.org/abs/hep-th/0611101}
  {\path{arXiv:hep-th/0611101}}.

\bibitem{Wesley:2005bd}
D.~H. Wesley, P.~J. Steinhardt, N.~Turok, {Controlling chaos through
  compactification in cosmological models with a collapsing phase}, Phys. Rev.
  D72 (2005) 063513.
\newblock \href {http://arxiv.org/abs/hep-th/0502108}
  {\path{arXiv:hep-th/0502108}}, \href
  {http://dx.doi.org/10.1103/PhysRevD.72.063513}
  {\path{doi:10.1103/PhysRevD.72.063513}}.

\bibitem{Brandenberger:2001:lpi}
R.~Brandenberger, D.~A. Easson, D.~Kimberly, Loitering phase in brane gas
  cosmology, Nucl. Phys. B623 (2002) 421--436.
\newblock \href {http://arxiv.org/abs/hep-th/0109165}
  {\path{arXiv:hep-th/0109165}}.

\bibitem{Battefeld:2005av}
T.~Battefeld, S.~Watson, {String gas cosmology}, Rev. Mod. Phys. 78 (2006)
  435--454.
\newblock \href {http://arxiv.org/abs/hep-th/0510022}
  {\path{arXiv:hep-th/0510022}}, \href
  {http://dx.doi.org/10.1103/RevModPhys.78.435}
  {\path{doi:10.1103/RevModPhys.78.435}}.

\bibitem{Brandenberger:2006vv}
R.~H. Brandenberger, A.~Nayeri, S.~P. Patil, C.~Vafa, {String gas cosmology and
  structure formation}, Int. J. Mod. Phys. A22 (2007) 3621--3642.
\newblock \href {http://arxiv.org/abs/hep-th/0608121}
  {\path{arXiv:hep-th/0608121}}, \href
  {http://dx.doi.org/10.1142/S0217751X07037159}
  {\path{doi:10.1142/S0217751X07037159}}.

\bibitem{Brandenberger:2007zz}
R.~H. Brandenberger, {String gas cosmology and structure formation: A brief
  review}, Mod. Phys. Lett. A22 (2007) 1875--1885.
\newblock \href {http://arxiv.org/abs/hep-th/0702001}
  {\path{arXiv:hep-th/0702001}}, \href
  {http://dx.doi.org/10.1142/S0217732307025091}
  {\path{doi:10.1142/S0217732307025091}}.

\bibitem{Brax:2004xh}
P.~Brax, C.~van~de Bruck, A.-C. Davis, {Brane world cosmology}, Rept. Prog.
  Phys. 67 (2004) 2183--2232.
\newblock \href {http://arxiv.org/abs/hep-th/0404011}
  {\path{arXiv:hep-th/0404011}}, \href
  {http://dx.doi.org/10.1088/0034-4885/67/12/R02}
  {\path{doi:10.1088/0034-4885/67/12/R02}}.

\bibitem{Alexander:2000:bgi}
S.~Alexander, R.~H. Brandenberger, D.~Easson, Brane gases in the early
  universe, Phys. Rev. D62 (2000) 103509.
\newblock \href {http://arxiv.org/abs/hep-th/0005212}
  {\path{arXiv:hep-th/0005212}}.

\bibitem{Tseytlin:1991xk}
A.~A. Tseytlin, C.~Vafa, {Elements of string cosmology}, Nucl. Phys. B372
  (1992) 443--466.
\newblock \href {http://arxiv.org/abs/hep-th/9109048}
  {\path{arXiv:hep-th/9109048}}, \href
  {http://dx.doi.org/10.1016/0550-3213(92)90327-8}
  {\path{doi:10.1016/0550-3213(92)90327-8}}.

\bibitem{Nayeri:2005ck}
A.~Nayeri, R.~H. Brandenberger, C.~Vafa, {Producing a scale-invariant spectrum
  of perturbations in a Hagedorn phase of string cosmology}, Phys. Rev. Lett.
  97 (2006) 021302.
\newblock \href {http://arxiv.org/abs/hep-th/0511140}
  {\path{arXiv:hep-th/0511140}}, \href
  {http://dx.doi.org/10.1103/PhysRevLett.97.021302}
  {\path{doi:10.1103/PhysRevLett.97.021302}}.

\bibitem{Carroll:1997ar}
S.~M. Carroll, {Lecture notes on general relativity}\href
  {http://arxiv.org/abs/gr-qc/9712019} {\path{arXiv:gr-qc/9712019}}.

\bibitem{Carroll:2004st}
S.~M. Carroll, {Spacetime and geometry: An introduction to general
  relativity}San Francisco, USA: Addison-Wesley (2004) 513 p.

\bibitem{waldgeneral}
R.~M. Wald, {General Relativity}, The University of Chicago Press, 1984.

\bibitem{Misner:1973cw}
C.~W. Misner, K.~Thorne, J.~A. Wheeler, Gravitation, W. H. Freeman \& Co., San
  Francisco, 1973.

\bibitem{Jackiw:2003pm}
R.~Jackiw, S.~Y. Pi, Chern-simons modification of general relativity, Phys.
  Rev. D68 (2003) 104012.
\newblock \href {http://arxiv.org/abs/gr-qc/0308071}
  {\path{arXiv:gr-qc/0308071}}.

\bibitem{Campbell:1990fu}
B.~A. Campbell, M.~J. Duncan, N.~Kaloper, K.~A. Olive, {Gravitational dynamics
  with Lorentz Chern-Simons terms}, Nucl. Phys. B351 (1991) 778--792.
\newblock \href {http://dx.doi.org/10.1016/S0550-3213(05)80045-8}
  {\path{doi:10.1016/S0550-3213(05)80045-8}}.

\bibitem{Campbell:1992hc}
B.~A. Campbell, N.~Kaloper, R.~Madden, K.~A. Olive, {Physical properties of
  four-dimensional superstring gravity black hole solutions}, Nucl. Phys. B399
  (1993) 137--168.
\newblock \href {http://arxiv.org/abs/hep-th/9301129}
  {\path{arXiv:hep-th/9301129}}, \href
  {http://dx.doi.org/10.1016/0550-3213(93)90620-5}
  {\path{doi:10.1016/0550-3213(93)90620-5}}.

\bibitem{Deser:1981wh}
S.~Deser, R.~Jackiw, S.~Templeton, {Topologically massive gauge theories}, Ann.
  Phys. 140 (1982) 372--411.
\newblock \href {http://dx.doi.org/10.1016/0003-4916(82)90164-6}
  {\path{doi:10.1016/0003-4916(82)90164-6}}.

\bibitem{Deser:1982vy}
S.~Deser, R.~Jackiw, S.~Templeton, {Three-Dimensional Massive Gauge Theories},
  Phys. Rev. Lett. 48 (1982) 975--978.
\newblock \href {http://dx.doi.org/10.1103/PhysRevLett.48.975}
  {\path{doi:10.1103/PhysRevLett.48.975}}.

\bibitem{Alexander:2004xd}
S.~H.~S. Alexander, J.~Gates, S.~James, Can the string scale be related to the
  cosmic baryon asymmetry?, JCAP 0606 (2006) 018.
\newblock \href {http://arxiv.org/abs/hep-th/0409014}
  {\path{arXiv:hep-th/0409014}}.

\bibitem{Alexander:2004us}
S.~H.-S. Alexander, M.~E. Peskin, M.~M. Sheikh-Jabbari, {Leptogenesis from
  gravity waves in models of inflation}, Phys. Rev. Lett. 96 (2006) 081301.
\newblock \href {http://arxiv.org/abs/hep-th/0403069}
  {\path{arXiv:hep-th/0403069}}, \href
  {http://dx.doi.org/10.1103/PhysRevLett.96.081301}
  {\path{doi:10.1103/PhysRevLett.96.081301}}.

\bibitem{Alexander:2004wk}
S.~Alexander, J.~Martin, Birefringent gravitational waves and the consistency
  check of inflation, Phys. Rev. D71 (2005) 063526.
\newblock \href {http://arxiv.org/abs/hep-th/0410230}
  {\path{arXiv:hep-th/0410230}}.

\bibitem{Alexander:2006mt}
S.~H.~S. Alexander, {Is cosmic parity violation responsible for the anomalies
  in the WMAP data?}\href {http://arxiv.org/abs/hep-th/0601034}
  {\path{arXiv:hep-th/0601034}}.

\bibitem{Alexander:2007qe}
S.~H.-S. Alexander, M.~E. Peskin, M.~M. Sheikh-Jabbari, {Gravi-leptogenesis:
  Leptogenesis from gravity waves in pseudo-scalar driven inflation
  models}\href {http://arxiv.org/abs/hep-ph/0701139}
  {\path{arXiv:hep-ph/0701139}}.

\bibitem{Alexander:2007zg}
S.~Alexander, N.~Yunes, {A new PPN parameter to test Chern-Simons gravity},
  Phys. Rev. Lett. 99 (2007) 241101.
\newblock \href {http://arxiv.org/abs/hep-th/0703265}
  {\path{arXiv:hep-th/0703265}}, \href
  {http://dx.doi.org/10.1103/PhysRevLett.99.241101}
  {\path{doi:10.1103/PhysRevLett.99.241101}}.

\bibitem{Alexander:2007kv}
S.~Alexander, L.~S. Finn, N.~Yunes, {A gravitational-wave probe of effective
  quantum gravity}\href {http://arxiv.org/abs/arXiv:0712.2542 [gr-qc]}
  {\path{arXiv:arXiv:0712.2542 [gr-qc]}}.

\bibitem{Yunes:2007ss}
N.~Yunes, C.~F. Sopuerta, {Perturbations of Schwarzschild Black Holes in
  Chern-Simons Modified Gravity}, Phys. Rev. D77 (2008) 064007.
\newblock \href {http://arxiv.org/abs/0712.1028} {\path{arXiv:0712.1028}},
  \href {http://dx.doi.org/10.1103/PhysRevD.77.064007}
  {\path{doi:10.1103/PhysRevD.77.064007}}.

\bibitem{Grumiller:2007rv}
D.~Grumiller, N.~Yunes, {How do Black Holes Spin in Chern-Simons Modified
  Gravity?}, Phys. Rev. D77 (2008) 044015.
\newblock \href {http://arxiv.org/abs/0711.1868} {\path{arXiv:0711.1868}},
  \href {http://dx.doi.org/10.1103/PhysRevD.77.044015}
  {\path{doi:10.1103/PhysRevD.77.044015}}.

\bibitem{pulsars}
S.~Alexander, , R.~O'Shaughnessy, B.~J. Owen, N.~Yunes, in preparation.

\bibitem{Smith:2007jm}
T.~L. Smith, A.~L. Erickcek, R.~R. Caldwell, M.~Kamionkowski, {The effects of
  Chern-Simons gravity on bodies orbiting the Earth}, Phys. Rev. D77 (2008)
  024015.
\newblock \href {http://arxiv.org/abs/0708.0001} {\path{arXiv:0708.0001}},
  \href {http://dx.doi.org/10.1103/PhysRevD.77.024015}
  {\path{doi:10.1103/PhysRevD.77.024015}}.

\bibitem{Garcia:2003bw}
A.~Garcia, F.~W. Hehl, C.~Heinicke, A.~Macias, {The Cotton tensor in Riemannian
  spacetimes}, Class. Quant. Grav. 21 (2004) 1099--1118.
\newblock \href {http://arxiv.org/abs/gr-qc/0309008}
  {\path{arXiv:gr-qc/0309008}}.

\bibitem{Grumiller:2008ie}
D.~Grumiller, R.~Mann, R.~McNees, {Dirichlet boundary value problem for
  Chern-Simons modified gravity}\href {http://arxiv.org/abs/0803.1485}
  {\path{arXiv:0803.1485}}.

\bibitem{Grumillerprivcomm}
D.~Grumiller, private communication.

\bibitem{Penrose:1986ca}
R.~Penrose, W.~Rindler, {Spinors and Space-Time II}, Cambridge University
  Press, 1986.

\bibitem{Stephani:2003tm}
H.~Stephani, D.~Kramer, M.~MacCallum, C.~Hoenselaers, E.~Herlt, Exact solutions
  of Einstein's field equations, Cambridge University Press, 2003.

\bibitem{Yunes:2005ve}
N.~Yunes, J.~A. Gonzalez, Metric of a tidally perturbed spinning black hole,
  Phys. Rev. D73 (2006) 024010.
\newblock \href {http://arxiv.org/abs/gr-qc/0510076}
  {\path{arXiv:gr-qc/0510076}}.

\bibitem{Cherubini:2003nj}
C.~Cherubini, D.~Bini, S.~Capozziello, R.~Ruffini, {Second Order Scalar
  Invariants of the Riemann Tensor: Applications to Black Hole Spacetimes},
  Int. J. Mod. Phys. D11 (2002) 827--841.
\newblock \href {http://arxiv.org/abs/gr-qc/0302095}
  {\path{arXiv:gr-qc/0302095}}.

\bibitem{Matte:1953}
A.~Matte, Sur de nouvelles solutions oscillatoires des {\'e}quations de la
  gravitation, Ca. J. Math. 5 (1953) 1.

\bibitem{Regge:1957rw}
T.~Regge, J.~A. {Wheeler}, Stability of a schwarzschild singularity, Phys. Rev.
  108 (1957) 1063--1069.
\newblock \href {http://dx.doi.org/10.1103/PhysRev.108.1063}
  {\path{doi:10.1103/PhysRev.108.1063}}.

\bibitem{Mann:2005yr}
R.~B. Mann, D.~Marolf, Holographic renormalization of asymptotically flat
  spacetimes, Class. Quant. Grav. 23 (2006) 2927--2950.
\newblock \href {http://arxiv.org/abs/hep-th/0511096}
  {\path{arXiv:hep-th/0511096}}.

\bibitem{Mann:2006bd}
R.~B. Mann, D.~Marolf, A.~Virmani, Covariant counterterms and conserved charges
  in asymptotically flat spacetimes, Class. Quant. Grav. 23 (2006) 6357--6378.
\newblock \href {http://arxiv.org/abs/gr-qc/0607041}
  {\path{arXiv:gr-qc/0607041}}.

\bibitem{Fujikawa:1979ay}
K.~Fujikawa, {Path Integral Measure for Gauge Invariant Fermion Theories},
  Phys. Rev. Lett. 42 (1979) 1195.
\newblock \href {http://dx.doi.org/10.1103/PhysRevLett.42.1195}
  {\path{doi:10.1103/PhysRevLett.42.1195}}.

\bibitem{Fujikawa:1980eg}
K.~Fujikawa, {Path Integral for Gauge Theories with Fermions}, Phys. Rev. D21
  (1980) 2848.
\newblock \href {http://dx.doi.org/10.1103/PhysRevD.21.2848}
  {\path{doi:10.1103/PhysRevD.21.2848}}.

\bibitem{1969PhRv..177.2426A}
S.~L. {Adler}, {Axial-Vector Vertex in Spinor Electrodynamics}, Physical Review
  177 (1969) 2426--2438.
\newblock \href {http://dx.doi.org/10.1103/PhysRev.177.2426}
  {\path{doi:10.1103/PhysRev.177.2426}}.

\bibitem{Bell:1969ts}
J.~S. Bell, R.~Jackiw, {A PCAC puzzle: pi0 $\to$ gamma gamma in the sigma
  model}, Nuovo Cim. A60 (1969) 47--61.

\bibitem{Mariz:2004cv}
T.~Mariz, J.~R. Nascimento, E.~Passos, R.~F. Ribeiro, {Chern-Simons-like action
  induced radiatively in general relativity}, Phys. Rev. D70 (2004) 024014.
\newblock \href {http://arxiv.org/abs/hep-th/0403205}
  {\path{arXiv:hep-th/0403205}}, \href
  {http://dx.doi.org/10.1103/PhysRevD.70.024014}
  {\path{doi:10.1103/PhysRevD.70.024014}}.

\bibitem{Mariz:2007gf}
T.~Mariz, J.~R. Nascimento, A.~Y. Petrov, L.~Y. Santos, A.~J. da~Silva,
  {Lorentz violation and the proper-time method}, Phys. Lett. B661 (2008)
  312--318.
\newblock \href {http://arxiv.org/abs/0708.3348} {\path{arXiv:0708.3348}},
  \href {http://dx.doi.org/10.1016/j.physletb.2007.10.089}
  {\path{doi:10.1016/j.physletb.2007.10.089}}.

\bibitem{Gomes:2008an}
M.~Gomes, et~al., {On the ambiguities in the effective action in Lorentz-
  violating gravity}, Phys. Rev. D78 (2008) 025029.
\newblock \href {http://arxiv.org/abs/0805.4409} {\path{arXiv:0805.4409}},
  \href {http://dx.doi.org/10.1103/PhysRevD.78.025029}
  {\path{doi:10.1103/PhysRevD.78.025029}}.

\bibitem{AlvarezGaume:1983ig}
L.~Alvarez-Gaume, E.~Witten, Gravitational anomalies, Nucl. Phys. B234 (1984)
  269.

\bibitem{Green:1987sp}
M.~B. Green, J.~H. Schwarz, E.~Witten, Superstring Theory, Cambridge University
  Press, 1987, vol. 1: {I}ntroduction.

\bibitem{Green:1987mn}
M.~B. Green, J.~H. Schwarz, E.~Witten, Superstring Theory. Vol. 2: Loop
  Amplitides, Anomalies and Phenomenology, Cambridge University Press
  (Cambridge Monographs On Mathematical Physics), Cambridge, Uk, 1987.

\bibitem{Polchinski:1998rr}
J.~Polchinski, {String theory. Vol. 2: Superstring theory and beyond}Cambridge,
  UK: Univ. Pr. (1998) 531 p.

\bibitem{Gukov:2003cy}
S.~Gukov, S.~Kachru, X.~Liu, L.~McAllister, {Heterotic moduli stabilization
  with fractional Chern- Simons invariants}, Phys. Rev. D69 (2004) 086008.
\newblock \href {http://arxiv.org/abs/hep-th/0310159}
  {\path{arXiv:hep-th/0310159}}, \href
  {http://dx.doi.org/10.1103/PhysRevD.69.086008}
  {\path{doi:10.1103/PhysRevD.69.086008}}.

\bibitem{Bergshoeff:1981um}
E.~Bergshoeff, M.~de~Roo, B.~de~Wit, P.~van Nieuwenhuizen, {Ten-Dimensional
  Maxwell-Einstein Supergravity, Its Currents, and the Issue of Its Auxiliary
  Fields}, Nucl. Phys. B195 (1982) 97--136.
\newblock \href {http://dx.doi.org/10.1016/0550-3213(82)90050-5}
  {\path{doi:10.1016/0550-3213(82)90050-5}}.

\bibitem{Green:1984sg}
M.~B. Green, J.~H. Schwarz, {Anomaly Cancellation in Supersymmetric D=10 Gauge
  Theory and Superstring Theory}, Phys. Lett. B149 (1984) 117--122.
\newblock \href {http://dx.doi.org/10.1016/0370-2693(84)91565-X}
  {\path{doi:10.1016/0370-2693(84)91565-X}}.

\bibitem{Ashtekar:2004eh}
A.~Ashtekar, J.~Lewandowski, {Background independent quantum gravity: A status
  report}, Class. Quant. Grav. 21 (2004) R53.
\newblock \href {http://arxiv.org/abs/gr-qc/0404018}
  {\path{arXiv:gr-qc/0404018}}, \href
  {http://dx.doi.org/10.1088/0264-9381/21/15/R01}
  {\path{doi:10.1088/0264-9381/21/15/R01}}.

\bibitem{Thiemann:2001yy}
T.~Thiemann, Introduction to modern canonical quantum general relativity\href
  {http://arxiv.org/abs/gr-qc/0110034} {\path{arXiv:gr-qc/0110034}}.

\bibitem{Rovelli:2004tv}
C.~Rovelli, {Quantum gravity}Cambridge, UK: Univ. Pr., 455 p.

\bibitem{Yang:1954ek}
C.-N. Yang, R.~L. Mills, {Conservation of isotopic spin and isotopic gauge
  invariance}, Phys. Rev. 96 (1954) 191--195.
\newblock \href {http://dx.doi.org/10.1103/PhysRev.96.191}
  {\path{doi:10.1103/PhysRev.96.191}}.

\bibitem{Ashtekar:1991hf}
A.~Ashtekar, {Lectures on nonperturbative canonical gravity}Singapore,
  Singapore: World Scientific, 334 p. (Advanced series in astrophysics and
  cosmology, 6).

\bibitem{Barbero:1994ap}
J.~F. Barbero, {Real Ashtekar variables for Lorentzian signature space times},
  Phys. Rev. D51 (1995) 5507--5510.
\newblock \href {http://arxiv.org/abs/gr-qc/9410014}
  {\path{arXiv:gr-qc/9410014}}, \href
  {http://dx.doi.org/10.1103/PhysRevD.51.5507}
  {\path{doi:10.1103/PhysRevD.51.5507}}.

\bibitem{Holst:1995pc}
S.~Holst, {Barbero's Hamiltonian derived from a generalized Hilbert- Palatini
  action}, Phys. Rev. D53 (1996) 5966--5969.
\newblock \href {http://arxiv.org/abs/gr-qc/9511026}
  {\path{arXiv:gr-qc/9511026}}, \href
  {http://dx.doi.org/10.1103/PhysRevD.53.5966}
  {\path{doi:10.1103/PhysRevD.53.5966}}.

\bibitem{Ashtekar:1988sw}
A.~Ashtekar, A.~P. Balachandran, S.~Jo, {THE CP PROBLEM IN QUANTUM GRAVITY},
  Int. J. Mod. Phys. A4 (1989) 1493.
\newblock \href {http://dx.doi.org/10.1142/S0217751X89000649}
  {\path{doi:10.1142/S0217751X89000649}}.

\bibitem{Taveras:2008yf}
V.~Taveras, N.~Yunes, {The Barbero-Immirzi Parameter as a Scalar Field: K-
  Inflation from Loop Quantum Gravity?}, Phys. Rev. D78 (2008) 064070.
\newblock \href {http://arxiv.org/abs/0807.2652} {\path{arXiv:0807.2652}},
  \href {http://dx.doi.org/10.1103/PhysRevD.78.064070}
  {\path{doi:10.1103/PhysRevD.78.064070}}.

\bibitem{Immirzi:1996di}
G.~Immirzi, {Real and complex connections for canonical gravity}, Class. Quant.
  Grav. 14 (1997) L177--L181.
\newblock \href {http://arxiv.org/abs/gr-qc/9612030}
  {\path{arXiv:gr-qc/9612030}}.

\bibitem{Perez:2005pm}
A.~Perez, C.~Rovelli, Physical effects of the immirzi parameter, Phys. Rev. D73
  (2006) 044013.
\newblock \href {http://arxiv.org/abs/gr-qc/0505081}
  {\path{arXiv:gr-qc/0505081}}.

\bibitem{Freidel:2005sn}
L.~Freidel, D.~Minic, T.~Takeuchi, {Quantum gravity, torsion, parity violation
  and all that}, Phys. Rev. D72 (2005) 104002.
\newblock \href {http://arxiv.org/abs/hep-th/0507253}
  {\path{arXiv:hep-th/0507253}}, \href
  {http://dx.doi.org/10.1103/PhysRevD.72.104002}
  {\path{doi:10.1103/PhysRevD.72.104002}}.

\bibitem{Randono:2005up}
A.~Randono, {A note on parity violation and the Immirzi parameter}\href
  {http://arxiv.org/abs/hep-th/0510001} {\path{arXiv:hep-th/0510001}}.

\bibitem{Mercuri:2007ki}
S.~Mercuri, {From the Einstein-Cartan to the Ashtekar-Barbero canonical
  constraints, passing through the Nieh-Yan functional}, Phys. Rev. D77 (2008)
  024036.
\newblock \href {http://arxiv.org/abs/0708.0037} {\path{arXiv:0708.0037}}.

\bibitem{Mercuri:2006wb}
S.~Mercuri, {Nieh-Yan invariant and fermions in Ashtekar-Barbero- Immirzi
  formalism}\href {http://arxiv.org/abs/gr-qc/0610026}
  {\path{arXiv:gr-qc/0610026}}.

\bibitem{Mercuri:2009zi}
S.~Mercuri, {Peccei--Quinn mechanism in gravity and the nature of the
  Barbero--Immirzi parameter}\href {http://arxiv.org/abs/0902.2764}
  {\path{arXiv:0902.2764}}.

\bibitem{Mercuri:2009vk}
S.~Mercuri, {A possible topological interpretation of the Barbero-- Immirzi
  parameter}\href {http://arxiv.org/abs/0903.2270} {\path{arXiv:0903.2270}}.

\bibitem{Mercuri:2009zt}
S.~Mercuri, V.~Taveras, {Interaction of the Barbero--Immirzi Field with Matter
  and Pseudo-Scalar Perturbations}\href {http://arxiv.org/abs/0903.4407}
  {\path{arXiv:0903.4407}}.

\bibitem{Guarrera:2007tu}
D.~Guarrera, A.~J. Hariton, {Papapetrou energy-momentum tensor for Chern-Simons
  modified gravity}, Phys. Rev. D76 (2007) 044011.
\newblock \href {http://arxiv.org/abs/gr-qc/0702029}
  {\path{arXiv:gr-qc/0702029}}.

\bibitem{Gerlach:1979rw}
U.~H. Gerlach, U.~K. Sengupta, Gauge invariant perturbations on most general
  spherically symmetric space-times, Phys. Rev. D19 (1979) 2268--2272.

\bibitem{Gerlach:1980tx}
U.~H. Gerlach, U.~K. Sengupta, Gauge invariant coupled gravitational,
  acoustical, and electromagnetic modes on most general spherical space- times,
  Phys. Rev. D22 (1980) 1300--1312.

\bibitem{slow-rot}
N.~Yunes, F.~Pretorius, {Dynamical Chern-Simons Modified Gravity I: Spinning
  Black Holes in the Slow-Rotation Approximation}, Phys. Rev. D79 (2008)
  084043.
\newblock \href {http://arxiv.org/abs/0902.4669} {\path{arXiv:0902.4669}},
  \href {http://dx.doi.org/10.1103/PhysRevD.79.084043}
  {\path{doi:10.1103/PhysRevD.79.084043}}.

\bibitem{Campbell:1990ai}
B.~A. Campbell, M.~J. Duncan, N.~Kaloper, K.~A. Olive, {Axion hair for Kerr
  black holes}, Phys. Lett. B251 (1990) 34--38.
\newblock \href {http://dx.doi.org/10.1016/0370-2693(90)90227-W}
  {\path{doi:10.1016/0370-2693(90)90227-W}}.

\bibitem{Campbell:1991rz}
B.~A. Campbell, N.~Kaloper, K.~A. Olive, {Axion hair for dyon black holes},
  Phys. Lett. B263 (1991) 364--370.
\newblock \href {http://dx.doi.org/10.1016/0370-2693(91)90474-5}
  {\path{doi:10.1016/0370-2693(91)90474-5}}.

\bibitem{Reuter:1991cb}
M.~Reuter, {A Mechanism generating axion hair for Kerr black holes}, Class.
  Quant. Grav. 9 (1992) 751--756.

\bibitem{Kaloper:1991rw}
N.~Kaloper, {Lorentz Chern-Simons terms in Bianchi cosmologies and the cosmic
  no hair conjecture}, Phys. Rev. D44 (1991) 2380--2387.
\newblock \href {http://dx.doi.org/10.1103/PhysRevD.44.2380}
  {\path{doi:10.1103/PhysRevD.44.2380}}.

\bibitem{Dolgov:1987yp}
A.~D. Dolgov, I.~B. Khriplovich, V.~I. Zakharov, {CHIRAL BOSON ANOMALY IN A
  GRAVITATIONAL FIELD}, JETP Lett. 45 (1987) 651--653.

\bibitem{Dolgov:1988qx}
A.~D. Dolgov, I.~B. Khriplovich, A.~I. Vainshtein, V.~I. Zakharov, {PHOTONIC
  CHIRAL CURRENT AND ITS ANOMALY IN A GRAVITATIONAL FIELD}, Nucl. Phys. B315
  (1989) 138.
\newblock \href {http://dx.doi.org/10.1016/0550-3213(89)90451-3}
  {\path{doi:10.1016/0550-3213(89)90451-3}}.

\bibitem{Lukacs:1982}
B.~Luk{\'a}cs, Z.~Perj{\'e}s, Note on conformastat vacuum space-times, Phys.
  Lett. A88 (1982) 267--268.

\bibitem{Furtado:2009ji}
C.~Furtado, T.~Mariz, J.~R. Nascimento, A.~Y. Petrov, A.~F. Santos, {The
  G\'{o}del solution in the modified gravity}, Phys. Rev. D79 (2009) 124039.
\newblock \href {http://arxiv.org/abs/0906.0554} {\path{arXiv:0906.0554}},
  \href {http://dx.doi.org/10.1103/PhysRevD.79.124039}
  {\path{doi:10.1103/PhysRevD.79.124039}}.

\bibitem{Jordan:1960}
P.~Jordan, J.~Ehlers, W.~Kundt, {Strenge L{\"o}sungen der Feldgleichungen der
  Allgemeinen Relativit{\"a}tstheorie}, Akad. Wiss. Lit. (Mainz) Abhandl.
  Math.-Nat. Kl. 2 (1960) 21.

\bibitem{Aichelburg:1971dh}
P.~C. Aichelburg, R.~U. Sexl, On the gravitational field of a massless
  particle, Gen. Rel. Grav. 2 (1971) 303--312.

\bibitem{Lousto:1992th}
C.~O. Lousto, N.~G. Sanchez, {The Ultrarelativistic limit of the boosted
  Kerr-Newman geometry and the scattering of spin 1/2 particles}, Nucl. Phys.
  B383 (1992) 377--394.

\bibitem{Balasin:1995tb}
H.~Balasin, H.~Nachbagauer, {The Ultrarelativistic Kerr geometry and its energy
  momentum tensor}, Class. Quant. Grav. 12 (1995) 707--714.
\newblock \href {http://arxiv.org/abs/gr-qc/9405053}
  {\path{arXiv:gr-qc/9405053}}.

\bibitem{Balasin:1994kf}
H.~Balasin, H.~Nachbagauer, {Distributional energy momentum tensor of the
  Kerr-Newman space-time family}, Class. Quant. Grav. 11 (1994) 1453--1462.
\newblock \href {http://arxiv.org/abs/gr-qc/9312028}
  {\path{arXiv:gr-qc/9312028}}.

\bibitem{Konno:2007ze}
K.~Konno, T.~Matsuyama, S.~Tanda, {Does a black hole rotate in Chern-Simons
  modified gravity?}, Phys. Rev. D76 (2007) 024009.
\newblock \href {http://arxiv.org/abs/arXiv:0706.3080 [gr-qc]}
  {\path{arXiv:arXiv:0706.3080 [gr-qc]}}.

\bibitem{Sopuerta:2009iy}
C.~F. Sopuerta, N.~Yunes, {Extreme- and Intermediate-Mass Ratio Inspirals in
  Dynamical Chern-Simons Modified Gravity}\href
  {http://arxiv.org/abs/0904.4501} {\path{arXiv:0904.4501}}.

\bibitem{Blanchet:2002av}
L.~Blanchet, Gravitational radiation from post-newtonian sources and
  inspiralling compact binaries, Living Rev. Rel. 9 (2006) 4, and references
  therein.
\newblock \href {http://arxiv.org/abs/gr-qc/0202016}
  {\path{arXiv:gr-qc/0202016}}.

\bibitem{Bender}
C.~M. Bender, S.~A. Orszag, Advanced mathematical methods for scientists and
  engineers 1, Asymptotic methods and perturbation theory, Springer, New York,
  1999.

\bibitem{Kevorkian}
J.~Kevorkian, J.~D. Cole, Multiple scale and singular perturbation methods,
  Springer, New York, 1991, and references therein.

\bibitem{Yunes:2005nn}
N.~Yunes, W.~Tichy, B.~J. Owen, B.~Br{\"u}gmann, Binary black hole initial data
  from matched asymptotic expansions, Phys. Rev. D74 (2006) 104011.
\newblock \href {http://arxiv.org/abs/gr-qc/0503011}
  {\path{arXiv:gr-qc/0503011}}.

\bibitem{Yunes:2006iw}
N.~Yunes, W.~Tichy, Improved initial data for black hole binaries by asymptotic
  matching of post-newtonian and perturbed black hole solutions, Phys. Rev. D74
  (2006) 064013.
\newblock \href {http://arxiv.org/abs/gr-qc/0601046}
  {\path{arXiv:gr-qc/0601046}}.

\bibitem{Will:1999dq}
C.~M. Will, {Generation of post-Newtonian gravitational radiation via direct
  integration of the relaxed Einstein equations}, Prog. Theor. Phys. Suppl. 136
  (1999) 158--167.
\newblock \href {http://arxiv.org/abs/gr-qc/9910057}
  {\path{arXiv:gr-qc/9910057}}, \href {http://dx.doi.org/10.1143/PTPS.136.158}
  {\path{doi:10.1143/PTPS.136.158}}.

\bibitem{damour-effecing}
T.~Damour, Gravitational Radiation, Amsterdam, North-Holland, 1983, pp 59-144.

\bibitem{Lanczos:1922}
C.~Lanczos, Bemerkung zur de sitterschen welt, Phys. Zeits. 23 (1922) 539.

\bibitem{Lanczos:1924}
C.~Lanczos, Fl\''{a}chenhafte verteilung der materie in der einsteinschen
  gravitationsheorie, Ann der Phys. 74 (1924) 528.

\bibitem{Darmois:1927}
G.~Darmois, Les {\'e}equations de la gravitation einsteinienne, M{\'e}morial
  des sciences math{\'e}matique XXV.

\bibitem{Lichnerowicz}
A.~Lichnerowicz, Th\'{e}ories Relativistes de la Gravitation et de
  l'Electromagn\'{e}tisme, Paris: Masson, 1955.

\bibitem{Misner:1964}
C.~W. Misner, D.~H. Sharp, Relativistic equations for adiabatic spherically
  symmetric gravitational collapse, Phys. Rev. B136 (1964) 571.

\bibitem{Barrabes:1991ng}
C.~Barrabes, W.~Israel, Thin shells in general relativity and cosmology: The
  lightlike limit, Phys. Rev. D43 (1991) 1129--1142.

\bibitem{Israel:1966nc}
W.~Israel, Singular hypersurfaces and thin shells in gr, Nuovo Cimento B44
  (1966) 1--14.

\bibitem{poisson}
E.~Poisson, A Relativist's Toolkit: The mathematics of black-hole mechanics,
  Cambridge, New York, United States, 2004, and references therein.

\bibitem{Will:1993ns}
C.~M. Will, Theory and experiment in gravitational physics, Cambridge
  University Press, Cambridge, UK, 1993.

\bibitem{Schiff:1960gi}
L.~I. Schiff, Motion of a gyroscope according to einstein's theory of
  gravitation, Proc. Nat. Acad. Sci. 46 (1960) 871.

\bibitem{Nordtvedt:1968qs}
K.~Nordtvedt, Equivalence principle for massive bodies. 2. theory, Phys. Rev.
  169 (1968) 1017--1025.

\bibitem{1972ApJ...177..775N}
K.~J. {Nordtvedt}, C.~M. {Will}, {Conservation Laws and Preferred Frames in
  Relativistic Gravity. II. Experimental Evidence to Rule Out Preferred-Frame
  Theories of Gravity}, apj 177 (1972) 775--+.

\bibitem{1971ApJ...163..611W}
C.~M. {Will}, {Theoretical Frameworks for Testing Relativistic Gravity. II.
  Parametrized Post-Newtonian Hydrodynamics, and the Nordtvedt Effect}, apj 163
  (1971) 611--+.

\bibitem{1973ApJ...185...31W}
C.~M. {Will}, {Relativistic Gravity tn the Solar System. 111. Experimental
  Disproof of a Class of Linear Theories of Gravitation}, apj 185 (1973)
  31--42.

\bibitem{Will:2005va}
C.~M. Will, The confrontation between general relativity and experiment, Living
  Rev. Relativity 9 (2005) 3.
\newblock \href {http://arxiv.org/abs/gr-qc/0510072}
  {\path{arXiv:gr-qc/0510072}}.

\bibitem{1986bhmp.book.....T}
K.~S. {Thorne}, R.~H. {Price}, D.~A. {MacDonald}, {Black holes: The membrane
  paradigm}, Black Holes: The Membrane Paradigm, 1986.

\bibitem{Mashhoon:2003ax}
B.~Mashhoon, Gravitoelectromagnetism: A brief review\href
  {http://arxiv.org/abs/gr-qc/0311030} {\path{arXiv:gr-qc/0311030}}.

\bibitem{lrr-2006-3}
C.~M. Will, \href{http://www.livingreviews.org/lrr-2006-3}{The confrontation
  between general relativity and experiment}, Living Reviews in Relativity
  9~(3).
\newline\urlprefix\url{http://www.livingreviews.org/lrr-2006-3}

\bibitem{Sopuerta:2006wj}
C.~F. Sopuerta, N.~Yunes, P.~Laguna, Gravitational recoil from binary black
  hole mergers: The close-limit approximation, Phys. Rev. D74 (2006) 124010.
\newblock \href {http://arxiv.org/abs/astro-ph/0608600}
  {\path{arXiv:astro-ph/0608600}}.

\bibitem{Cunningham:1978cp}
C.~T. {Cunningham}, R.~H. {Price}, V.~{Moncrief}, {Radiation from collapsing
  relativistic stars. I - Linearized odd-parity radiation}, apj 224 (1978)
  643--667.

\bibitem{Zerilli:1970fj}
F.~J. Zerilli, Effective potential for even-parity regge-wheeler gravitational
  perturbation equations, prl 24 (1970) 737--738.

\bibitem{Moncrief:1974vm}
V.~Moncrief, Ann. Phys. (N.Y.) 88 (1974) 323.

\bibitem{Konno:2008np}
K.~Konno, T.~Matsuyama, Y.~Asano, S.~Tanda, {Flat rotation curves in
  Chern-Simons modified gravity}\href {http://arxiv.org/abs/0807.0679}
  {\path{arXiv:0807.0679}}.

\bibitem{Ciufolini:2007wx}
I.~Ciufolini, {Gravitomagnetism, Frame-Dragging and Lunar Laser Ranging}\href
  {http://arxiv.org/abs/0704.3338} {\path{arXiv:0704.3338}}.

\bibitem{Ciufolini:2004rq}
I.~Ciufolini, E.~C. Pavlis, {A confirmation of the general relativistic
  prediction of the Lense-Thirring effect}, Nature 431 (2004) 958--960.
\newblock \href {http://dx.doi.org/10.1038/nature03007}
  {\path{doi:10.1038/nature03007}}.

\bibitem{Ciufolini:2004gp}
I.~Ciufolini, A.~Paolozzi, D.~G. Currie, E.~C. Pavlis, {LARES/WEBER-SAT,
  frame-dragging and fundamental physics}Prepared for International Workshop on
  Frontier Science: Physics and Astrophysics in Space, Frascati and Rome,
  Italy, 14-19 Jun 2004.

\bibitem{Konno:2009kg}
K.~Konno, T.~Matsuyama, S.~Tanda, {Rotating black hole in extended Chern-Simons
  modified gravity}\href {http://arxiv.org/abs/0902.4767}
  {\path{arXiv:0902.4767}}.

\bibitem{Lue:1998mq}
A.~Lue, L.-M. Wang, M.~Kamionkowski, Cosmological signature of new
  parity-violating interactions, Phys. Rev. Lett. 83 (1999) 1506--1509.
\newblock \href {http://arxiv.org/abs/astro-ph/9812088}
  {\path{arXiv:astro-ph/9812088}}.

\bibitem{Cantcheff:2008qn}
M.~B. Cantcheff, {Einstein-Cartan formulation of Chern-Simons Lorentz Violating
  Gravity and Holographic Structure}, Phys. Rev. D78 (2008) 025002.
\newblock \href {http://arxiv.org/abs/0801.0067} {\path{arXiv:0801.0067}},
  \href {http://dx.doi.org/10.1103/PhysRevD.78.025002}
  {\path{doi:10.1103/PhysRevD.78.025002}}.

\bibitem{Romano:1991up}
J.~D. Romano, {Geometrodynamics versus connection dynamics (in the context of
  (2+1) and (3+1) gravity}, Gen. Rel. Grav. 25 (1993) 759--854.
\newblock \href {http://arxiv.org/abs/gr-qc/9303032}
  {\path{arXiv:gr-qc/9303032}}.

\bibitem{Randono:2007mr}
A.~C. Randono, {In Search of Quantum de Sitter Space: Generalizing the Kodama
  State}\href {http://arxiv.org/abs/0709.2905} {\path{arXiv:0709.2905}}.

\bibitem{Kostelecky:2003fs}
V.~A. Kostelecky, Gravity, lorentz violation, and the standard model, Phys.
  Rev. D69 (2004) 105009.
\newblock \href {http://arxiv.org/abs/hep-th/0312310}
  {\path{arXiv:hep-th/0312310}}.

\bibitem{Belyaev:2007fn}
A.~S. Belyaev, I.~L. Shapiro, M.~A.~B. do~Vale, {Torsion phenomenology at the
  LHC}, Phys. Rev. D75 (2007) 034014.
\newblock \href {http://arxiv.org/abs/hep-ph/0701002}
  {\path{arXiv:hep-ph/0701002}}, \href
  {http://dx.doi.org/10.1103/PhysRevD.75.034014}
  {\path{doi:10.1103/PhysRevD.75.034014}}.

\bibitem{Kostelecky:2007kx}
V.~A. Kostelecky, N.~Russell, J.~Tasson, {New Constraints on Torsion from
  Lorentz Violation}, Phys. Rev. Lett. 100 (2008) 111102.
\newblock \href {http://arxiv.org/abs/0712.4393} {\path{arXiv:0712.4393}}.

\bibitem{GPBwebsite}
{\href{http://einstein.stanford.edu}{{\tt http://einstein.stanford.edu}}}.

\bibitem{Will:2002ma}
C.~M. Will, {Covariant Calculation of General Relativistic Effects in an
  Orbiting Gyroscope Experiment}, Phys. Rev. D67 (2003) 062003.
\newblock \href {http://arxiv.org/abs/gr-qc/0212069}
  {\path{arXiv:gr-qc/0212069}}, \href
  {http://dx.doi.org/10.1103/PhysRevD.67.062003}
  {\path{doi:10.1103/PhysRevD.67.062003}}.

\bibitem{Yunes:2008ua}
N.~Yunes, D.~N. Spergel, {Double Binary Pulsar Test of Dynamical Chern-Simons
  Modified Gravity}\href {http://arxiv.org/abs/0810.5541}
  {\path{arXiv:0810.5541}}.

\bibitem{Jones:2001yg}
D.~I. Jones, N.~Andersson, {Gravitational waves from freely precessing neutron
  stars}, Mon. Not. Roy. Astron. Soc. 331 (2002) 203.
\newblock \href {http://arxiv.org/abs/gr-qc/0106094}
  {\path{arXiv:gr-qc/0106094}}, \href
  {http://dx.doi.org/10.1046/j.1365-8711.2002.05180.x}
  {\path{doi:10.1046/j.1365-8711.2002.05180.x}}.

\bibitem{tegpriv}
M.~Tegmark, S.~Hughes, private communication.

\bibitem{Iorio:1999ru}
L.~Iorio, {An alternative derivation of the periodical and secular
  Lense-Thirring effect on the orbit of a test body}, Nuovo Cim. B116 (2001)
  777--789.
\newblock \href {http://arxiv.org/abs/gr-qc/9908080}
  {\path{arXiv:gr-qc/9908080}}.

\bibitem{Murray-Book}
C.~D. Murray, S.~F. Dermott, Solar System Dynamics, Cambridge University Press,
  Cambridge, 1999.

\bibitem{Richard}
R.~O'Shaughnessy, private communication.

\bibitem{Brunfiel}
G.~Brunfiel, Nature 444 (2006) 978.

\bibitem{Burgay:2003jj}
M.~Burgay, et~al., {An increased estimate of the merger rate of double neutron
  stars from observations of a highly relativistic system}, Nature. 426 (2003)
  531--533.
\newblock \href {http://arxiv.org/abs/astro-ph/0312071}
  {\path{arXiv:astro-ph/0312071}}, \href
  {http://dx.doi.org/10.1038/nature02124} {\path{doi:10.1038/nature02124}}.

\bibitem{Hulse:1974eb}
R.~A. Hulse, J.~H. Taylor, {Discovery of a pulsar in a binary system},
  Astrophys. J. 195 (1975) L51--L53.

\bibitem{satbook}
O.~Montenbruck, E.~Gill, Satellite Orbits. Models, methods, applications.,
  Springer, Germany, 2000.

\bibitem{Kramer:2006nb}
M.~Kramer, et~al., {Tests of general relativity from timing the double pulsar},
  Science 314 (2006) 97--102.
\newblock \href {http://arxiv.org/abs/astro-ph/0609417}
  {\path{arXiv:astro-ph/0609417}}, \href
  {http://dx.doi.org/10.1126/science.1132305}
  {\path{doi:10.1126/science.1132305}}.

\bibitem{1970ApJ...159..379R}
V.~C. {Rubin}, W.~K.~J. {Ford}, {Rotation of the Andromeda Nebula from a
  Spectroscopic Survey of Emission Regions}, apj 159 (1970) 379--+.

\bibitem{Rubin:1980zd}
V.~C. Rubin, N.~Thonnard, W.~K. Ford, Jr., {Rotational properties of 21 SC
  galaxies with a large range of luminosities and radii, from NGC 4605 /R =
  4kpc/ to UGC 2885 /R = 122 kpc/}, Astrophys. J. 238 (1980) 471.

\bibitem{Bender:1998}
P.~Bender, et~al., {Laser Interferometer Space Antenna for the Detection and
  Observation of Gravitational Waves: An International Project in the Field of
  Fundamental Physics in Space}, {LISA Pre-Phase A Report}, Max-Planck-Institut
  f{\"u}r Quantenoptik, Garching, mPQ~233 (July 1998).

\bibitem{Danzmann:1997hm}
K.~Danzmann, {LISA - an ESA cornerstone mission for a gravitational wave
  observatory}, Class. Quant. Grav. 14 (1997) 1399--1404.
\newblock \href {http://dx.doi.org/10.1088/0264-9381/14/6/002}
  {\path{doi:10.1088/0264-9381/14/6/002}}.

\bibitem{Danzmann:2003tv}
K.~Danzmann, A.~Rudiger, {LISA technology - Concept, status, prospects}, Class.
  Quant. Grav. 20 (2003) S1--S9.

\bibitem{Sumner:2004}
T.~J. Sumner, D.~N.~A. Shaul, {The Observations of Gravitational Waves from
  Space using LISA}, MPLA 19~(11) (2004) 785--800.

\bibitem{Merkowitz:2006yh}
e.~. Merkowitz, S.~M., e.~. Livas, J.~C., {Laser interferometer space antenna.
  Proceedings, 6th International LISA Symposium, Greenbelt, USA, June 19-23,
  2006}Prepared for 6th International LISA Symposium, Greenbelt, Maryland,
  19-23 Jun 2006.

\bibitem{Finn:1992wt}
L.~S. Finn, {Detection, measurement and gravitational radiation}, Phys. Rev.
  D46 (1992) 5236--5249.
\newblock \href {http://arxiv.org/abs/gr-qc/9209010}
  {\path{arXiv:gr-qc/9209010}}, \href
  {http://dx.doi.org/10.1103/PhysRevD.46.5236}
  {\path{doi:10.1103/PhysRevD.46.5236}}.

\bibitem{Finn:1992xs}
L.~S. Finn, D.~F. Chernoff, {Observing binary inspiral in gravitational
  radiation: One interferometer}, Phys. Rev. D47 (1993) 2198--2219.
\newblock \href {http://arxiv.org/abs/gr-qc/9301003}
  {\path{arXiv:gr-qc/9301003}}, \href
  {http://dx.doi.org/10.1103/PhysRevD.47.2198}
  {\path{doi:10.1103/PhysRevD.47.2198}}.

\bibitem{Carroll:1990vb}
S.~M. Carroll, G.~B. Field, R.~Jackiw, Limits on a {L}orentz and parity
  violating modification of electrodynamics, Phys. Rev. D41 (1990) 1231.

\bibitem{Carroll:1998zi}
S.~M. Carroll, Quintessence and the rest of the world, Phys. Rev. Lett. 81
  (1998) 3067--3070.
\newblock \href {http://arxiv.org/abs/arXiv:astro-ph/9806099}
  {\path{arXiv:arXiv:astro-ph/9806099}}.

\bibitem{Saito:2007kt}
S.~Saito, K.~Ichiki, A.~Taruya, {Probing polarization states of primordial
  gravitational waves with CMB anisotropies}, JCAP 0709 (2007) 002.
\newblock \href {http://arxiv.org/abs/0705.3701} {\path{arXiv:0705.3701}},
  \href {http://dx.doi.org/10.1088/1475-7516/2007/09/002}
  {\path{doi:10.1088/1475-7516/2007/09/002}}.

\bibitem{Mukhanov:1990me}
V.~F. Mukhanov, H.~A. Feldman, R.~H. Brandenberger, {Theory of cosmological
  perturbations. Part 1. Classical perturbations. Part 2. Quantum theory of
  perturbations. Part 3. Extensions}, Phys. Rept. 215 (1992) 203--333.
\newblock \href {http://dx.doi.org/10.1016/0370-1573(92)90044-Z}
  {\path{doi:10.1016/0370-1573(92)90044-Z}}.

\bibitem{Martin:1999wa}
J.~Martin, D.~J. Schwarz, {The precision of slow-roll predictions for the CMBR
  anisotropies}, Phys. Rev. D62 (2000) 103520.
\newblock \href {http://arxiv.org/abs/astro-ph/9911225}
  {\path{arXiv:astro-ph/9911225}}, \href
  {http://dx.doi.org/10.1103/PhysRevD.62.103520}
  {\path{doi:10.1103/PhysRevD.62.103520}}.

\bibitem{Grad}
I.~S. Gradshteyn, J.~M. Ryshik, {Table of Integrals, Series and Products},
  Academic Press, New York, 1980.

\bibitem{Martin:2000ei}
J.~Martin, D.~J. Schwarz, {New exact solutions for inflationary cosmological
  perturbations}, Phys. Lett. B500 (2001) 1--7.
\newblock \href {http://arxiv.org/abs/astro-ph/0005542}
  {\path{arXiv:astro-ph/0005542}}, \href
  {http://dx.doi.org/10.1016/S0370-2693(01)00071-5}
  {\path{doi:10.1016/S0370-2693(01)00071-5}}.

\bibitem{Martin:2000xs}
J.~Martin, R.~H. Brandenberger, {The trans-Planckian problem of inflationary
  cosmology}, Phys. Rev. D63 (2001) 123501.
\newblock \href {http://arxiv.org/abs/hep-th/0005209}
  {\path{arXiv:hep-th/0005209}}, \href
  {http://dx.doi.org/10.1103/PhysRevD.63.123501}
  {\path{doi:10.1103/PhysRevD.63.123501}}.

\bibitem{Lemoine:2001ar}
M.~Lemoine, M.~Lubo, J.~Martin, J.-P. Uzan, {The stress-energy tensor for
  trans-Planckian cosmology}, Phys. Rev. D65 (2002) 023510.
\newblock \href {http://arxiv.org/abs/hep-th/0109128}
  {\path{arXiv:hep-th/0109128}}, \href
  {http://dx.doi.org/10.1103/PhysRevD.65.023510}
  {\path{doi:10.1103/PhysRevD.65.023510}}.

\bibitem{Choi:1999zy}
K.~Choi, J.-c. Hwang, K.~W. Hwang, {String theoretic axion coupling and the
  evolution of cosmic structures}, Phys. Rev. D61 (2000) 084026.
\newblock \href {http://arxiv.org/abs/hep-ph/9907244}
  {\path{arXiv:hep-ph/9907244}}, \href
  {http://dx.doi.org/10.1103/PhysRevD.61.084026}
  {\path{doi:10.1103/PhysRevD.61.084026}}.

\bibitem{Kamionkowski:1996ks}
M.~Kamionkowski, A.~Kosowsky, A.~Stebbins, {Statistics of Cosmic Microwave
  Background Polarization}, Phys. Rev. D55 (1997) 7368--7388.
\newblock \href {http://arxiv.org/abs/astro-ph/9611125}
  {\path{arXiv:astro-ph/9611125}}, \href
  {http://dx.doi.org/10.1103/PhysRevD.55.7368}
  {\path{doi:10.1103/PhysRevD.55.7368}}.

\bibitem{Miller:2009pt}
N.~J. Miller, M.~Shimon, B.~G. Keating, {CMB Polarization Systematics Due to
  Beam Asymmetry: Impact on Cosmological Birefringence}\href
  {http://arxiv.org/abs/0903.1116} {\path{arXiv:0903.1116}}.

\bibitem{Cabella:2007br}
P.~Cabella, P.~Natoli, J.~Silk, {Constraints on CPT violation from WMAP three
  year polarization data: A wavelet analysis}, Phys. Rev. D76 (2007) 123014.
\newblock \href {http://arxiv.org/abs/0705.0810} {\path{arXiv:0705.0810}},
  \href {http://dx.doi.org/10.1103/PhysRevD.76.123014}
  {\path{doi:10.1103/PhysRevD.76.123014}}.

\bibitem{Xia:2007qs}
J.-Q. Xia, H.~Li, X.-l. Wang, X.-m. Zhang, {Testing CPT Symmetry with CMB
  Measurements}, Astron. Astrophys. 483 (2008) 715--718.
\newblock \href {http://arxiv.org/abs/0710.3325} {\path{arXiv:0710.3325}},
  \href {http://dx.doi.org/10.1051/0004-6361:200809410}
  {\path{doi:10.1051/0004-6361:200809410}}.

\bibitem{Feng:2006dp}
B.~Feng, M.~Li, J.-Q. Xia, X.~Chen, X.~Zhang, {Searching for CPT violation with
  WMAP and BOOMERANG}, Phys. Rev. Lett. 96 (2006) 221302.
\newblock \href {http://arxiv.org/abs/astro-ph/0601095}
  {\path{arXiv:astro-ph/0601095}}, \href
  {http://dx.doi.org/10.1103/PhysRevLett.96.221302}
  {\path{doi:10.1103/PhysRevLett.96.221302}}.

\bibitem{Kamionkowski:2008fp}
M.~Kamionkowski, {How to De-Rotate the Cosmic Microwave Background
  Polarization}, Phys. Rev. Lett. 102 (2009) 111302.
\newblock \href {http://arxiv.org/abs/0810.1286} {\path{arXiv:0810.1286}},
  \href {http://dx.doi.org/10.1103/PhysRevLett.102.111302}
  {\path{doi:10.1103/PhysRevLett.102.111302}}.

\bibitem{Spergel:2003cb}
D.~N. Spergel, et~al., {First Year Wilkinson Microwave Anisotropy Probe (WMAP)
  Observations: Determination of Cosmological Parameters}, Astrophys. J. Suppl.
  148 (2003) 175.
\newblock \href {http://arxiv.org/abs/astro-ph/0302209}
  {\path{arXiv:astro-ph/0302209}}, \href {http://dx.doi.org/10.1086/377226}
  {\path{doi:10.1086/377226}}.

\bibitem{Huet:1994jb}
P.~Huet, E.~Sather, {Electroweak baryogenesis and standard model CP violation},
  Phys. Rev. D51 (1995) 379--394.
\newblock \href {http://arxiv.org/abs/hep-ph/9404302}
  {\path{arXiv:hep-ph/9404302}}, \href
  {http://dx.doi.org/10.1103/PhysRevD.51.379}
  {\path{doi:10.1103/PhysRevD.51.379}}.

\bibitem{Kuzmin:1985mm}
V.~A. Kuzmin, V.~A. Rubakov, M.~E. Shaposhnikov, {On the Anomalous Electroweak
  Baryon Number Nonconservation in the Early Universe}, Phys. Lett. B155 (1985)
  36.
\newblock \href {http://dx.doi.org/10.1016/0370-2693(85)91028-7}
  {\path{doi:10.1016/0370-2693(85)91028-7}}.

\bibitem{Fukugita:1986hr}
M.~Fukugita, T.~Yanagida, {Baryogenesis Without Grand Unification}, Phys. Lett.
  B174 (1986) 45.
\newblock \href {http://dx.doi.org/10.1016/0370-2693(86)91126-3}
  {\path{doi:10.1016/0370-2693(86)91126-3}}.

\bibitem{Sakharov:1967dj}
A.~D. Sakharov, {Violation of CP Invariance, c Asymmetry, and Baryon Asymmetry
  of the Universe}, Pisma Zh. Eksp. Teor. Fiz. 5 (1967) 32--35.

\bibitem{Dimopoulos:2005ac}
S.~Dimopoulos, S.~Kachru, J.~McGreevy, J.~G. Wacker, {N-flation}, JCAP 0808
  (2008) 003.
\newblock \href {http://arxiv.org/abs/hep-th/0507205}
  {\path{arXiv:hep-th/0507205}}, \href
  {http://dx.doi.org/10.1088/1475-7516/2008/08/003}
  {\path{doi:10.1088/1475-7516/2008/08/003}}.

\bibitem{McAllister:2007bg}
L.~McAllister, E.~Silverstein, {String Cosmology: A Review}, Gen. Rel. Grav. 40
  (2008) 565--605.
\newblock \href {http://arxiv.org/abs/0710.2951} {\path{arXiv:0710.2951}},
  \href {http://dx.doi.org/10.1007/s10714-007-0556-6}
  {\path{doi:10.1007/s10714-007-0556-6}}.

\bibitem{LL}
R.~Lyddle, D.~H. Lyth, {Cosmological Inflation and Large-Scale Structure},
  Cambridge University Press, Cambridge, 2000.

\bibitem{Gibbons:1993hg}
G.~W. Gibbons, A.~R. Steif, {Anomalous fermion production in gravitational
  collapse}, Phys. Lett. B314 (1993) 13--20.
\newblock \href {http://arxiv.org/abs/gr-qc/9305018}
  {\path{arXiv:gr-qc/9305018}}, \href
  {http://dx.doi.org/10.1016/0370-2693(93)91315-E}
  {\path{doi:10.1016/0370-2693(93)91315-E}}.

\bibitem{Linde}
A.~Linde, private communication.

\bibitem{Lyth:2005jf}
D.~H. Lyth, C.~Quimbay, Y.~Rodriguez, {Leptogenesis and tensor polarisation
  from a gravitational Chern-Simons term}, JHEP 03 (2005) 016.
\newblock \href {http://arxiv.org/abs/hep-th/0501153}
  {\path{arXiv:hep-th/0501153}}, \href
  {http://dx.doi.org/10.1088/1126-6708/2005/03/016}
  {\path{doi:10.1088/1126-6708/2005/03/016}}.

\bibitem{Fischler:2007tj}
W.~Fischler, S.~Paban, {Leptogenesis from Pseudo-Scalar Driven Inflation}, JHEP
  10 (2007) 066.
\newblock \href {http://arxiv.org/abs/0708.3828} {\path{arXiv:0708.3828}},
  \href {http://dx.doi.org/10.1088/1126-6708/2007/10/066}
  {\path{doi:10.1088/1126-6708/2007/10/066}}.

\bibitem{Richter:2006fa}
M.~C. Richter, G.~B. Tupper, R.~D. Viollier, {A Symbiotic Scenario for the
  Rapid Formation of Supermassive Black Holes}, JCAP 0612 (2006) 015.
\newblock \href {http://arxiv.org/abs/astro-ph/0611552}
  {\path{arXiv:astro-ph/0611552}}.

\end{thebibliography}

\end{document}